# Stratigraphy of Aeolis Dorsa, Mars: stratigraphic context of the great river deposits

Edwin S. Kite[1,2,3], Alan D. Howard[4], Antoine S. Lucas[5], John C. Armstrong[6], Oded Aharonson[7], and Michael P. Lamb[8]

[1] Princeton University, Geoscience Department. [2]Princeton University, Astrophysics Department. [3]University of Chicago (kite@uchicago.edu). [4] University of Virginia. [5] Université Paris 7. [6] Weber State University.[7] Weizmann Institute of Science. [8] California Institute of Technology.

## Abstract.

Unraveling the stratigraphic record is the key to understanding ancient climate and past climate changes on Mars (Grotzinger et al. 2011). Stratigraphic records of river deposits hold particular promise because rain or snowmelt must exceed infiltration plus evaporation to allow sediment transport by rivers. Therefore, river deposits when placed in stratigraphic order could constrain the number, magnitudes, and durations of the wettest (and presumably most habitable) climates in Mars history. We use crosscutting relationships to establish the stratigraphic context of river and alluvial-fan deposits in the Aeolis Dorsa sedimentary basin, 10° E of Gale crater. At Aeolis Dorsa, wind erosion has exhumed a stratigraphic section of sedimentary rocks consisting of at least four unconformity-bounded rock packages, recording three or more distinct episodes of surface runoff. Early deposits (>700m thick) are embayed by river deposits (>400m thick), which are in turn unconformably draped by fan-shaped deposits (<100m thick) which we interpret as alluvial fans. Yardang-forming layered deposits (>900 m thick) unconformably drape all previous deposits.

River deposits embay a dissected landscape formed of sedimentary rock. The river deposits are eroding out of at least two distinguishable units. There is evidence for pulses of erosion during the interval of river deposition. The total interval spanned by river deposits is >($1 \times 10^6 – 2 \times 10^7$) yr, and this is extended if we include alluvial-fan deposits. Alluvial-fan deposits unconformably postdate thrust faults which crosscut the river deposits. This relationship suggests a relatively dry interval of >$4 \times 10^7$ yr after the river deposits formed and before the fan-shaped deposits formed, based on probability arguments. Yardang-forming layered deposits unconformably postdate all of the earlier deposits. They contain rhythmite and their induration suggests a damp or wet (near-) surface environment. The time gap between the end of river deposition and the onset of yardang-forming layered deposits is constrained to >$1 \times 10^8$ yr by the high density of impact craters embedded at the unconformity. The time gap between the end of alluvial-fan deposition and the onset of yardang-forming layered deposits was at least long enough for wind-induced saltation abrasion to erode 20-30 m into the alluvial-fan deposits. We correlate the yardang-forming layered deposits to the upper layers of Gale crater's mound (Mt. Sharp / Aeolis Mons), and the fan-shaped deposits to Peace Vallis fan in Gale crater. Alternations between periods of low mean obliquity and periods of high mean obliquity may have modulated erosion-deposition cycling in Aeolis. This is consistent with the results from an ensemble of simulations of Solar System orbital evolution and the resulting history of the obliquity of Mars. Almost all of our simulations produce one or more intervals of continuously low mean Mars obliquity that are long enough to match our Aeolis Dorsa unconformity data.

**Contents**

# 1. Introduction.

An exceptionally dense concentration of exceptionally well-preserved river deposits exists in the Aeolis Dorsa sedimentary basin, 10° E of Gale crater (Burr et al. 2010). The range of river-deposit styles, the high frequency of interbedded impact craters, and evidence for major erosional episodes interspersed with deposition all suggest that Aeolis Dorsa's time series of constraints on climate is unusually long and complete (Kite et al. 2013a, Kite et al. 2014). Within a single sedimentary basin such as Aeolis Dorsa, wet-climate events can be placed in time order using crosscutting relationships. This is an advantage, because it sidesteps the uncertainties involved in using crater retention ages and lithostratigraphy to correlate small climate-sensitive deposits between sedimentary basins. However, the stratigraphy of Aeolis Dorsa has not been clearly characterized at the ~100m stratigraphic scale needed to isolate river-forming climate episodes. The number and relative timing of river-forming climate episodes therefore remains an open question. Nor has it been clearly demonstrated that the river deposits and the rocks that surround them are of comparable age. The fill of incised valleys can greatly postdate the age of the rocks into which the valleys are cut (Christie-Blick et al. 1990, Cardenas & Mohrig 2014, Johnson 2009). Are the Aeolis Dorsa river deposits embedded within the basin stratigraphy, or are they much younger incised-valley fill? To recover a detailed paleohydrologic record from orbiter observations of Aeolis Dorsa, we need to place these deposits in a stratigraphic framework.

Therefore, our aim in this study is to identify river-deposit-containing basin-scale geologic units and characterize their distribution in space and time.

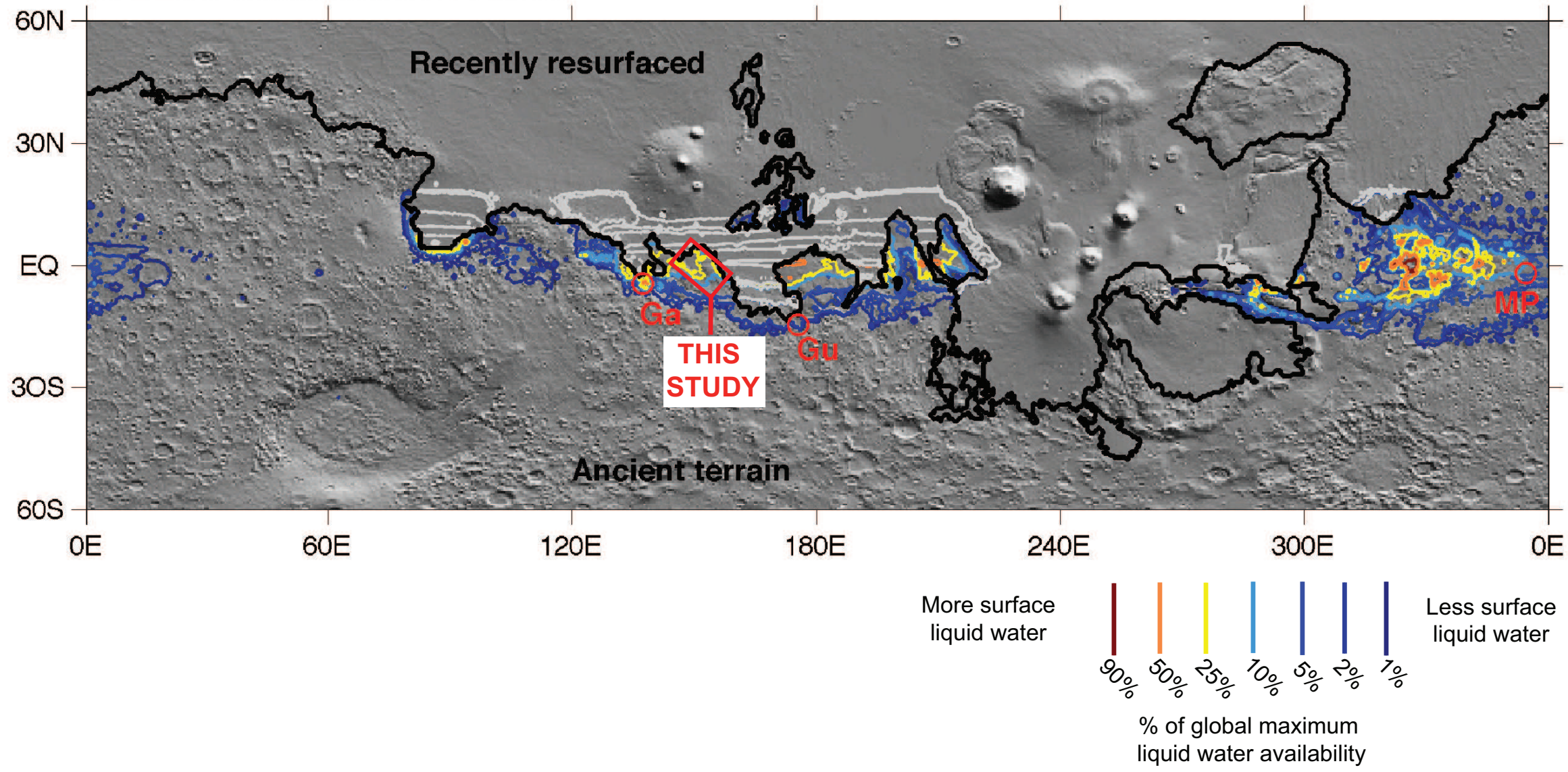

**Figure 1.** Our study area (red rectangle) in global context. Background is shaded relief Mars Orbiter Laser Altimeter (MOLA) topography, illuminated from top right. Colored contours show surface liquid water predictions from a seasonal-melting model (Kite et al. 2013b). Colors correspond to percentages of the global maximum liquid-water availability: 1% (dark blue), 5%, 10%, 25%, 50%, 75% and 90% (red). The black line marks the border of the area of recently-resurfaced terrain, and approximately corresponds to the hemispheric dichotomy. Within the area of recently-resurfaced terrain, liquid-water availability contours are grayed out. Landing sites of long-range rovers are shown by red circles: **Ga** = Gale (Mars Science Laboratory Curiosity rover); **Gu** = Gusev (Spirit, Mars Exploration Rover MER-A); **MP** = Meridiani Planum (Opportunity, Mars Exploration Rover MER-B). This figure is modified from Figure 14b in Kite et al. 2013b.

## 1.1. Geologic setting.

Aeolis Dorsa's river deposits form part of a sedimentary wedge that thins northward away from the equatorial dichotomy boundary scarp (Fig. 1) (Irwin et al. 2004, Irwin & Watters 2010). The dichotomy boundary (~5 km relief) was formed very early in Mars history (Andrews-Hanna et al. 2008b, Marinova et al. 2008, Nimmo et al. 2008, Irwin & Watters 2010, Andrews-Hanna 2012). Aeolis Dorsa sediments are thickest towards the south (near where they contact the equatorial dichotomy boundary scarp), and thin northward (where they contact basaltic lava plains), defining a sedimentary wedge. Sediments in our study area consist of two rises (Aeolis and Zephyria Plana), separated by a trough with abundant river deposits (Aeolis Dorsa) (Fig. 2a). The deposits are not older than Late Noachian based on regional geologic mapping by Irwin & Watters (2010). Our focus on the geologic context of the rivers is complementary to broader regional studies (Zimbelman & Scheidt 2012).

# 2. Photogeologic lithofacies.

In §2 we describe photogeologic lithofacies within Aeolis Dorsa in the order of their timing relative to key unconformities that are described in §3. Correlations are made on the basis of deposit characteristics, but in most cases deposits with similar characteristics have similar elevations or similar relative elevations. Details of our approach are given in the Supplementary Methods.

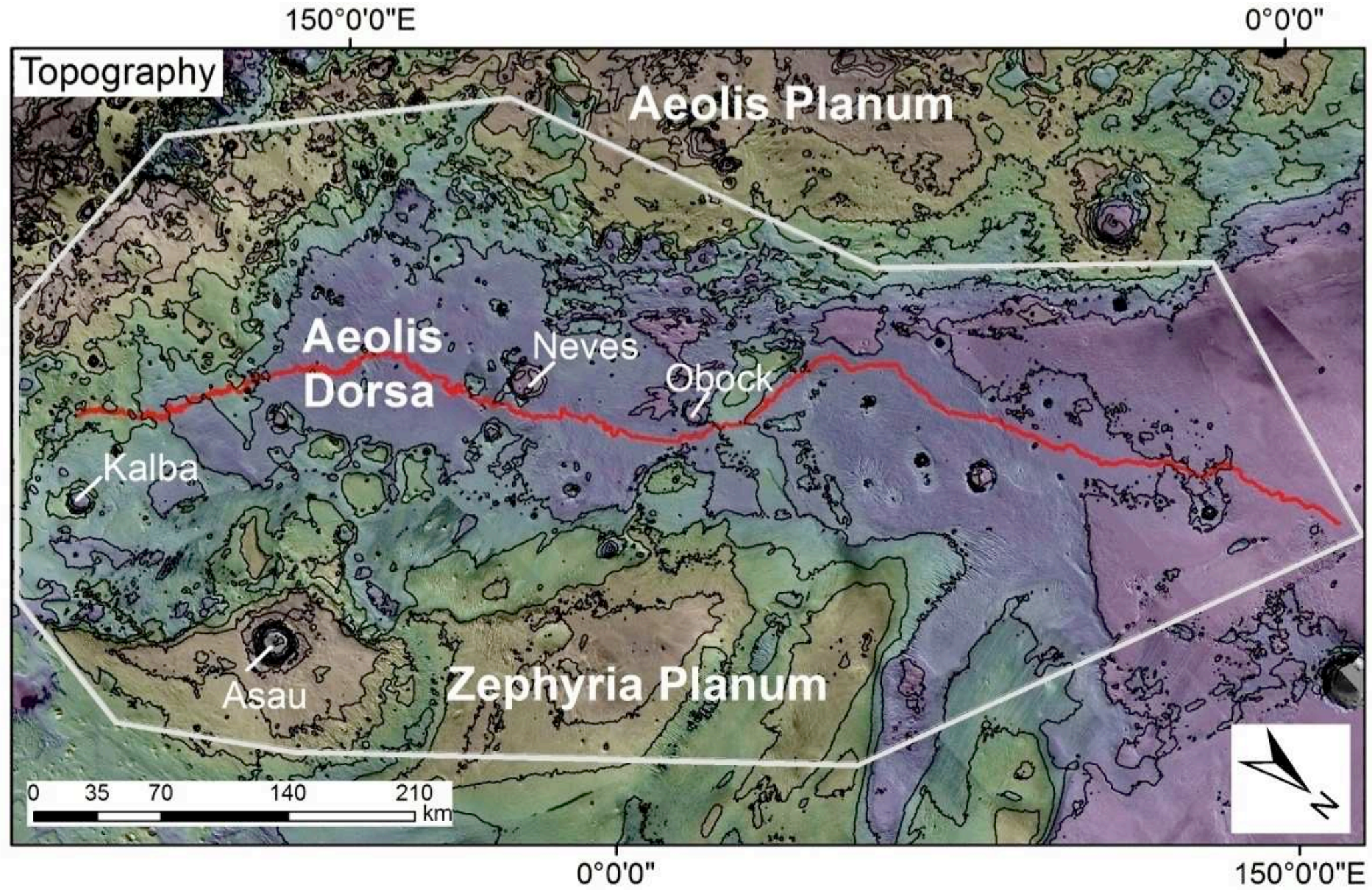

a)

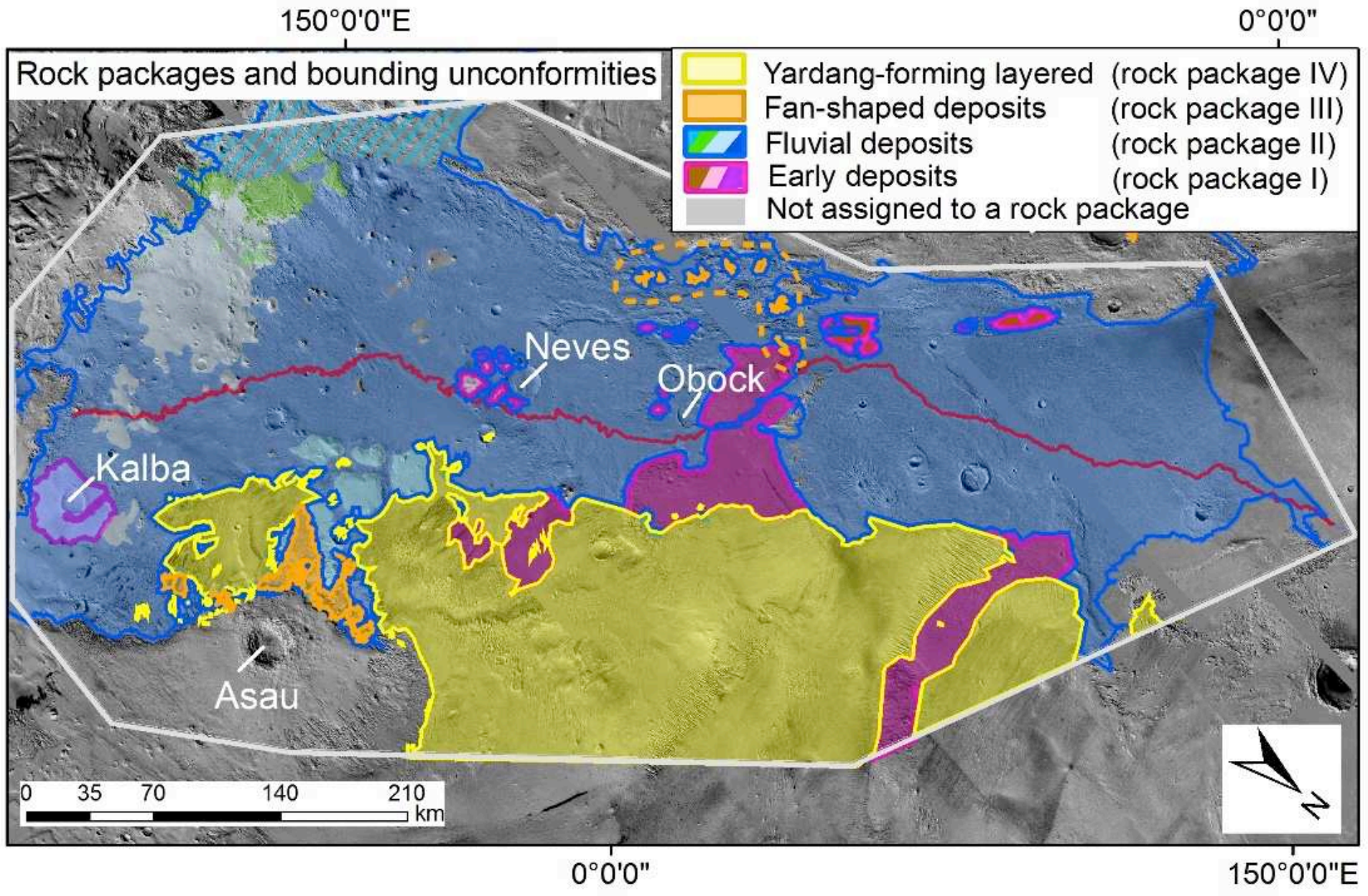
150°0'0"E
0°0'0"
Rock packages and bounding unconformities
Yardang-forming layered (rock package IV)
Fan-shaped deposits (rock package III)
Fluvial deposits (rock package II)
Early deposits (rock package I)
Not assigned to a rock package
Neves
Obock
Kalba
Asau
0 35 70 140 210 km
0°0'0"
150°0'0"E

b)

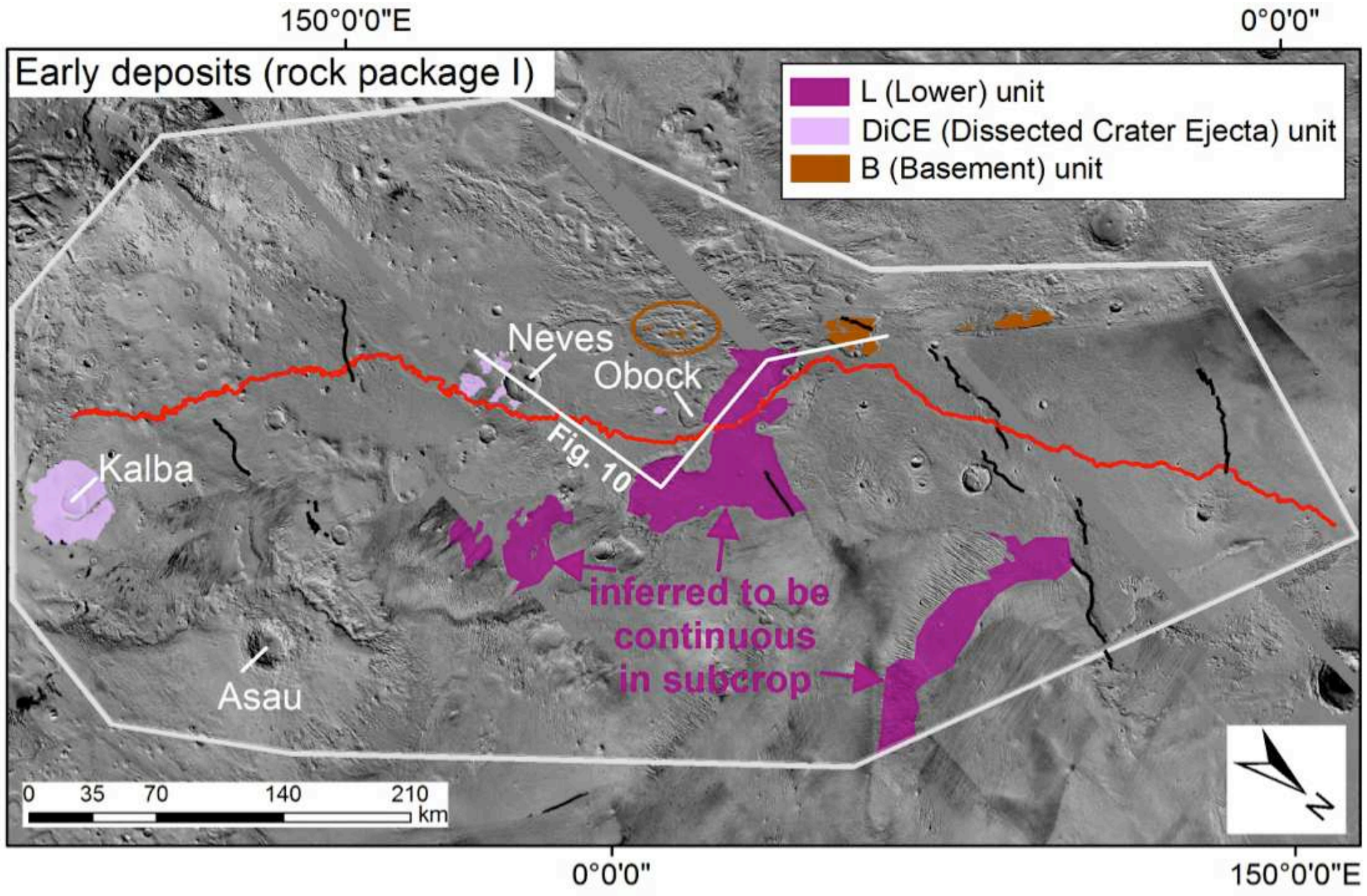
150°0'0"E
0°0'0"
Early deposits (rock package I)
L (Lower) unit
DiCE (Dissected Crater Ejecta) unit
B (Basement) unit
Neves
Obock
Fig. 10
Kalba
inferred to be continuous in subcrop
Asau
0 35 70 140 210 km
0°0'0"
150°0'0"E

c)

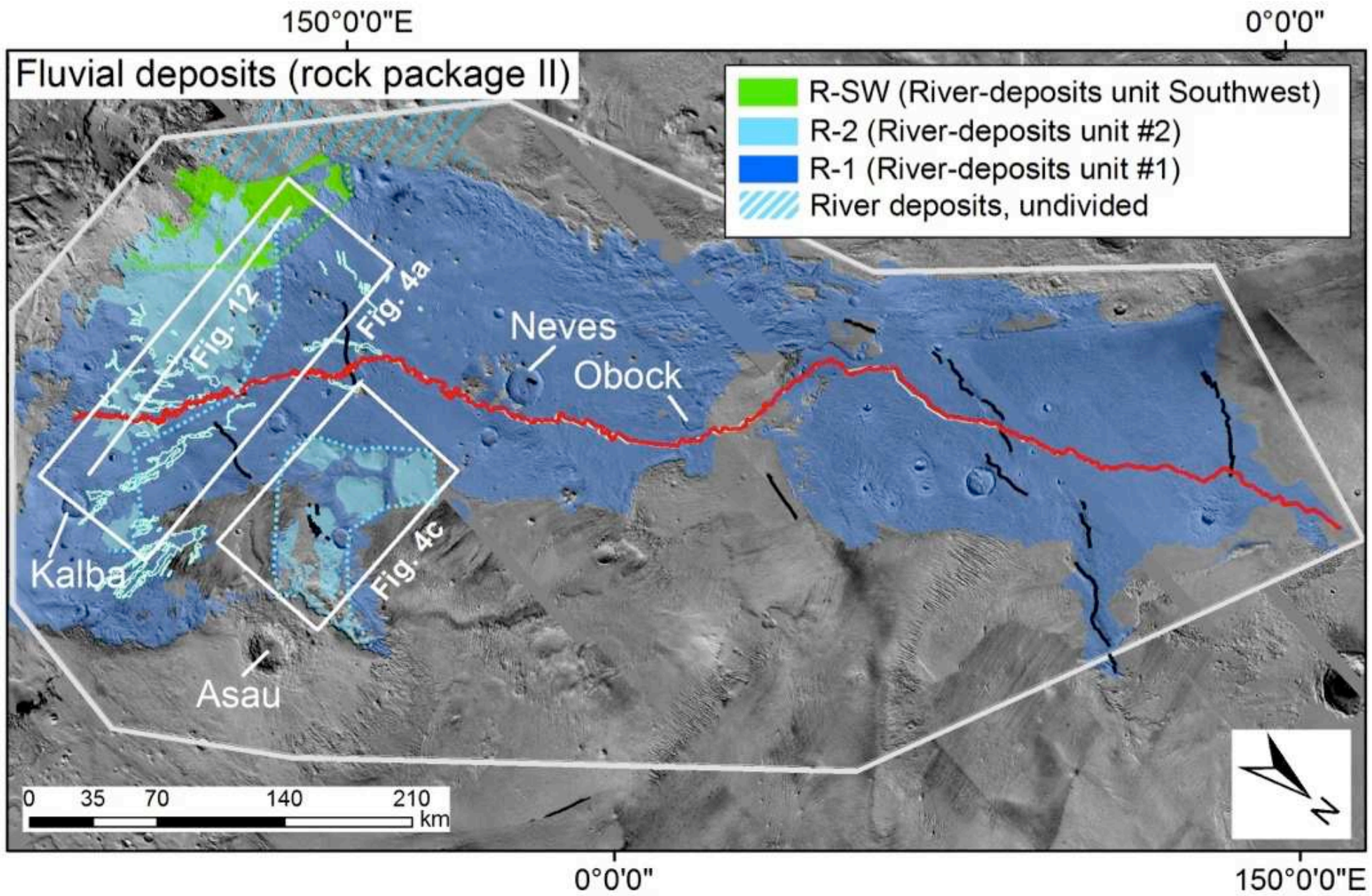
150°0'0"E
0°0'0"
Fluvial deposits (rock package II)
R-SW (River-deposits unit Southwest)
R-2 (River-deposits unit #2)
R-1 (River-deposits unit #1)
River deposits, undivided
Fig. 12
Fig. 4a
Fig. 4c
Neves
Obock
Kalba
Asau
0 35 70 140 210
km
0°0'0"
150°0'0"E

d)

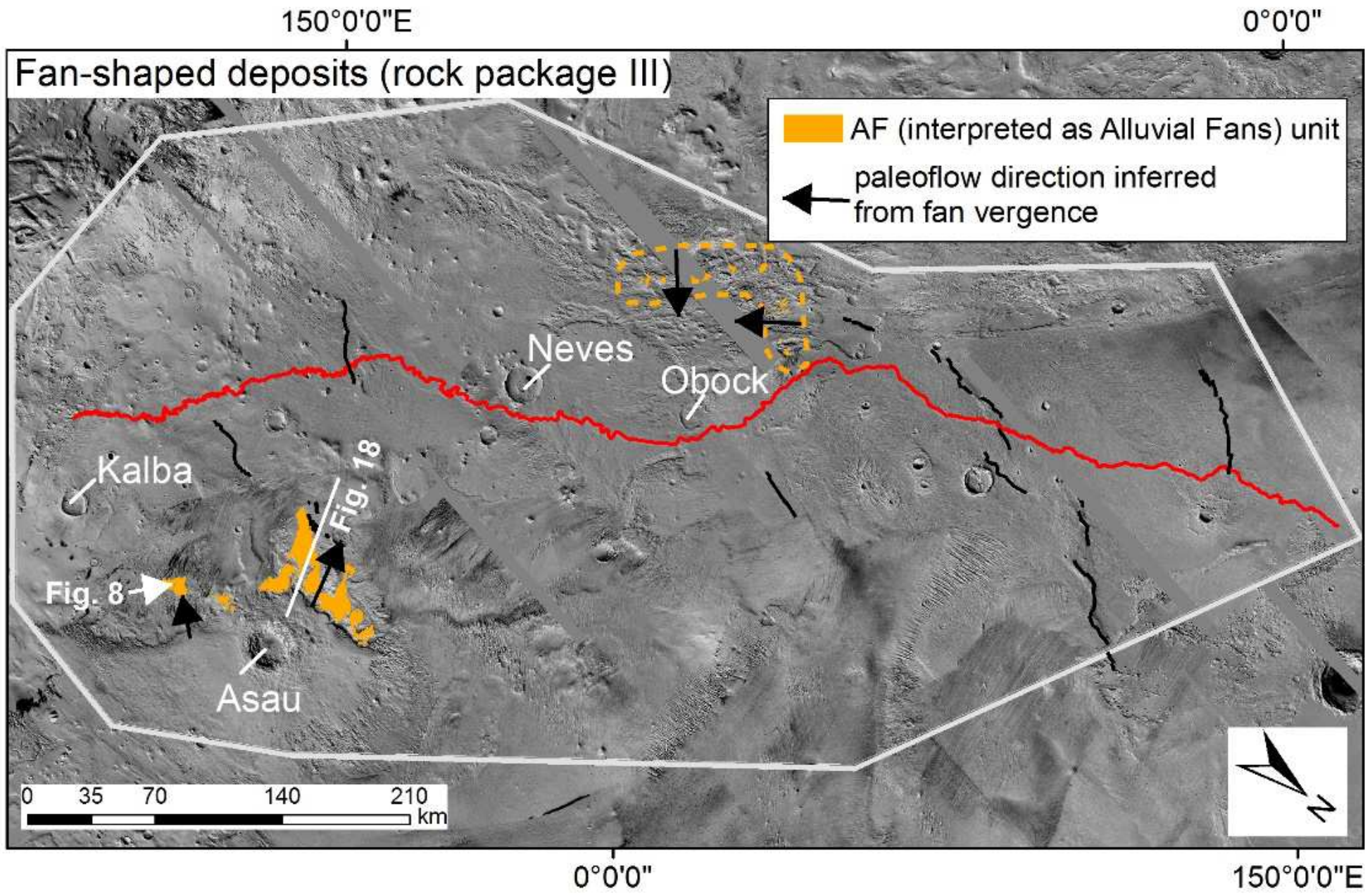

e)

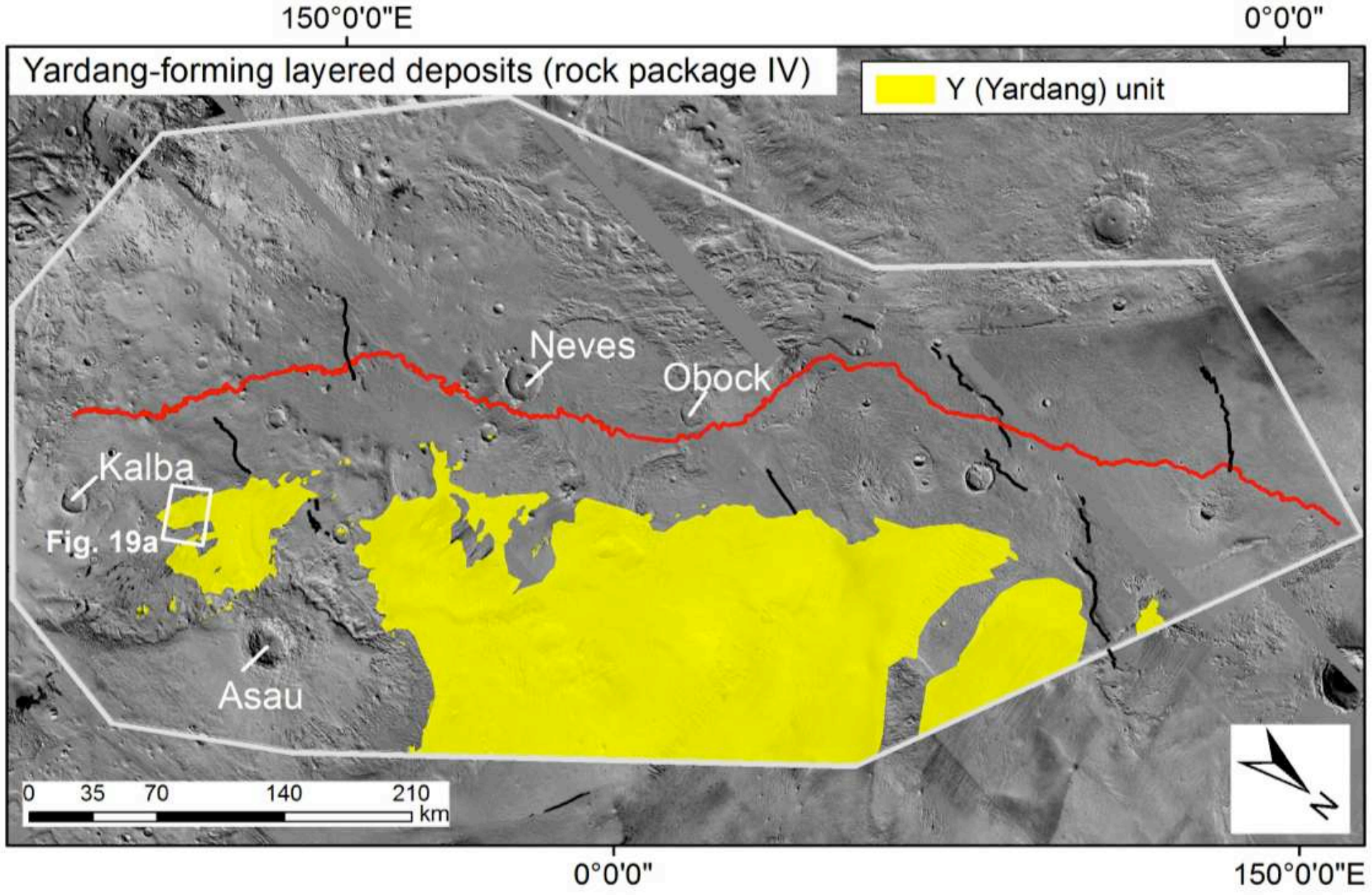

f)

**Figure 2:** Distribution of unconformity-bounded sedimentary-rock rock packages in Aeolis Dorsa. The white line shows the outer limits of our study area, the red line traces the longest known chain of Martian channel deposits formed by sustained flows with distributed tributary inputs, and the lack lines show mapped wrinkle ridges (which are the surface expression of thrust faults; Golombek et al. 2001). (a) Topography: contour interval 200 m. (b) Overview map showing all rock packages. (c) Early deposits (rock package I). (d) Fluvial deposits (rock package II). Dashed cyan lines link outer boundaries of R-2 outliers, and dashed green line links outer boundaries of R-SW outliers. (e) Fan-shaped deposits (rock package III). (f) Yardang-forming layered deposits (rock package IV) – note 3 main lobes, plus small scattered outliers. (Fig. S1 shows distribution of material not assigned to a rock package.)

## 2.1 Early deposits (rock package I).

We define early materials as rock package I (Fig. 2c). These early materials consist of three informally-defined units: the B (Basement), L (Lower), and DiCE (Dissected Crater Ejecta) units. All three units represent patchy remnants of the hummocky landscape that were (fluvially?) eroded prior to embayment by fluvial deposits (Figs. 2c, 2d). We found no evidence that the early materials contain river deposits.

<u>B (Basement) unit.</u> B outcrops in inliers bounded by inward-facing scarps of R-1 (Fig. S2). B lacks yardangs, lacks layers, and retains a large number of intact-rimmed craters. These attributes suggest erosional resistance, and are consistent with volcanic materials or resistant cemented sedimentary rocks.

<u>L: Lower Unit.</u> L consists of material that is topographically high-standing relative to the river deposits, with a high density of variably-degraded km-sized craters. Interpolating the topography of the currently-exposed outcrops of L implies that L continues as subcrop under northern

Zephyria Planum (the area draped by the central lobe of Y: Fig. 2f). In turn, this implies a paleotopographic high in the area of present-day Zephyria Planum rise and northern Aeolis Dorsa (Holt et al., 2010). Erosionally resistant ridges and/or knobs at 0.5km-1km scale are common. Layering is expressed by breaks-in slope, reflecting serial retreat of <10 m-high scarps due to wind-induced saltation abrasion. We interpret L as sedimentary rocks of unknown origin.[1] Near 149.5°E 1.72°S, L rises 700m above surrounding terrain (along a profile which rises ~700m in <14 km). Because the strata within L appear horizontal when comparing MOLA elevations to georeferenced CTX images, we interpret this elevation difference as a stratigraphic thickness.

DiCE: Dissected Crater Ejecta. DiCE consists of the remnant ejecta of Neves crater (diameter, $\phi$ = 21 km), Obock crater ($\phi$ = 15 km), and Kalba crater ($\phi$ = 14 km). The river deposits onlap the ejecta, and river channels also cut into the ejecta. All three craters show post-impact sedimentary infill.

*Neves crater*: Crater ejecta, marked by ejecta sculpture including both radial grooves and circumferential terminal ramparts, forms isolated, locally high-standing broad domes (Fig. 3). Steep-sided mounds that form the highest points on these broad domes may also represent Neves ejecta, or alternatively they may represent outliers of later sedimentary materials. The ejecta of Neves is not visible to the N and W of the Neves rim. Neves ejecta is crosscut by branching, sinuous valleys (Fig. S3a). These channels are oriented N-S rather than draining radially away from the crater, and run across ramparts of Neves ejecta without deflection (Fig. 3a, inset), and crosscut a $\phi$ = 1.5 km crater on the ejecta (Fig. S3a). These relationships suggest that the channels formed after fluvial deposits had accumulated around Neves up to the level of the ramparts, and significantly postdate the Neves impact.

*Obock crater*: Two lobes of Obock ejecta are visible in an erosional inlier beneath finely-layered materials. They can be identified by impact sculpture. Both terminate 14km from the Obock rim.

*Kalba crater*: Meander-belt deposits overlie Kalba ejecta; the flow pattern of these rivers diverts around Kalba's rim. The northernmost meander belt deposit is broken in two by the distal rampart of Kalba's ejecta; the two parts of the meander belt deposit are joined by a valley cutting across the rampart (inset of Fig. S3b).

[1] Polygonal ridges in L near 150.7°E 1.1°N resemble the ridge-forming unit reported in Meridiani stratigraphy by Edgett (2005) (his Figs. 25-29).

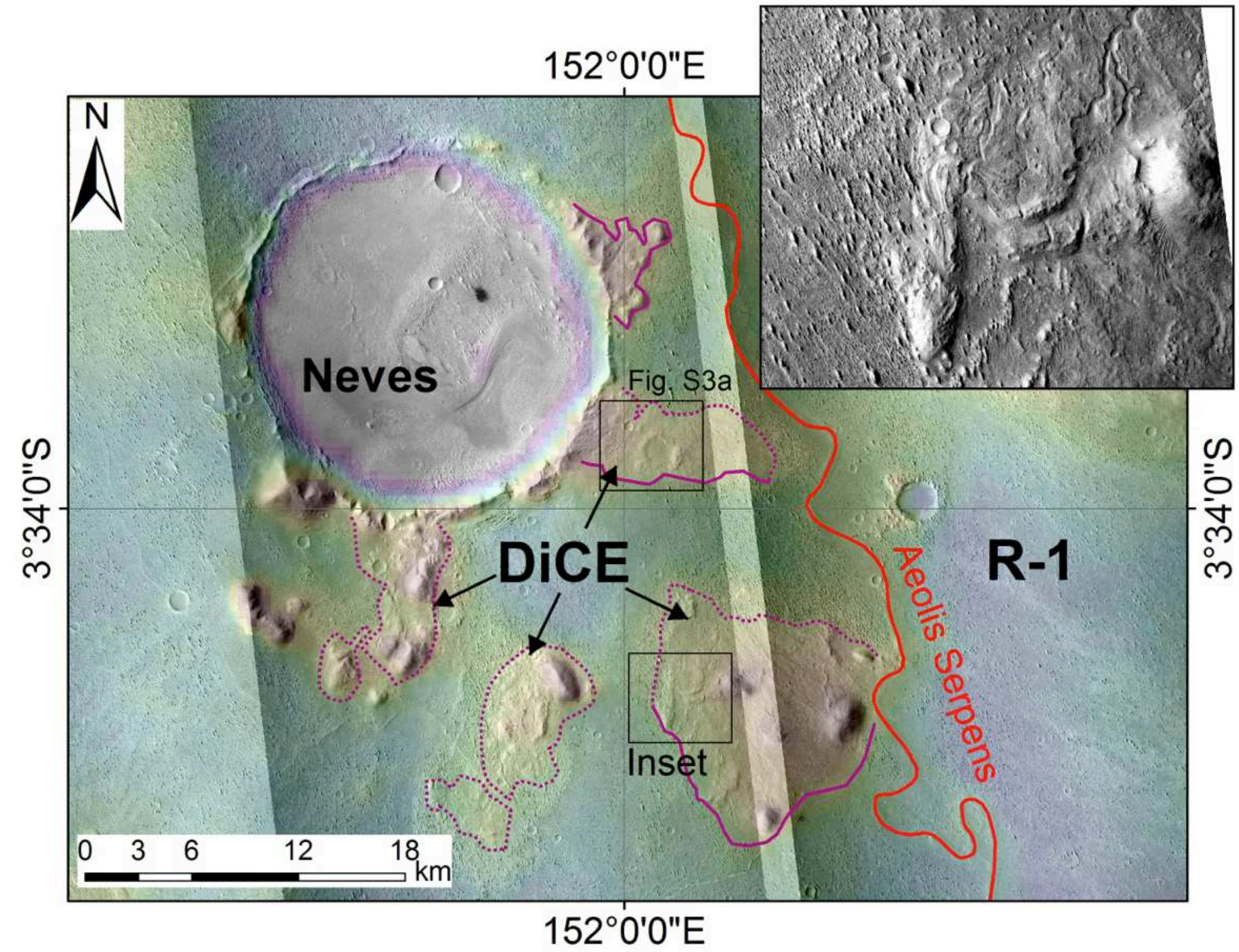

**Figure 3.** Distribution of dissected crater ejecta (DiCE) around Neves crater. Color ramp is MOLA topography (red is high). Purple line corresponds to the contact of DiCE (dashed where inferred or gradational), which is onlapped by river deposits. Red line highlights the Aeolis Serpens river deposit (Williams et al. 2013a). Inset shows rampart margin of Neves ejecta, now incised by channels (Fig. S3). (P15_006894_1744_XI_05S208W, P05_003136_1746_XN_05S208W).

## 2.2 Fluvial deposits (rock package II).

Rock package II contains the main river deposits of Aeolis Dorsa (Burr et al. 2009, Williams et al. 2013a) (Fig. 4). The river deposits are identified using sinuosity, continuity, location downslope from topographic highs, and association with fan-shaped forms, following Burr et al. (2009, 2010) and Williams et al. (2013a). We identify 3 informally-defined units within rock package II (R-1, R-2, and R-SW), differentiating between the units using the density of yardangs, outcrop topography, and channel morphology.[2] The total interval spanned by river deposits was shown by Kite et al. (2013a) using the frequency of interbedded craters to be >($1 \times 10^6 - 2 \times 10^7$) yr.

[2] "R-1" is the "F1" of Kite et al. 2014, and "R-2" is the "F2" of Kite et al. 2014.

a)

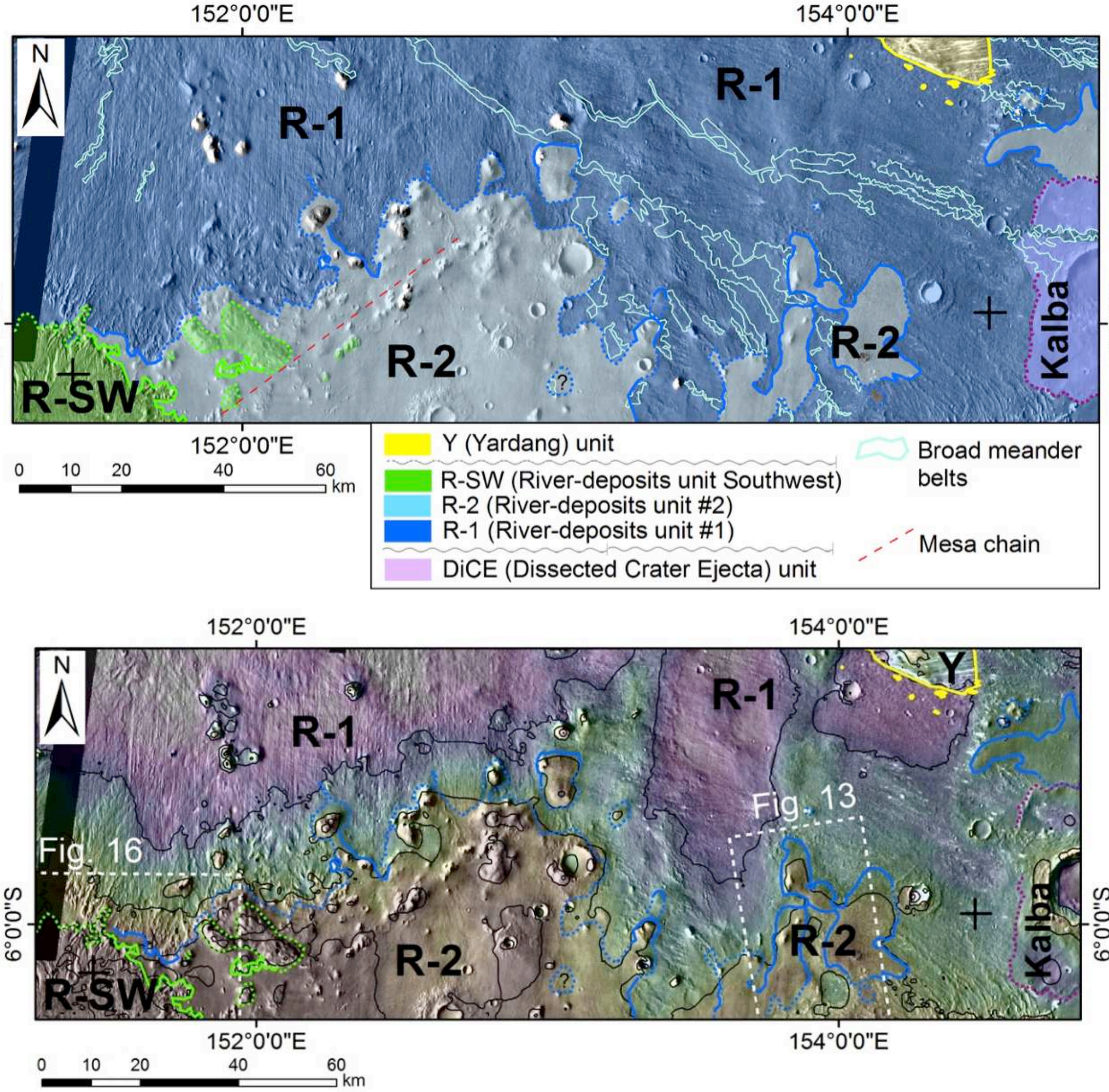
152°0'0"E
154°0'0"E
N
R-1
R-1
R-2
R-2
R-SW
Kalba
6°0'0"S
6°0'0"S
?
0 10 20 40 60 km
Y (Yardang) unit
R-SW (River-deposits unit Southwest)
R-2 (River-deposits unit #2)
R-1 (River-deposits unit #1)
DiCE (Dissected Crater Ejecta) unit
Broad meander belts
Mesa chain
Y
Fig. 13
Fig. 16

b)

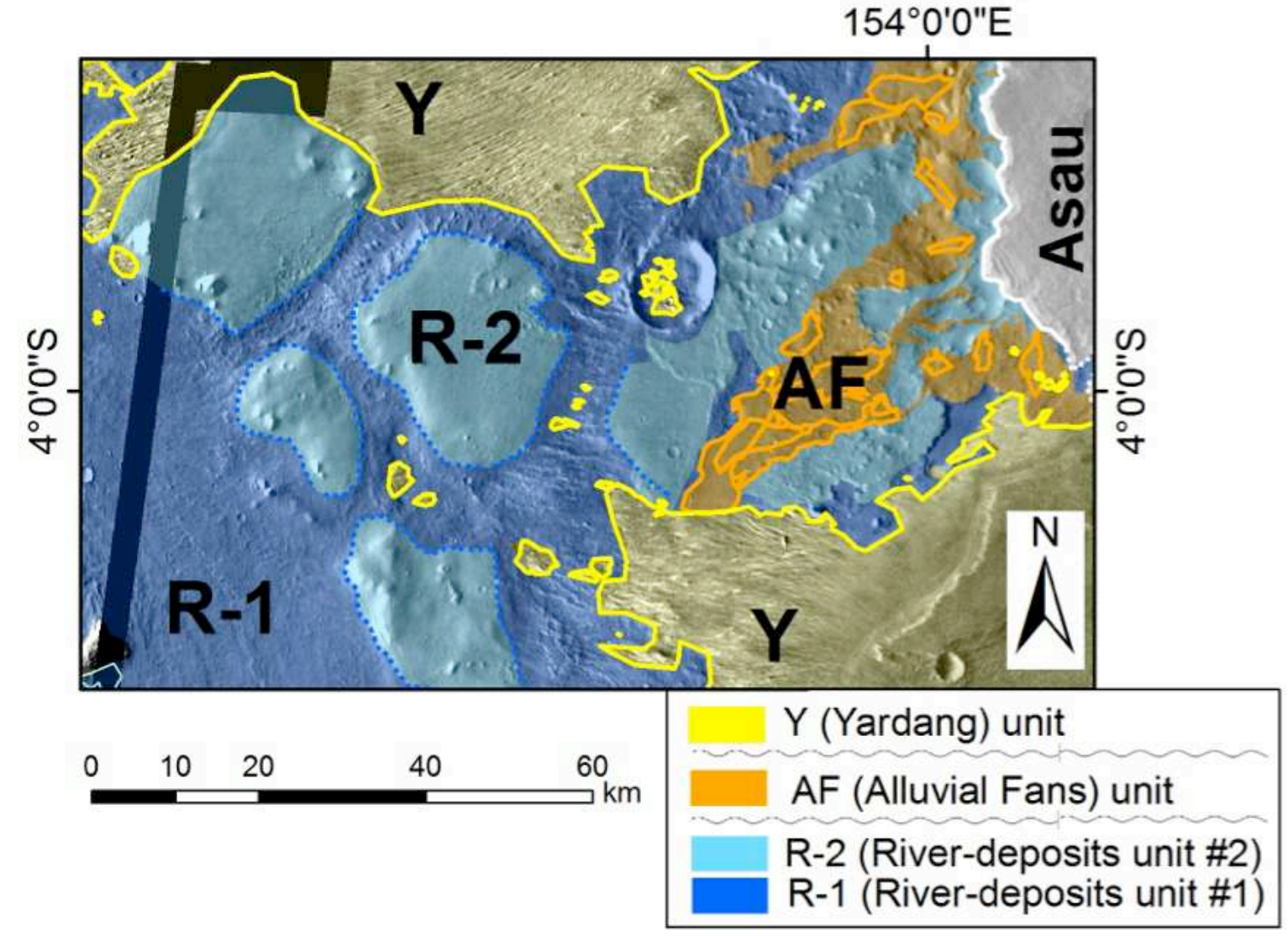

c)

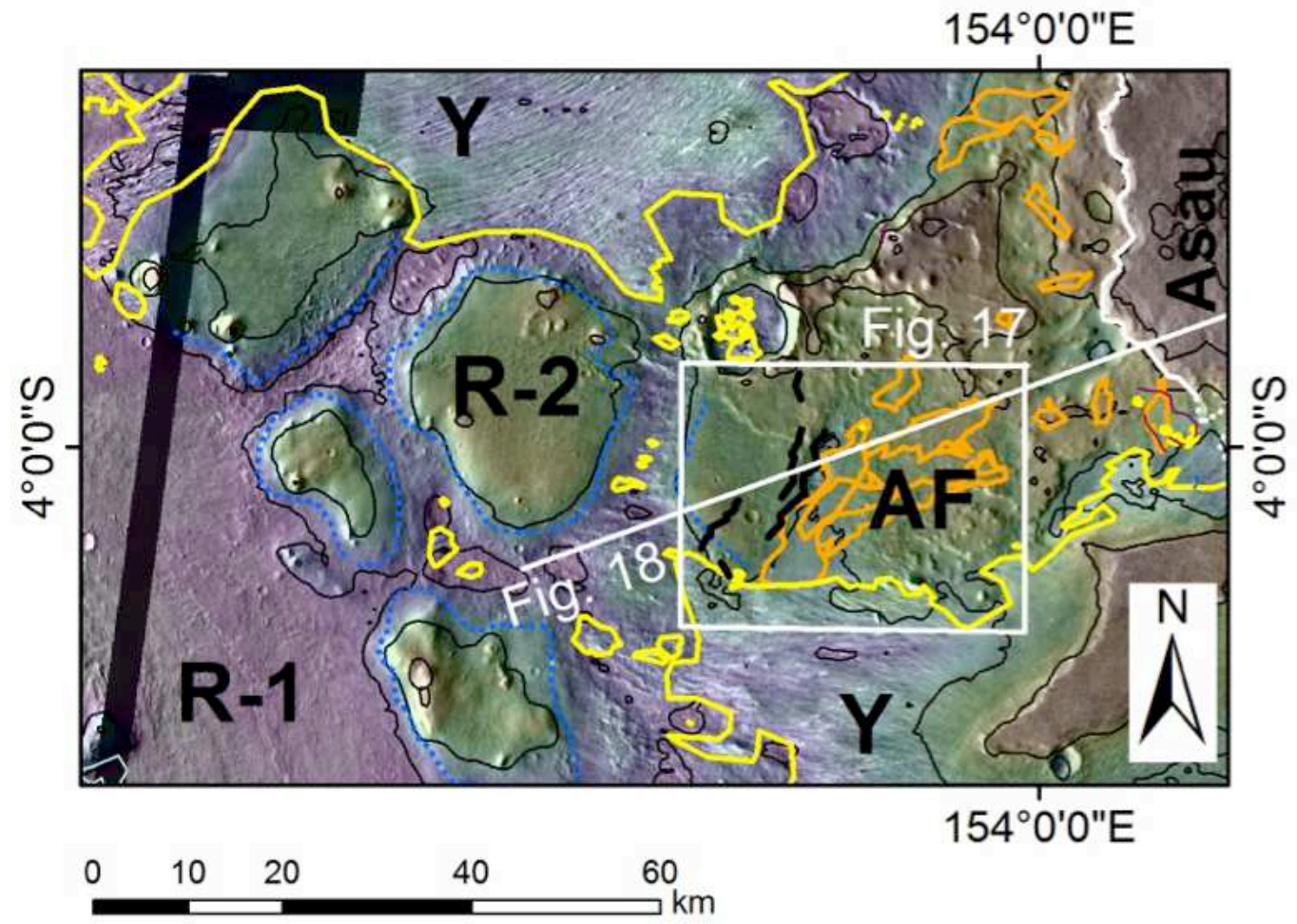

d)

**Figure 4.** Topographic distribution of fluvial units in two areas of exceptionally good preservation (Fig. 2d shows basin-wide context). Background image is Thermal Emission Imaging System (THEMIS) Visible (VIS) mosaic. **(a)** Topographically lower fluvial unit (R-1) erodes to form yardangs, and contains broad meander belts (cyan outlines) and a high density of sinuous ridges. The overlying R-2 unit is smoothly eroding and contains few meander belts. R-SW may postdate R-2, or alternatively it may be stratigraphically equivalent to R-1. Contacts are solid where mapped with high confidence, and dotted where inferred. Early materials include Kalba ejecta (lilac tint with red edge), and post-river materials include yardang-forming rhythmite (yellow tint). Crosses mark the ends of the Fig. 12 cross-section. **(b)** Topography of (a), with contours at 200m intervals. **(c)** To the N of (a), R-2 unit forms broad domes superposed on the inverted meandering channels that are common in R-1. Individual fan-shaped deposits, which superpose R units, are shown by orange outlines. Black lines correspond to thrust faults (Golombek et al. 2001). **(d)** Topography of (c), with contours at 200m intervals.

R-1 (River-deposits unit #1). R-1 consists of large exhumed meander-belts, sinuous ridges, and the materials that surround them (Figs. 4, 5) (Burr et al. 2009). One R-1 channel (Aeolis Serpens, red line in Fig. 2) is the longest known chain of Martian channel deposits formed by sustained flows with distributed tributary inputs (Williams et al. 2013a). DiBiase et al. (2013) interpret some of the R-1 deposits as deltas. We find river deposits exposed at all stratigraphic levels within R-1, and the channel-deposit proportion is high, so we infer that most of the material in R-1 has been transported by rivers. R-1 forms yardangs and preserves few small impact craters. The best-preserved meandering channels define a cliff-capping member – perhaps a coarse-grained layer (Armitage et al. 2011). North of 2°S, R-1 is nearly flat, and traversed by numerous narrow channels of diverse orientation. This is consistent with deposition by alluvial-plain distributaries. South of 2°S, R-1 usually outcrops as semi-parallel channels (that have low junction angles) and associated floodplain deposits. These channels and meander-belts are generally preserved in inverted relief. The poorly-preserved zones between the R-1 meander-belts and channels, which are now low-standing, may have originally been the sites of drainage divides between the meander belts (Cardenas & Mohrig 2014). Alternatively, these poorly-preserved zones may represent the deposits of earlier avulsions, or floodplain deposits.

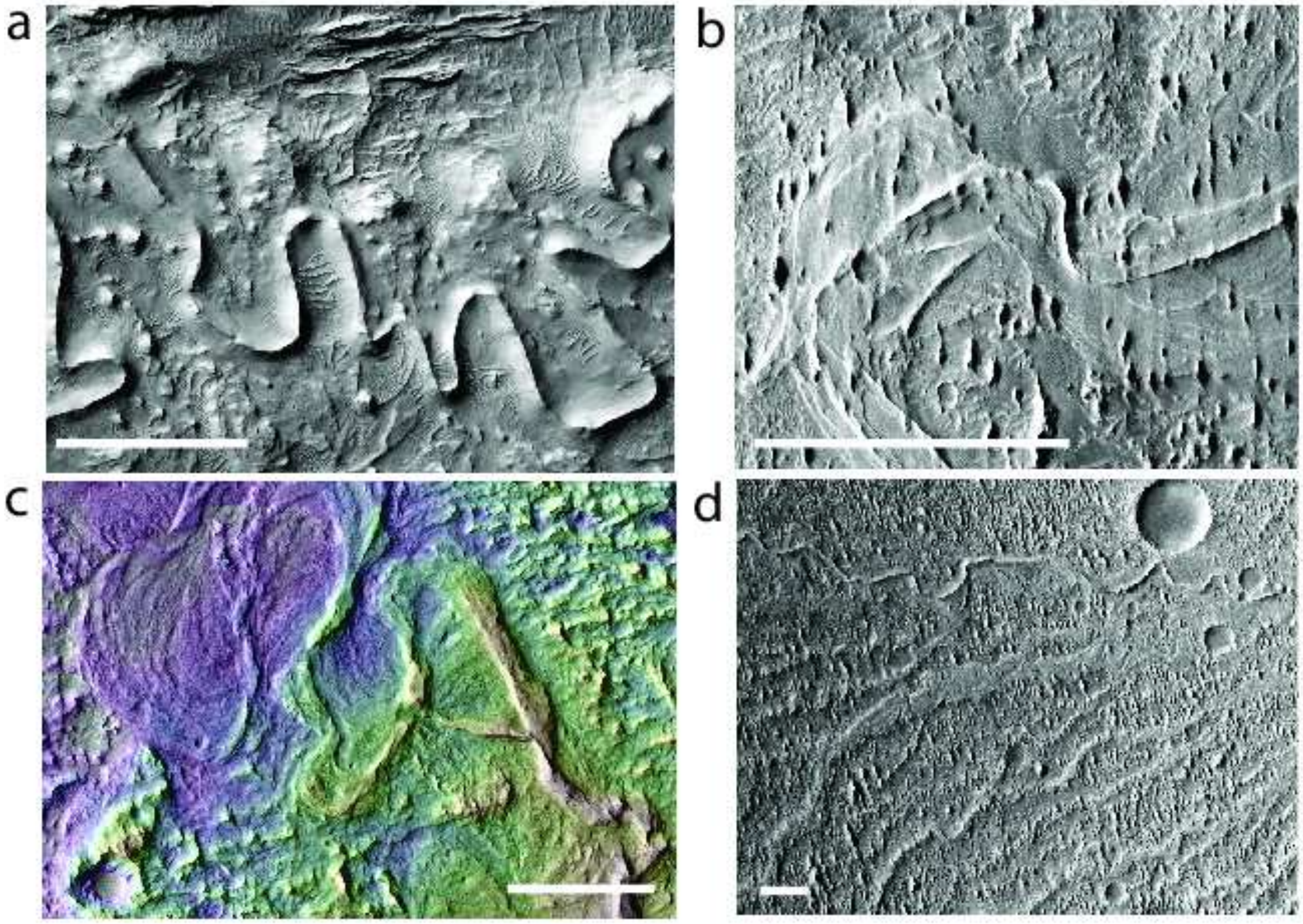

**Figure 5.** River deposits in unit R-1. Scale bars correspond to 500m and N is up. Details: a) Near 154.6°E 4.9°S, ESP_016269_1750. b) Near 151.6°E 5.0°S, ESP_024985_1750. c) Near 153.6°E 5.9°S, ESP_019882_1740. d) Near 150.6°E 4.2°S, P16_007105_1774_XN_02S209W.

R-2. (River-deposits unit #2). R-2 consists of smoothly eroding materials that form broad domes superposed on R-1 (Fig. 4a). Yardangs are uncommon, and recent aeolian bedforms are common. R-2 retains many small craters (Kite et al. 2014), and exhibits fine-scale channels that are preserved both in inverted relief and as depressions (Fig. 6). R-2 channels branch more frequently than R-1 channels branch (Fig. 6). Evidence for lateral accretion is uncommon in R-2 channels – in contrast to R-1 channels. The sinuous ridges in R-2 trend locally subparallel to the channels in underlying R-1, and do not follow the present-day local slope of the R-2 domes. For this reason, we interpret the present-day broad domes of R-2 as wind-eroded outliers of formerly more extensive outcrops of R-2 (the two areas outlined by dotted blue lines in Fig. 2d). It is possible that R-2 was once a single continuous outcrop across southern Aeolis Dorsa.

R-SW. (River-deposits unit - Southwest). R-SW may represent the same wet episode as R-1, or alternatively it may represent a post-R-2 wet event (§3.2). R-SW is defined by a cliff-capping, boulder-forming fluvial member, with many sinuous ridges, topographically >200m above the Aeolis Serpens (Fig. 2d, Fig. 7). We group the material mantled by talus from the cliff-capping member with R-SW. Evidence for lateral migration of the channel during aggradation is again found (Fig. 7).

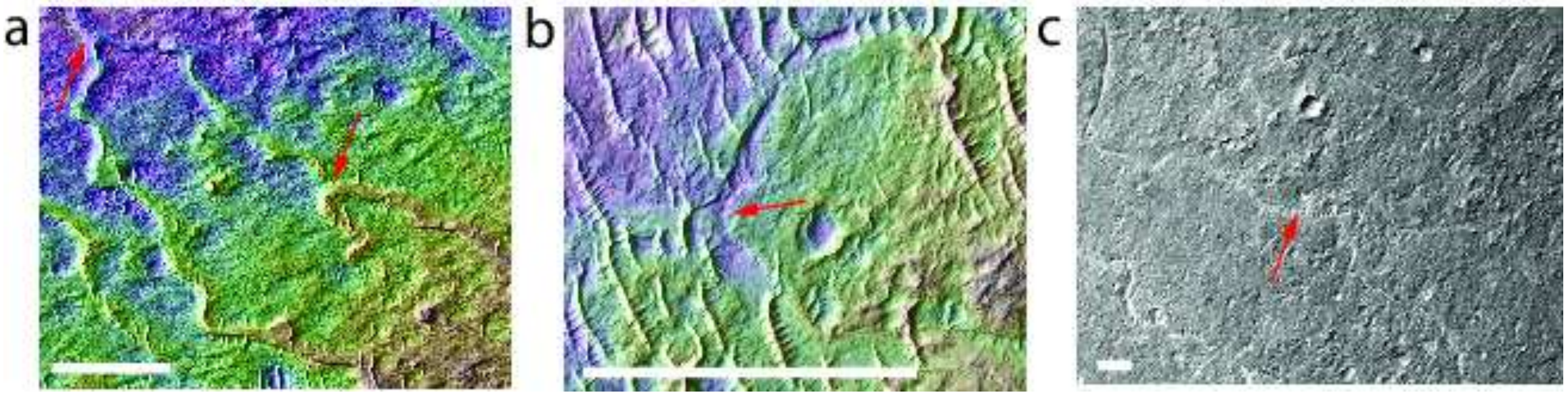

**Figure 6.** River deposits in unit R-2. Scale bars correspond to 200m. Red arrows highlight junctions between channels. a) Near 154.6°E, 5.4°S. PSP_007474_1745. b) Near 153.7°E, 6.1°S. ESP_019882_1740. c) Near 154.0°E, 6.0°S. B20_017548_1739_XI_06S206W.

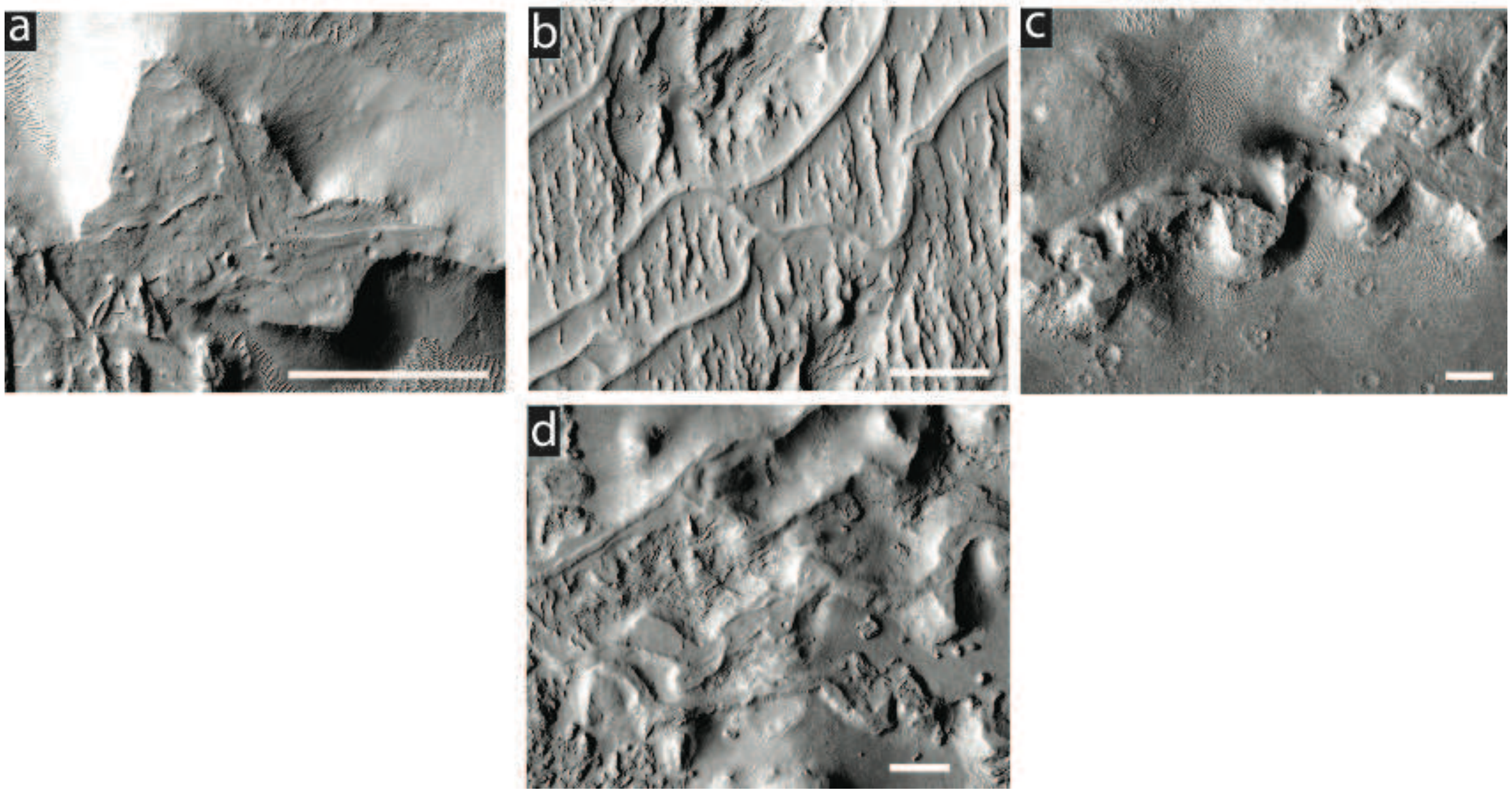

**Figure 7.** River deposits in unit R-SW. Scale bars are 500m across. a) Near 152.3°E, 6.8°S; ESP_018181_1730. b) Near 151.4°E, 6.2°S. PSP_002279_1735. c) Near 152.0°E, 6.1°S. P05_003136_1746_XN_05S208W. d) Near 151.7°E, 6.3°S. G05_02034_1735_XI_06S206W.

Channel-deposit proportion. A lower bound on the channel-deposit proportion is set by the fractional area of outcrop that is occupied by channel deposits that are sufficiently well-exposed to be mapped from orbit. At some stratigraphic levels, especially within R-2, river-deposit-containing units have a small channel-deposit proportion. Where the channel-deposit proportion is small, the river-deposit-containing materials could represent fluvial deposits with a high proportion of overbank (floodplain) materials (Bridge 2003), atmospherically transported material with limited fluvial reworking (Haberlah et al. 2010, Schiller et al. 2014, Ewing et al. 2006, Grotzinger et al. 2006), or some combination. At other stratigraphic levels, especially within R-1, river-deposit-containing units have a channel-deposit proportion exceeding 30% (Kite et al., submitted).

## 2.3 Fan-shaped deposits (Rock package III).

Rock package III consists of branching networks of low-sinuosity channel-form ridges radiating downslope away from a locally high-standing fan apex (Fig. 2e, Fig. 8). We interpret these as wind-eroded alluvial fan deposits (e.g. Moore & Howard 2005, Lefort et al. 2012, Williams et al. 2013b, Palucis et al. 2014, Morgan et al. 2014). In Aeolis Dorsa, deposits interpreted as alluvial-fans (mapped as AF) are only found overlying large-river deposits, and always widen with increasing distance away from highs in the modern topography (Aeolis Planum Rise, Zephyria rise; Fig. 2e). All alluvial fans in Aeolis Dorsa have hollows or shallow moats at their apex (Kite 2012). Alluvial fan outcrops total <1% of Aeolis Dorsa's area (Fig. 2). Alluvial fans require surface liquid water production in excess of infiltration plus evaporation. A candidate liquid water source is snowmelt (Kite et al. 2013b, Morgan et al. 2014, Palucis et al. 2014).

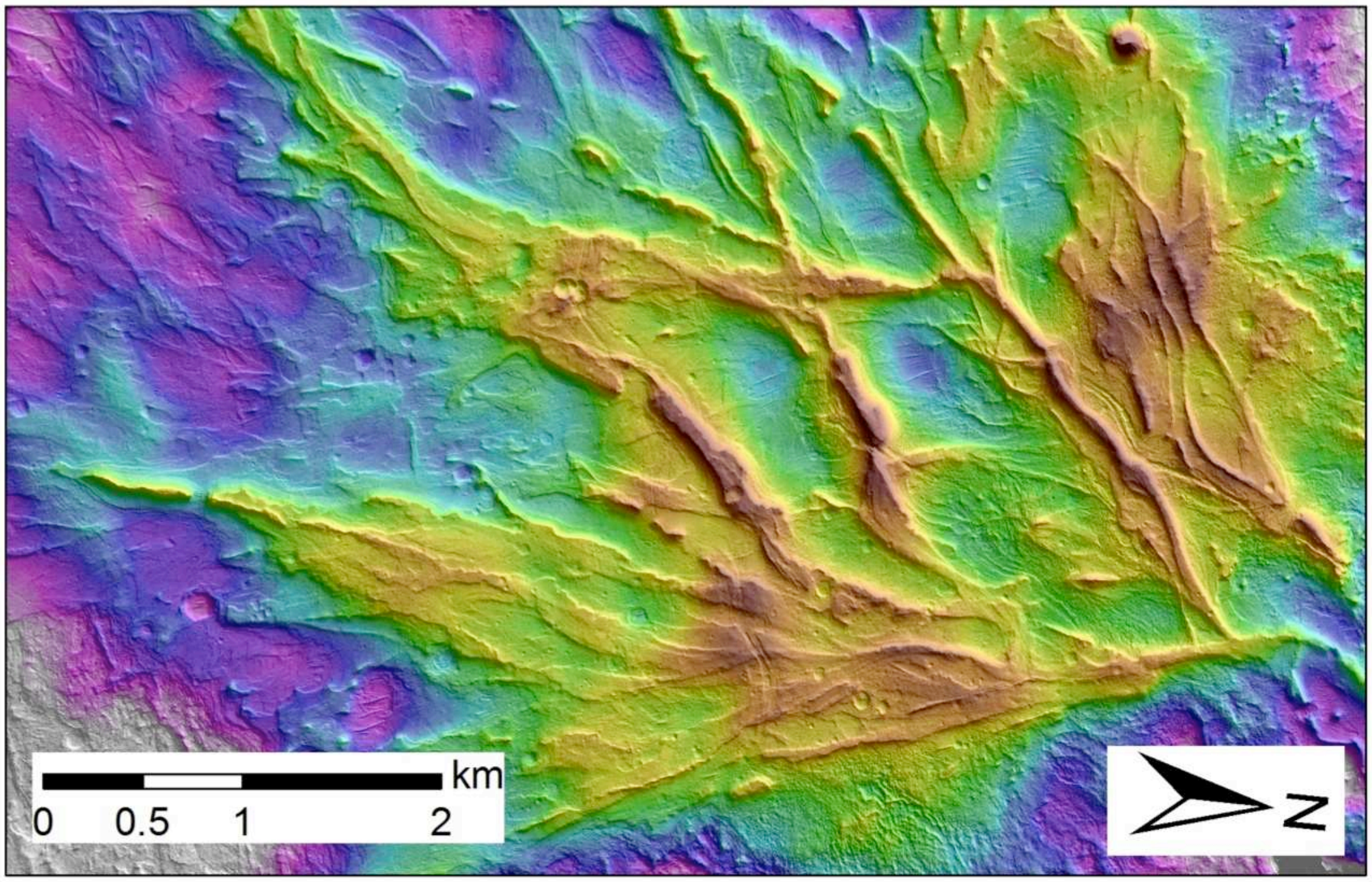

**Figure 8.** Example of Aeolis Dorsa fan-shaped deposit (near 154.8°E 4.6°S), interpreted as an eroded alluvial-fan deposit. Full range of topography is 100m (red is high and white is low). Significant wind erosion of the fan surface has occurred (locally >30m; Fig. 19). HiRISE DTM (PSP_009795_1755/PSP_009623_1755) is available at `http://geosci.uchicago.edu/~kite/stereo`.

## 2.4 Yardang-Forming Layered Deposits (Rock package IV).

The youngest deposits in Aeolis Dorsa (rock package IV, consisting of a single unit Y, Fig. 2f) are defined by steep-sided outcrops of finely-layered materials that are densely grooved by yardangs (e.g. Ward 1979). Y forms three main lobes and numerous smaller outliers (Fig. 2f; see also Harrison et al. 2010). The northernmost lobe of Y is >900m thick.

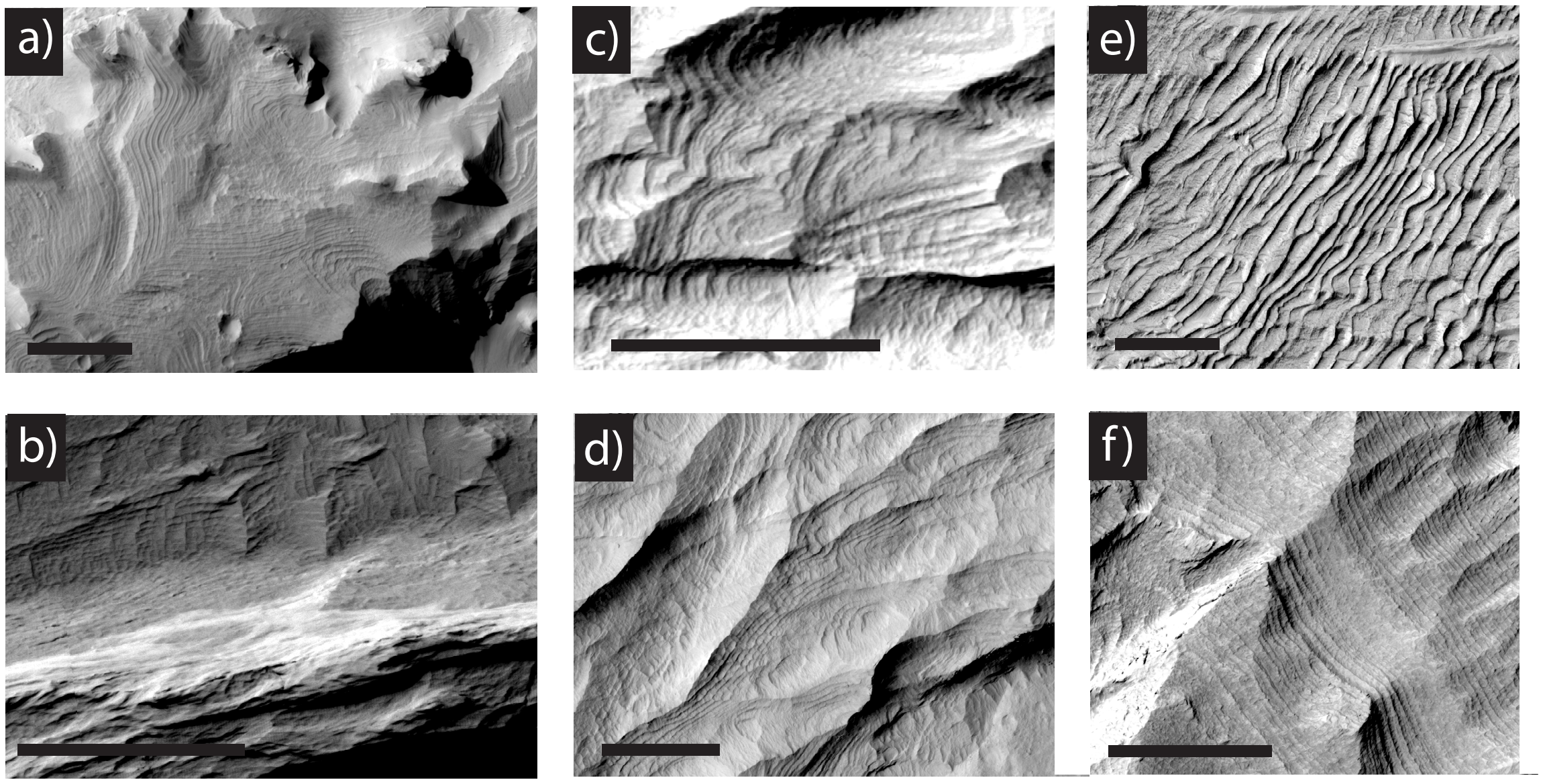

**Figure 9.** Rhythmite on Mars. (a)-(d), from within our study area; (e-f), previously described rhythmite from Grotzinger & Milliken (2012b). Scale bars correspond to 100m, N is up, and illumination is from the left. Details: (a) near 153.5°E 3.8°S, ESP_027807_1765; (b) near 154.7°E 4.9°S, ESP_016269_1750; (c) near 154.7°E 4.4°S, ESP_026897_1755; (d) near 154.8°E 4.2°S, ESP 019605_1755; (e) near 24.1°E 9.9°N, PSP_010353_1900 – Henry Crater; (f) near 137.7°E 4.9°S, PSP_008002_1750 – upper formation of Gale crater's mound (Mt. Sharp / Aeolis Mons.) Layers of constant stratigraphic thickness will be constantly spaced on outcrops of constant slope.

Y contains many rhythmite layers and (taking into account poor preservation) may be entirely rhythmite (Fig. 9). In the definition of Grotzinger & Milliken (2012a): "Primary attributes of the rhythmite facies are very thin (~1-5m) beds that exhibit a repeatable thickness (are rhythmic) within a vertical sequence […] Planar bedded stratal geometries dominate, perhaps uniquely so." On Mars, rhythmite-facies materials commonly form the youngest materials in a sedimentary succession, draping older sedimentary rocks.

Rhythmite formation probably involved orbitally-paced accumulation of atmospherically-transported dust, silt or sand (Lewis et al. 2008, Lewis & Aharonson 2014), perhaps similar to loess on Earth. Consistent with the loess interpretation, radar sounding of Y shows a permittivity of ~3, indicating high porosity or a high percentage of buried water ice (Carter et al., 2009; Mouginot et al., 2010).

Y sustains locally >60° slopes and resists deflation to allow yardangs to form. These observations require that the sediments are indurated. This induration probably required some (near-)surface liquid water for cementation (Andrews-Hanna & Lewis 2011, Head & Kreslavsky 2001, Kite et al. 2013b, Lewis et al. 2008, Moore 1990, Nickling 1984). Therefore, orbitally-paced surface liquid water availability some time after the rivers flowed is suggested by the rhythmic bedding and induration of the youngest deposits in Aeolis Dorsa (Lewis et al. 2008).

Y drapes all older materials unconformably (Fig. 2f, §3.4) and is rhythmically bedded, light-toned, and deeply grooved. These attributes are similar to (and thus suggest correlation with) the rhythmites overlying the major unconformity at Gale crater (§5.3, Edgett & Malin 2001, Milliken et al. 2010, Anderson & Bell 2010, Thomson et al. 2011, Wray 2013, Le Deit et al. 2013).

# 3. Major stratigraphic relationships.

Three main unconformities sub-divide the sedimentary rock packages of Aeolis Dorsa. These unconformities represent the following geologic events:

- Erosion of early sediments (rock package I) followed by embayment of this eroded landscape by fluvial units (rock package II) (§3.1, §3.2);
- Tectonic deformation and probable erosion of the fluvial units (rock package II) followed by the formation of fan-shaped deposits (rock package III) (§3.3);
- A lengthy period of region-wide erosion followed by deposition of materials that contain rhythmite (rock package IV) and which recently eroded to form yardangs (§3.4).

Evidence for additional unconformities within the fluvial units is discussed in §3.2 and in the Supplementary Discussion. We now describe the observations that underpin these interpretations.

## 3.1 River deposits embay a dissected landscape formed of sedimentary rock.

Fig. 10 summarizes our interpretation of the contact between rock packages I and II.

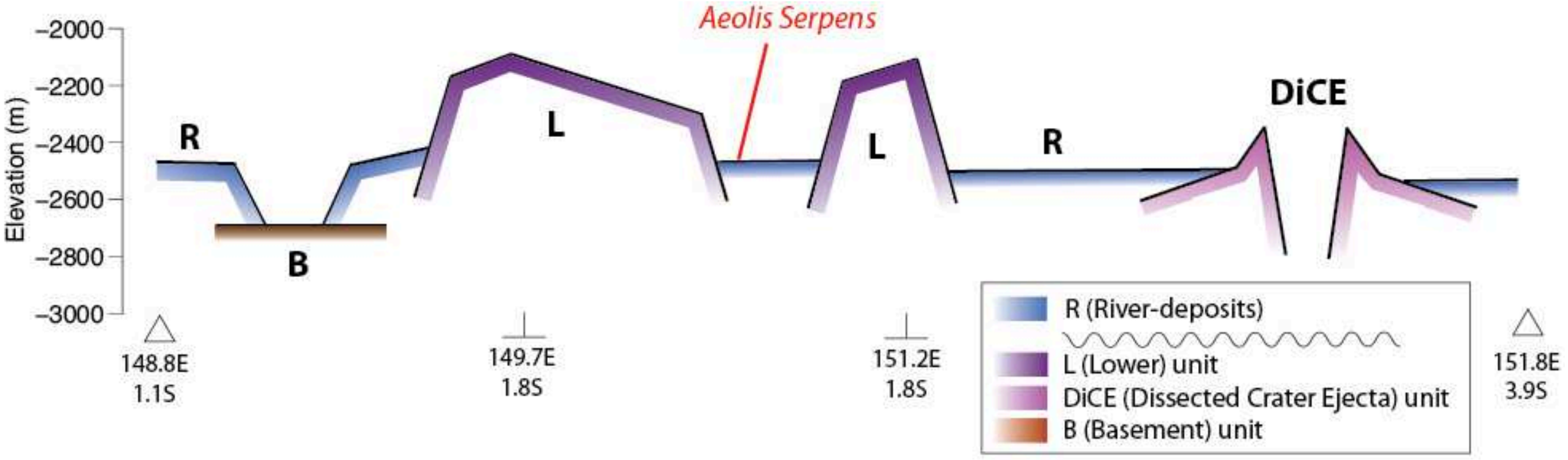

**Figure 10.** Interpretative cross-section showing the relationship between fluvial units (R) and early deposits (L, DiCE, B). Fig. 2c shows line of section.

The evidence for this interpretation is as follows.

L and DiCE were eroded before R was deposited. This erosion can be reconstructed assuming that L originally formed a nearly-horizontal sheet, and that craters eject axisymmetric ejecta. Especially around Neves, erosion was patchy: most of Neves' ejecta was removed, but the remnants show detailed ejecta sculpture (Fig. 3, Fig. S3). This pattern of nonuniform erosion is most consistent with erosion by rivers, glaciers, or subsurface piping (Rose et al. 2013, Pederson & Head 2011). River deposits are found within, surrounding and incising all three DiCE craters. Therefore, erosion by rivers simply explains the erosion of L and DiCE to form the sub-river-deposits unconformity surface.

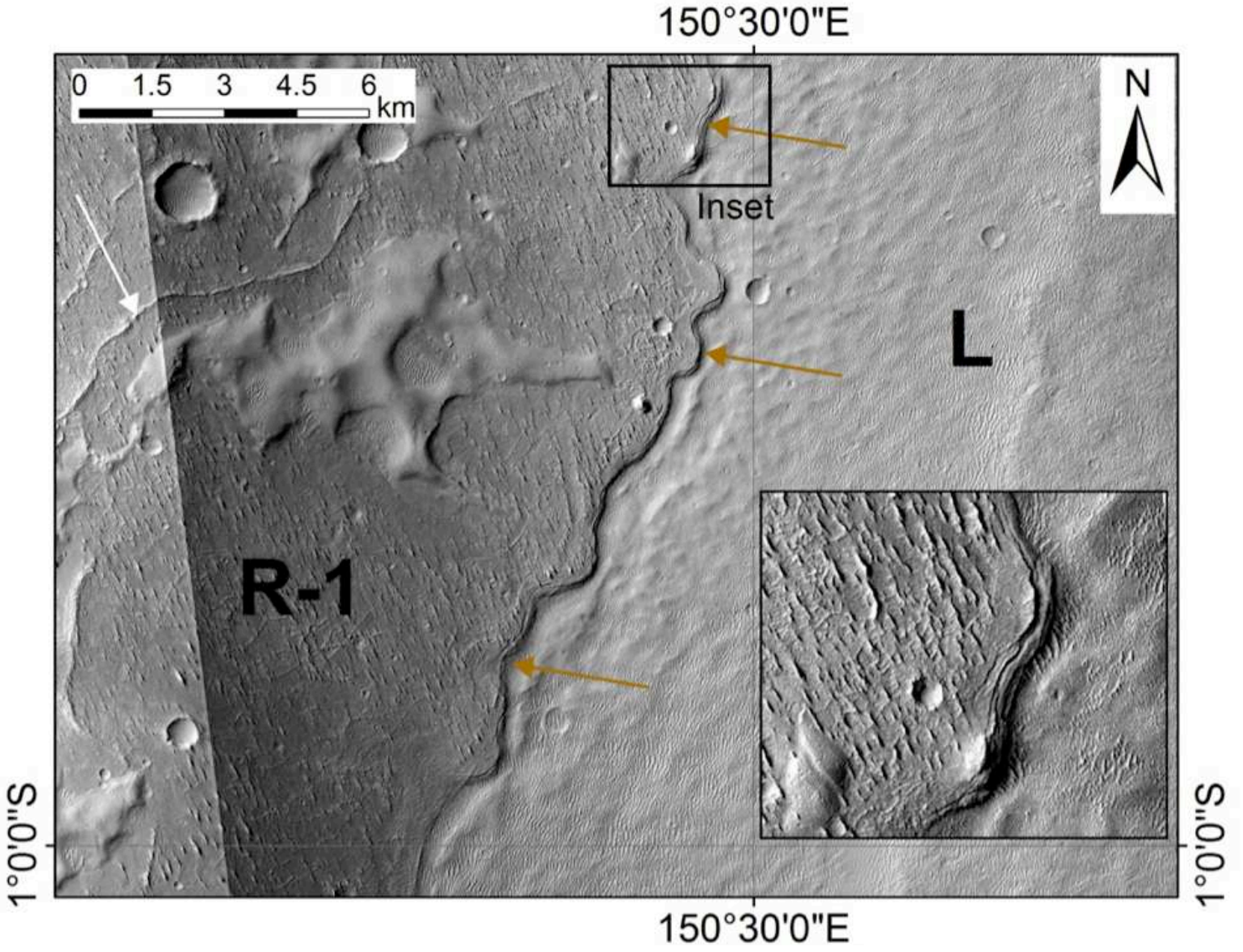

a)

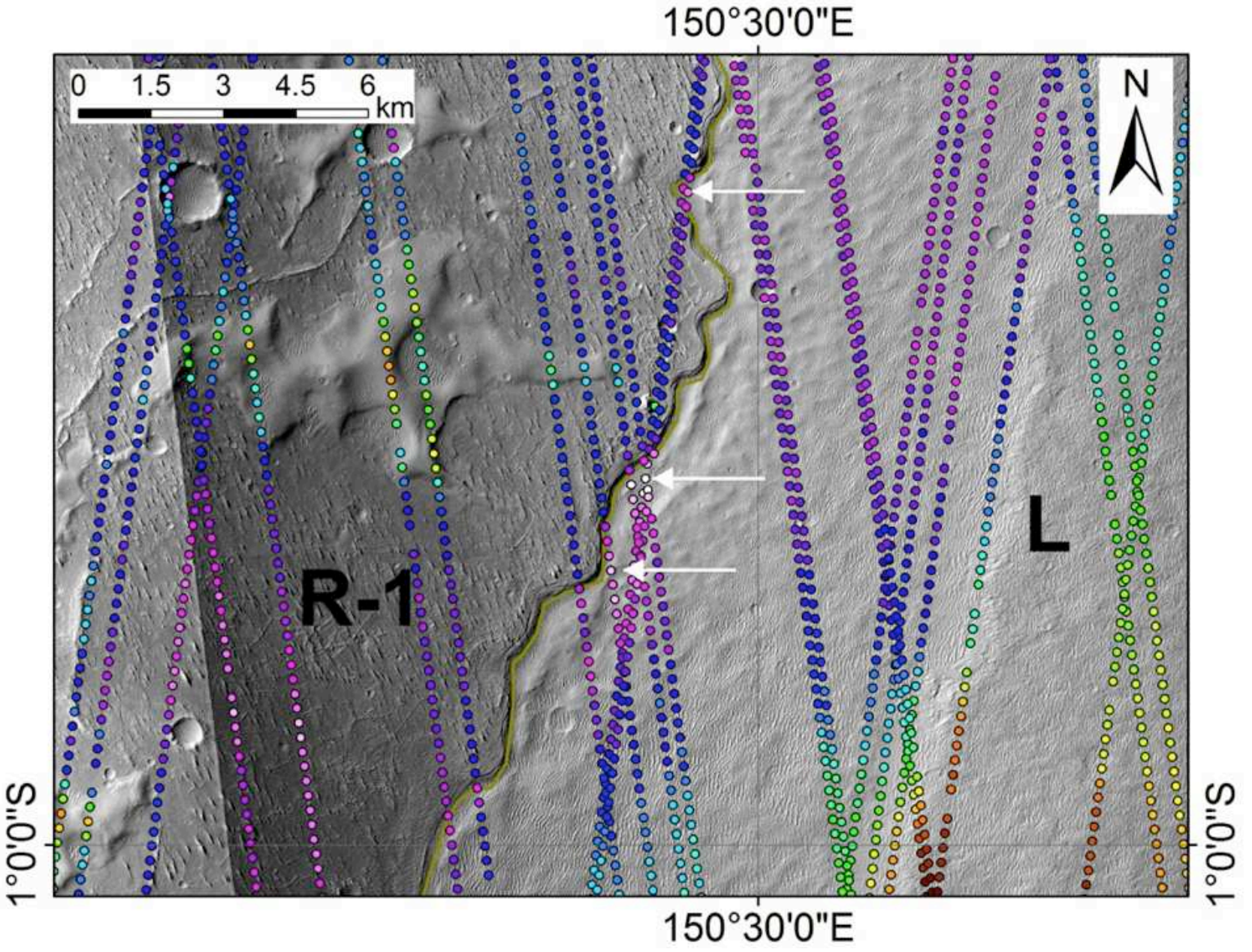

b)

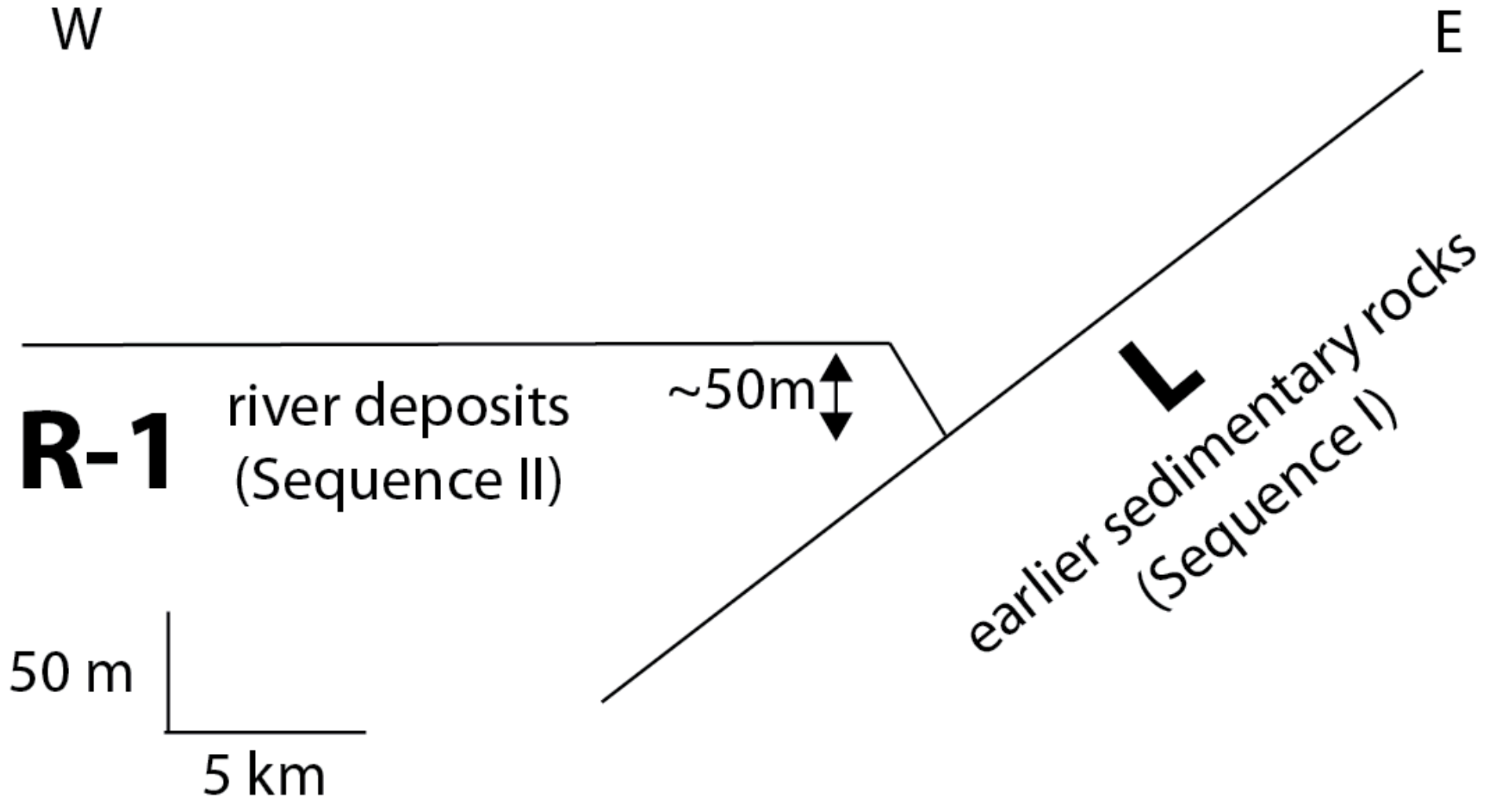

c)

**Figure 11.** Showing location where R-1 onlaps L (P19_008397_1791_XI_00S209W, P16_007105_1774_XN_02S209W; Fig. S6 shows context). (a) showing undulating contact (brown arrows) between L and R-1, with R-1 embaying L. Inset (black rectangle) shows detail. Sinuous ridges (example: white arrow) interpreted as inverted channels superpose sheet of yardang-forming material. (b) shows topographic relationships near L/R-1 contact (brown line). White arrows highlight points along the contact where R-1 locally overlies L. Color range of PEDR spots is -2250m (red) to -2470m (white). (c) Interpretative cross-section. Layers are not visible in L at this location, so the relative orientation of the layers in L and the layers in R-1 is not constrained.

Fig. 11 shows that the river deposits embay an erosionally-dissected paleotopographic high built up of L (as previously noted by Williams et al. 2013a). Fig. 11a shows an undulating contact between an outcrop of L to the east that rises to >-2250m, and a topographically low-standing sheet of material containing sinuous ridges (interpreted as river channels)[3] to the west that never rises above -2380m. However, Precision Experimental Data Record (PEDR) tracks prove that everywhere along the contact, the river-deposit-containing unit is *higher* than L (Fig. 11b), typically by 50m. These topographic relationships indicate that the river-channel-containing material embays L. Figs. S4-S6 show similar relationships.

Fluvial deposits also onlap the ejecta of Neves, Obock and Kalba (the DiCE unit). Kalba ejecta is superposed by meander belts. Aeolis Serpens flowed after the crater ejecta was dissected: for example, Aeolis Serpens approaches within 6 km of the Neves rim without visible deflection, even though Neves ejecta extends up to 24 km from the rim. The ejecta of Obock and Kalba have been dissected by valleys at 100m scale (Fig. 3).

B only outcrops in the deepest pits within Aeolis Dorsa, forming inliers within R-1 (Fig. S2). Because the pit floors are smooth and within 100m of each other in elevation, we interpret B as forming a smooth basement surface underlying the river deposits. Because B is never observably wind-eroded, we interpret the depth of wind-cut troughs within the river deposits that do not expose B as a lower limit on the thickness of river deposits.

## 3.2 River deposits at different stratigraphic levels record fluvial activity at different times.

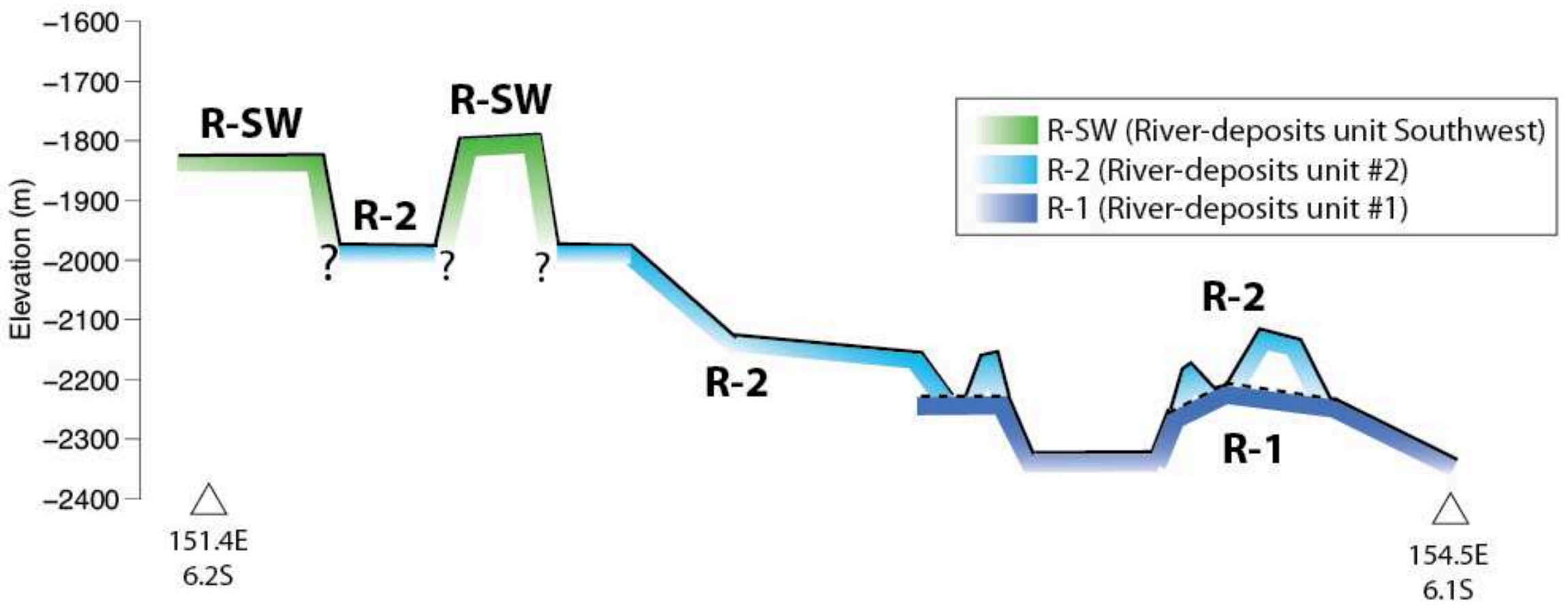

**Figure 12.** Topographic relationships between wind-eroded outcrops of R-2, R-1, and R-SW. Line of section corresponds to black crosses in Fig. 4a, and white line in Fig. 2d. R-SW is visible mostly in the SW of Aeolis Dorsa. R-2 postdates R-1, although the contact (dashed lines) may represent an erosion surface. R-SW may postdate R-2, or alternatively R-2 may drape both R-SW and R-1 ("?" symbols).

3 This river-channel-containing material is here itself superposed by hummocks of smooth, concave-margined material of poorly constrained age.

Rock Package II comprises three units containing river deposits – R-1, R-2 and R-SW. R-1 and R-2 always have a consistent topographic relationship (R-2 above R-1). After testing for the possibility that the meander belts were deposited in deep incised valleys, we conclude that this topographic relationship corresponds to a stratigraphic relationship (R-2 postdates R-1; Fig. 12). In other words, the rock units stratigraphically encapsulate the river deposits – the relative time order of the river deposits in R-1, R-2 and R-SW is the same as the stratigraphic order of the rock units that host them. The time-ordering of R-SW is uncertain; R-SW may postdate R-2, or alternatively correlate with R-1 (Fig. 16). The evidence for these interpretations is as follows.

R-1. Erosionally-resistant meander belts disappear beneath and reappear from underneath smooth lobes of R-2 at >10 locations along the R-1/R-2 contact (Figs. 13-14). In Fig. 14, the superposing R-2 lobes are much wider than the superposed R-1 meander belts. If the R-1 meander belts are the fill of incised valleys (Cardenas & Mohrig 2014, Christie-Blick et al. 1990), then this relationship constrains the timing of incision that predated the fill: incision must have occurred before R-2 or the valleys would be visible as cuts into R-2, which they are not. The stratigraphic tie between river deposits and hosting strata is even tighter if the meander belts were not confined within incised valleys, but instead bordered by floodplains much wider than the meander belts. In an aggrading floodplain, the maximum elevation of the channel deposits above the floodplain is limited to $\lesssim$1 channel depth ($\lesssim$ 10m) by levee breach and avulsion (Slingerland & Smith 1998, Mohrig et al. 2000, Jerolmack & Paola 2007).

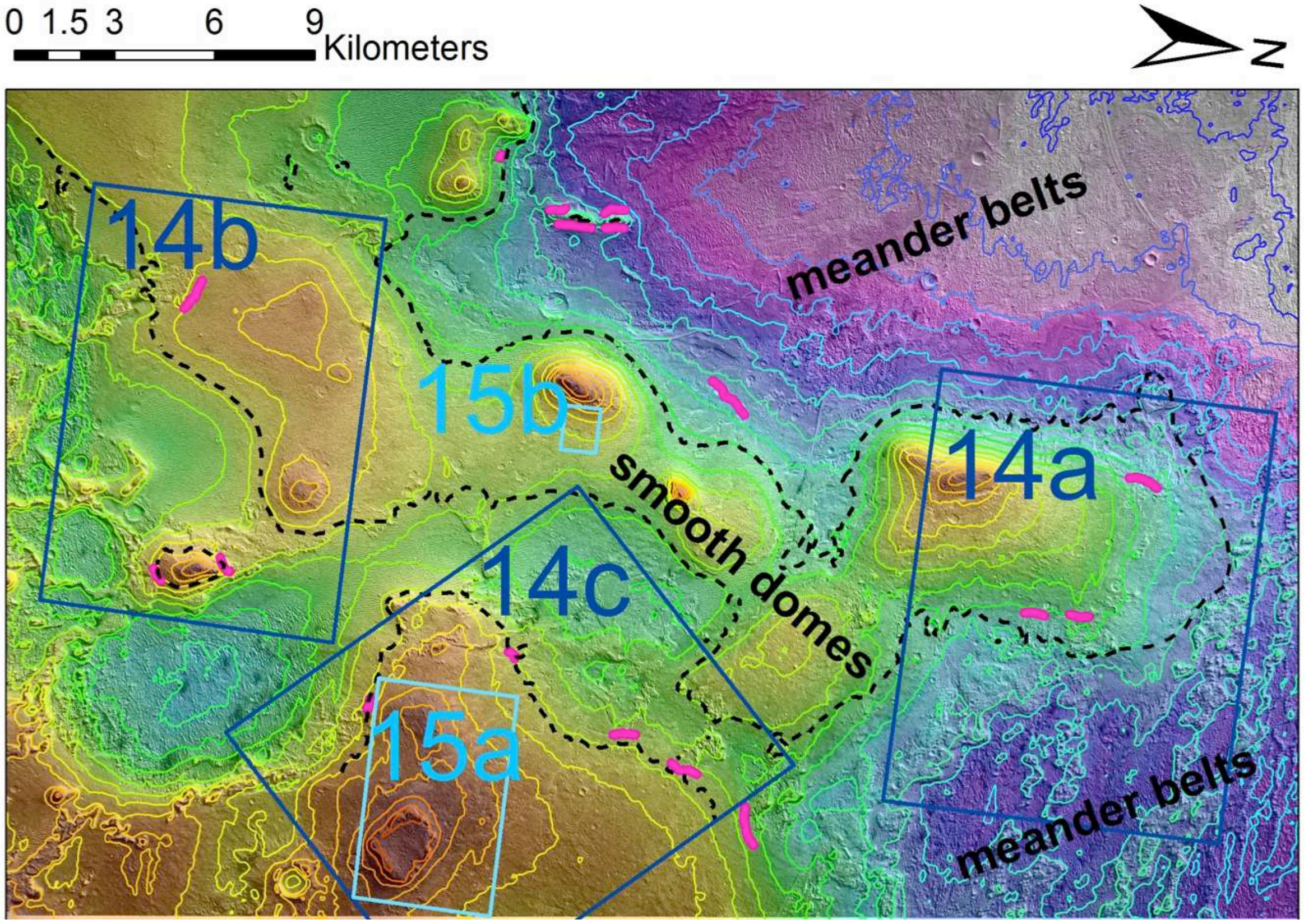

**Figure 13.** Context for locations of river deposit - host rock relationships detailed in Figs. 14-15. Dark blue: R-1 meander belt – host rock relationships shown in Fig. 14. Light blue: R-2 channel-deposit relationships shown in Fig. 15. Pink highlights places where meander belts disappear beneath overlying layered river-deposit-containing sediments. Dashed black line shows interpolated (smoothed) contact between R-1 (above) and R-2 (below). Contours (25m intervals) and background colors from CTX DTM (B20_017548_1739_XI_06S206W/ G02_019104_1740_XI_06S206W). Center is near 153.9°E 6.0°S.

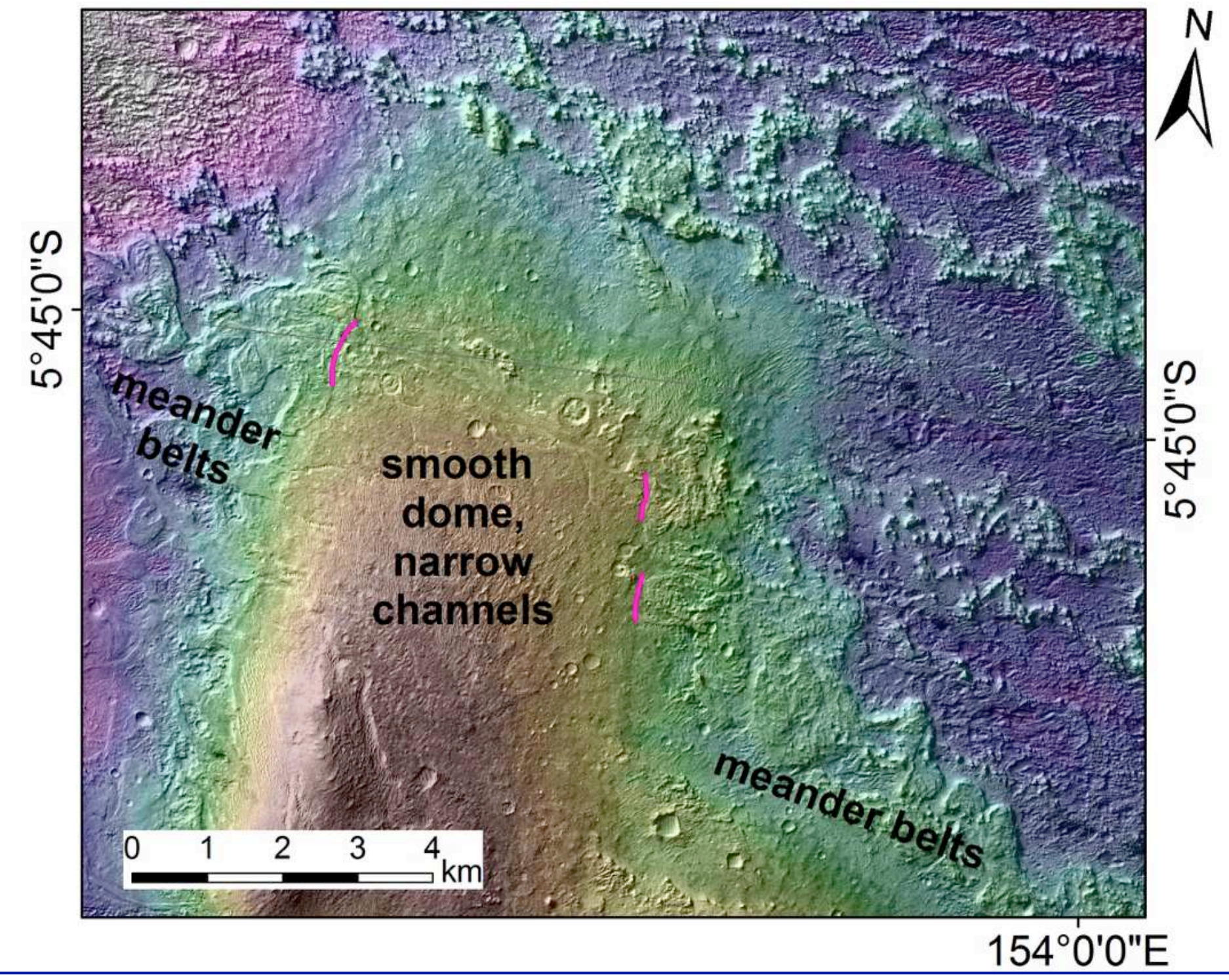

a)

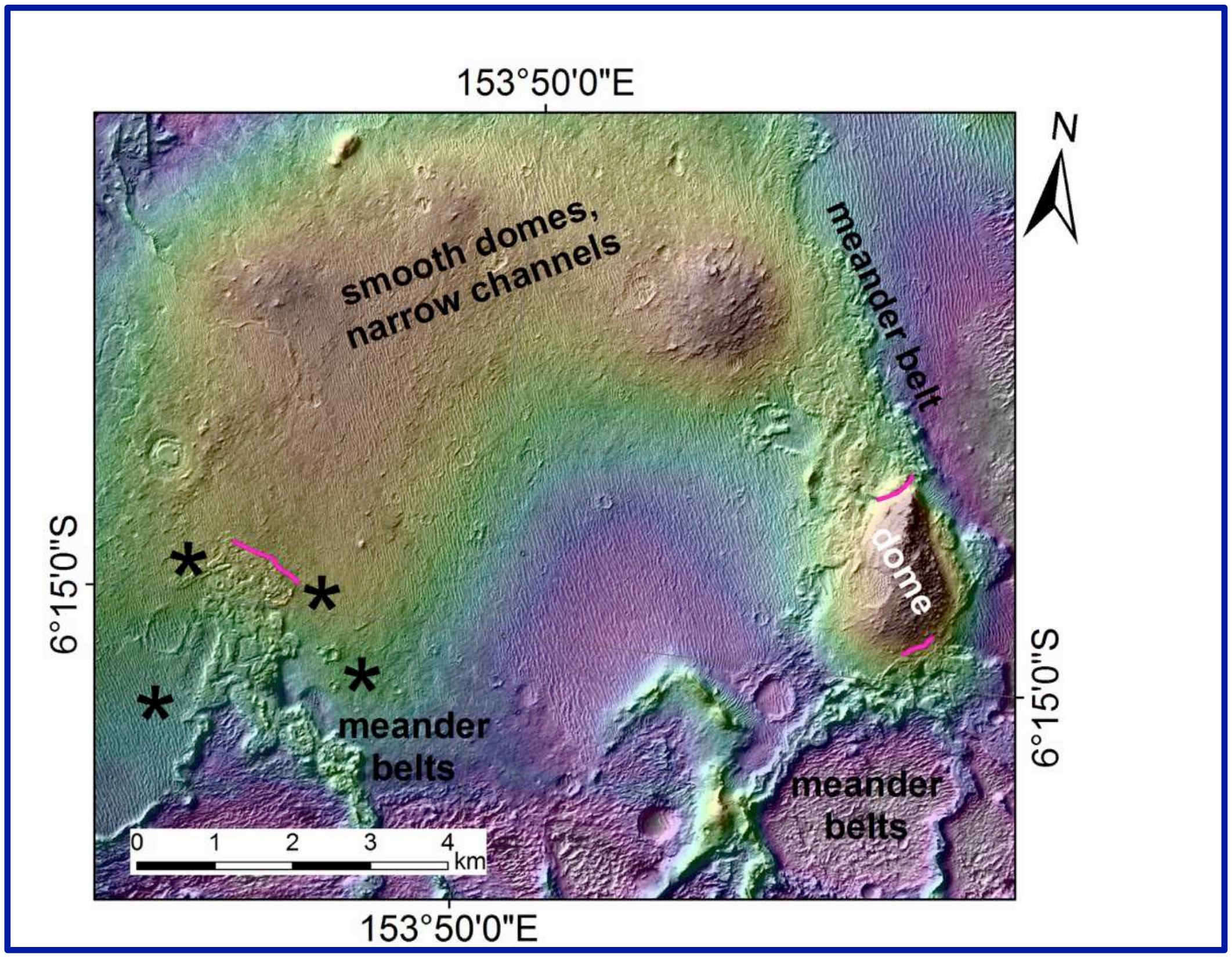
b)

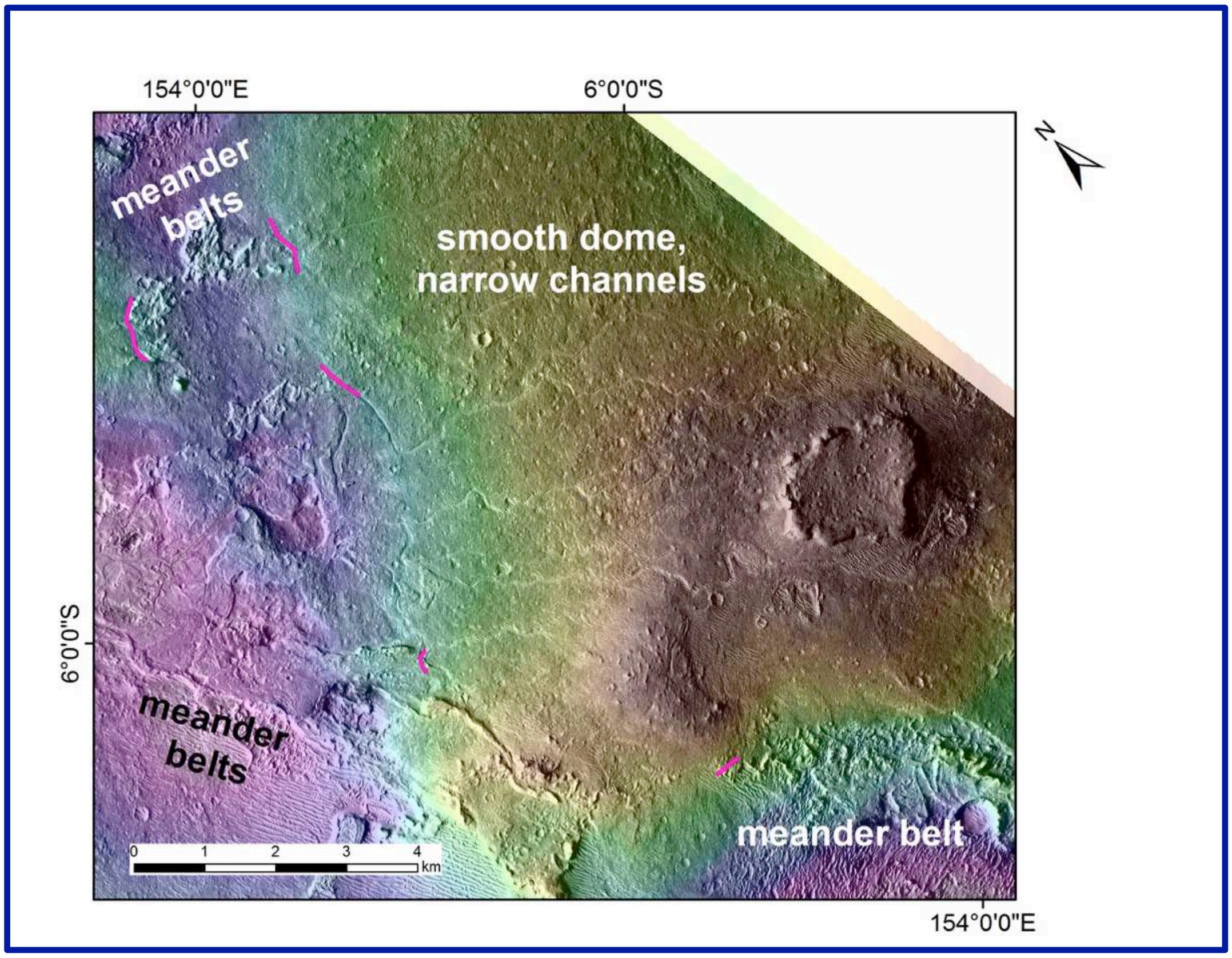
c)

**Figure 14.** Showing locations (pink highlights) where meander belts within R-1 disappear beneath overlying river-deposit containing unit (R-2). (a) Color ramp is -2335 to -2068m. (b) Color ramp is -2226m to -2036m. Black asterisks highlight areas where R-2 material may embay R-1 meander-belts. (c) Color ramp is -2245 to -2050m. Background colors from CTX DTM (B20_017548_1739_XI_06S206W/ G02_019104_1740_XI_06S206W).

R-2. We infer that R-2 channels are encapsulated within R-2 materials – as opposed to being incised into modern topography – because of four observations. (1) Many R-2 channels are preserved via topographic inversion of channel-fill. Inversion requires at least one post-incision cycle of deposition (to fill the channels) and erosion (to invert the fill). (2) River channels in R-2 to the E of 153°E generally trend SSE-NNW or SE-NW (an example is the PSP_007474_1745/ ESP_024497_1745 stereo-pair). This trend matches the trend of the underlying R-1 rivers (Figs. 2d, 13, 14). Consistency of paleochannel orientation between R-2 and R-1 implies that the R-2 outcrops are outliers of a sheet of R-2 that was once more continuous and which sloped in the same direction as R-1, and that the R-2 rivers formed prior to the erosion of that sheet back to leave patchy outliers (Fig. 2d). (3) Channels and channel deposits in R-2 frequently run obliquely or sub-parallel to topography (Fig. 15). This observation rules out the channels forming on close-to-modern topography, and suggests instead that these river deposits are eroding out of the deposit. (4) Differential compaction – which requires overburden – probably affected river deposits in R-2. Where channels cross erosionally resistant features such as embedded craters, channel fill bows upward (e.g. Kite et al. 2014, their Figure 1f). Additionally, R-2 river-channel fill is often high along the channel margins buts sags along the channel centerline. Both observations suggest differential compaction (e.g. Lefort et al. 2012). Differential compaction shows that the R-2 river deposits have been exhumed from beneath overburden as opposed to being incised into modern topography.

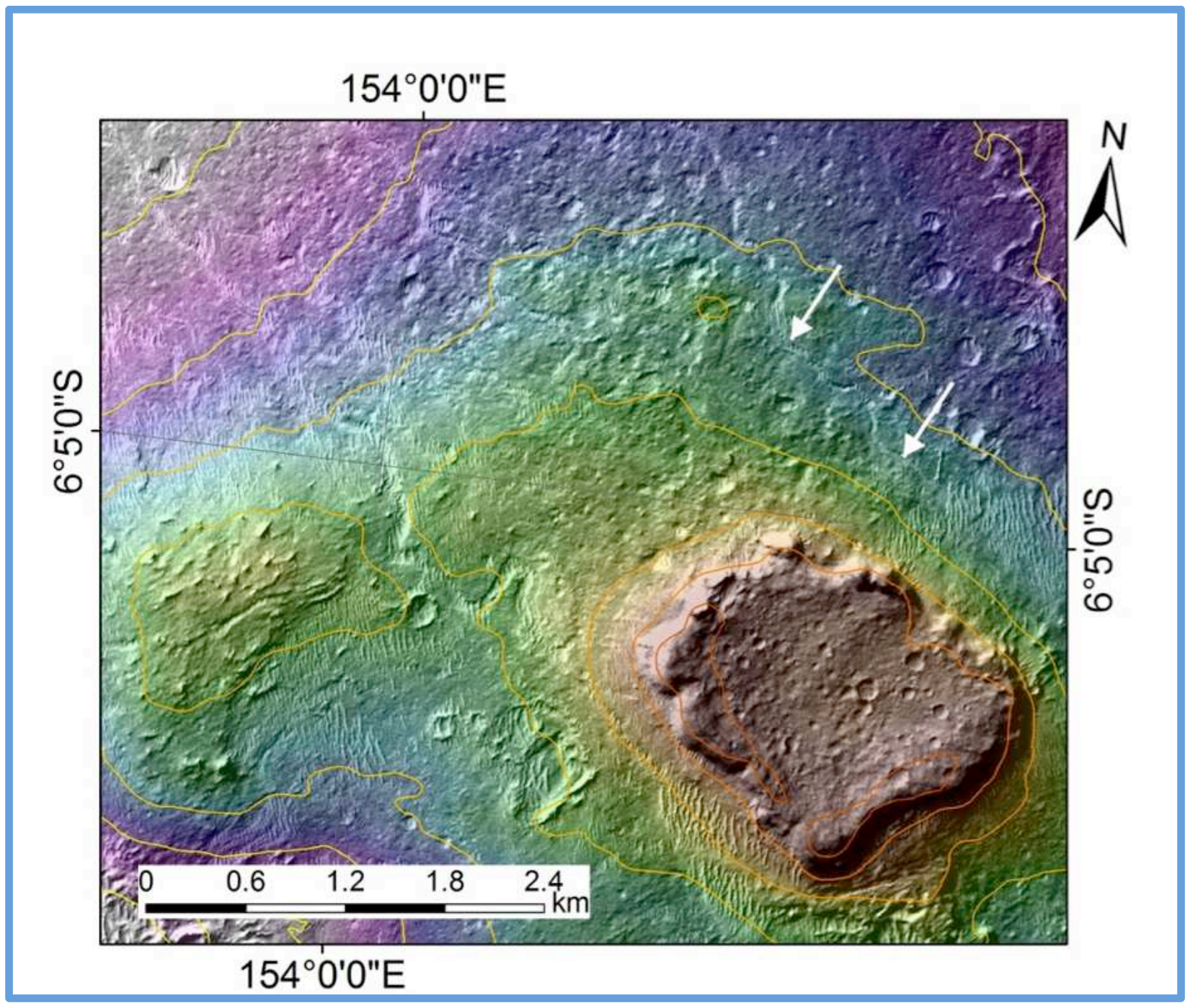

a)

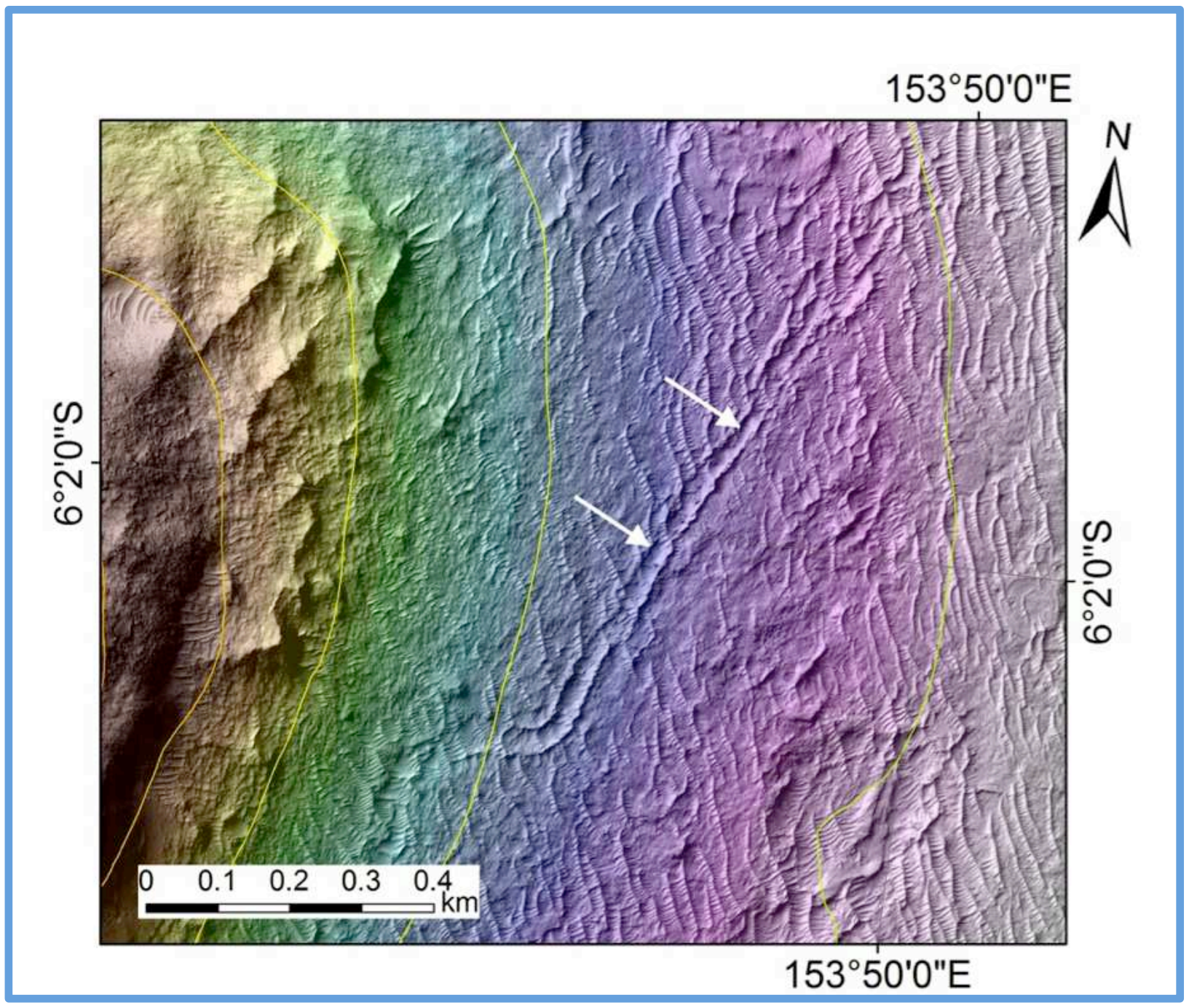

b)

**Figure 15.** Locations where river channels within R-2 run parallel to modern topographic contours (drawn at 25m intervals). Examples highlighted by white arrows.

Is R-2 concordant on R-1? The common trend of R-1 and R-2 paleochannels indicates that there was little change in downslope directions between R-1 time and R-2 time. The R-2/R-1 contact varies smoothly around an elevation of -2300m on long wavelenths, consistent with postfluvial long-wavelength tectonic deformation (Nimmo 2005, Lefort et al. 2012, Lefort et al. 2015). This is in dramatic contrast to the sub-R-1 paleosurface, the sub-Y paleosurface, and the modern erosion surface, all of which have hundreds of m of relief on short horizontal length scales indicating deep erosion. Therefore, if there was deep erosion of R-1 before R-2 was deposited, it was parallel to bedding (a regional planation surface) and very unlike other erosion surfaces in Aeolis Dorsa and elsewhere on Mars (e.g. Milkovich & Plaut 2008, Holt et al. 2010).

There is some evidence for ~10m of erosion after R-1 deposition but before R-2 deposition. For example, between the locations marked "*" in Fig. 14b, the R-1 meander belt is in high inverted relief. As the meander-belt continues to the N towards the highest portions of the R-2 deposits (the smooth-textured domal surface), the apparent inversion of the meander-belts diminishes (near the "*" symbols) and the R-1 meander belt appears to become surrounded and submerged within R-2 deposits. Although bedforms cover almost all of the R-2 material, complicating the interpretation, this is consistent with erosion of R-1 to invert the channel belts (height ~10m) followed by draping deposition of R-2 onto the eroded surface.

R-SW. R-SW is defined by erosionally resistant member with a high proportion of channel deposits, often expressed as mesa-chains. These mesa-chains are generally oriented NE-SW. There are two ways to interpret R-SW (Fig. 16a): either R-SW postdates R-2 and R-1 (Fig. 16b-c), or R-SW is correlative with R-1 (Fig 16d-e).

(1) *R-SW postdates R-2*: In this view, R-SW is ~200m topographically higher than R-1 because R-SW is ~200m stratigraphically higher than R-1 (Fig. 16b). The pattern of cliff-forming units separated by bench-forming units is interpreted as a pattern of differential erosional resistance, with cliff-forming (coarse-grained?) members representing the wettest conditions (Fig. 16c). This interpretation does not make a prediction about paleoflow direction. Paleoflow directions can change drastically during a period of aggradation (a location where this occurs, the Paleozoic/Mesozoic Grand Staircase of S. Utah, is shown in Fig. 16c). A complication with the interpretation that R-SW postdates R-2 is that materials matching the description of R-2 (and containing small channel deposits) superpose R-SW near 151.7°E 6.7°S (CTX image G05_020304_1735_XI_06S208W). If R-SW postdates R-2, then this material cannot have formed at the same time as R-2 elsewhere; it must instead represent an additional (later) return of conditions suitable for forming R-2 like deposits. Another difficulty with the interpretation that R-SW postdates R-2 is the circular depressions containing inverted channels at 152.3°E 6.8°S ("W circular depression") and 152.7°E 7.0°S ("E circular depression"). Based on channel continuity and the orientations of the channels, the "W circular depression" channels correlate with R-SW and superpose R-2 materials. The "E circular depression" channels appear to be superposed by R-2 materials, but are <100m topographically below the "W circular depression" channels (i.e., at comparable topographic elevations). This is difficult to understand if R-SW postdates R-2, but easy to understand if R-1 and R-SW are stratigraphically equivalent and R-2 is a later drape.

**Figure 16. a)** Distribution of fluvial units in SW Aeolis Dorsa (colored outlines show contacts). Channels in R-SW ramify to the NE, and channels in R-1 ramify to the SW. Topography (color ramp) is from inverse-distance-weighted interpolation between PEDR points. Color ramp is linear from -2450m (white) to -1700m (red) (range = 750m). Background is CTX mosaic. **b)** Sketch cross-section of a) showing the scenario where R-SW postdates R-2: the stratigraphic-offset hypothesis. **c)** A terrestrial analog for the stratigraphic-offset hypothesis: Grand Staircase, SW Utah, USA. Topography and shaded relief are from the 1-arcsecond National Elevation Dataset. Triassic through Early Jurassic paleochannels (not resolved in the DEM, but outcropping in the lower part of the Staircase) drain to the NW; Late Jurassic and Cretaceous paleochannels (not resolved in the DEM, but outcropping in the upper part of the Grand Staircase) drain to the NE (Blakey & Ranney 2008). Neither paleo-drainage direction matches modern topography. Color ramp is linear from 1250m (white) to 2000m (red) (range = 750m). **d)** Sketch cross-section of a) showing the scenario where R-SW is contemporaneous with R-1: the drainage-divide hypothesis. **e)** A terrestrial analog for the drainage-divide hypothesis: Blue Ridge Scarp, North Carolina, USA (Willett et al. 2014). Topography and shaded relief are from the 1-arcsecond National Elevation Dataset. Color ramp is linear from 300m (white) to 1050m (red) (range = 750m).

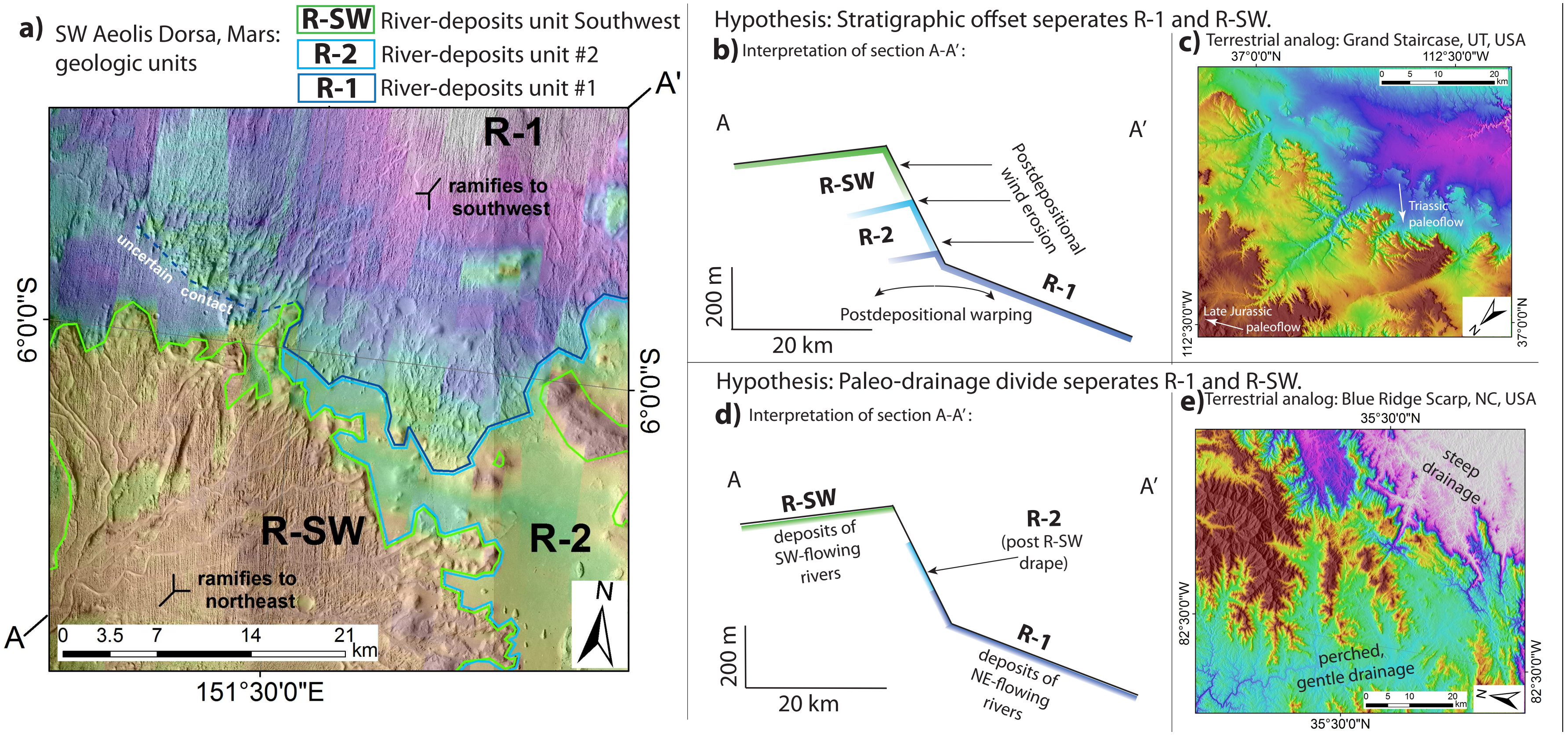

a) SW Aeolis Dorsa, Mars: geologic units
R-SW River-deposits unit Southwest
R-2 River-deposits unit #2
R-1 River-deposits unit #1
A'
R-1
ramifies to southwest
uncertain contact
6°0'0"S
6°0'0"S
R-SW
R-2
ramifies to northeast
A
0 3.5 7 14 21 km
N
151°30'0"E
Hypothesis: Stratigraphic offset seperates R-1 and R-SW.
b) Interpretation of section A-A′:
A
A′
R-SW
R-2
R-1
Postdepositional wind erosion
Postdepositional warping
200 m
20 km
c) Terrestrial analog: Grand Staircase, UT, USA
37°0'0"N
112°30'0"W
Triassic paleoflow
Late Jurassic paleoflow
112°30'0"W
37°0'0"N
Hypothesis: Paleo-drainage divide seperates R-1 and R-SW.
d) Interpretation of section A-A′:
A
A′
R-SW
deposits of SW-flowing rivers
R-2
(post R-SW drape)
R-1
deposits of NE-flowing rivers
200 m
20 km
e) Terrestrial analog: Blue Ridge Scarp, NC, USA
35°30'0"N
steep drainage
perched, gentle drainage
82°30'0"W
82°30'0"W
35°30'0"N

(2) *R-1 and R-SW are stratigraphically equivalent*: R-2 drapes and partly buries both R-1 and R-SW. In this interpretation, R-SW is separated from R-1 by a drainage divide, with R-SW draining to the SW and R-1 draining to the NE. The topographic offset in between R-SW and R-1 is then caused by a difference in base level, with the eastern side being steeper, analogous to the asymmetric Eastern Continental Drainage Divide in the Blue Ridge Mountains of Appalachia (Willett et al. 2014) (Fig. 16e). Because both R-1 and R-SW are preserved as deposits <3 km from the inferred drainage divide, mass input by atmospheric transport is required to exceed divergence by fluvial transport (a terrestrial analog for this fluvial-eolian recycling is discussed in Blakey, 1994). A problem with the interpretation that the R-SW / R-1 contact zone is a drainage divide is to account for the large widths and large wavelengths of sinuous ridges within R-SW and <3 km from the inferred drainage divide (Fig. 7b). One possibility is that these sinuous ridges represent indurated hyporheic zones, in which case their width could greatly exceed the width of the paleochannels.

Future work might test the drainage-divide hypothesis by using meander-migration directions or meander asymmetry to constrain paleoflow directions in R-1 and R-SW.

## 3.3 Fan-shaped deposits unconformably postdate thrust faults which crosscut thick river deposits.

Fig. 17 shows that fan-shaped deposits are deflected by thrust faults and that the thrust faults crosscut large-river deposits. Specifically, wrinkle ridges (which are the surface expression of thrust faults; Golombek et al. 2001) deflect fan-shaped deposits, but inverted channels underlying the fan-shaped deposits can be traced across the wrinkle ridge. Therefore thrusting occurred after inverted channels and before the fan-shaped deposits (interpreted as alluvial fans). Therefore, the alluvial fans must postdate the large-river deposits. Does this change represent merely a climatic pulse of high sediment yield causing fan buildup (without a long spell of dry conditions), or does it represent a significant time gap in the record of surface runoff?

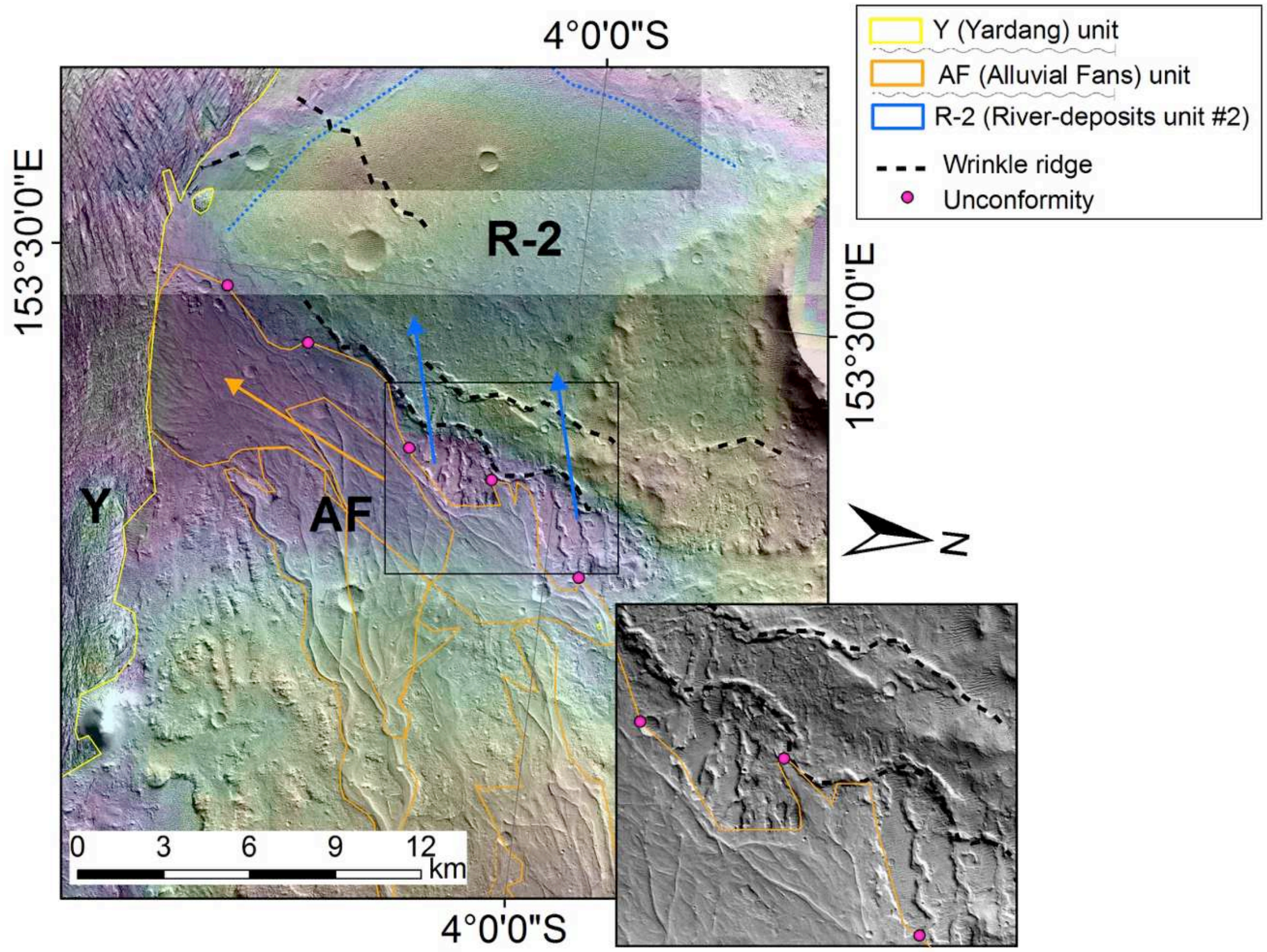

**Figure 17.** Showing unconformity between fan-shaped deposits (interpreted as alluvial fans) and large-river deposits. Wrinkle ridges (black dashed lines) deflect fan-shaped deposits (orange arrows) but inverted channels underlying the alluvial fans (blue arrows) continue across the fault. Thin black rectangle outlines location of inset (lower right; P17_007830_1754_XI_04S206W). Wrinkle ridges in Aeolis Dorsa can be clearly distinguished from inverted channels because wrinkle ridges have a preferred orientation (NNE-SSW), are steeper on one side, show backthrusts, postdate river deposits, are more linear than inverted channels, and have more rugged km-scale surface topography than do inverted channels. Yardang-forming material overlies fan-shaped deposits.

The low likelihood of the change in river deposit style occurring at the same time as thrusting favors a long time gap on the unconformity. Thrusting must have occurred after the river deposits. This is because faulting in the presence of crosscutting rivers leaves distinctive offsets (Burbank & Anderson, 2011) that are not seen here – and because there is no evidence for fluvial erosion of the wrinkle ridges. However, none of the alluvial fans flowed across the fault. A conservative estimate of the time gap for this sequence of events is > 4 $x10^7$ yr. This is based on a probability argument, which has the following steps: (i) There is no direct causal relationship between the evolving state of stress in the Mars lithosphere (which sets the timing of new faults) and the evolving climate of Mars (which dictates the transition between rivers and alluvial fans) (e.g. Andrews-Hanna et al. 2008a, Solomon et al. 2005). (ii) Therefore, the event "a new fault breaks the surface" and the event "climate shifts to favor alluvial fans" can be approximated as independent and uncorrelated. Lack of correlation of independent and random events implies that the most likely time gap is ½ the duration of the era within which large-scale climate-driven surface runoff could have occurred. (iii) Planetary contraction and runoff episodes both appear to have concentrated in the Late Noachian, Hesperian, and Early Amazonian (Nahm & Schultz 2011, Fassett & Head 2008, Grant & Wilson 2012). (iv) We assume that the runoff episodes and the major shift in faulting style could only have occurred in the Hesperian (which is conservative in terms of setting a lower limit on the time gap at the unconformity). We assume further that there were only 2 intervals of climate-driven runoff and 1 major shift in faulting style during the Hesperian, that all 3 events were brief in comparison to the duration of the Hesperian, and that they were independent random processes[4]. (v) With these assumptions, the most likely time gap is ½ the duration of the Hesperian (i.e., $\sim 1.5 \times 10^8$ yr), and a conservative estimate of the time gap at the unconformity is $>4 \times 10^7$ yr, which corresponds to the $2\sigma$ lower limit on the time gap[5].

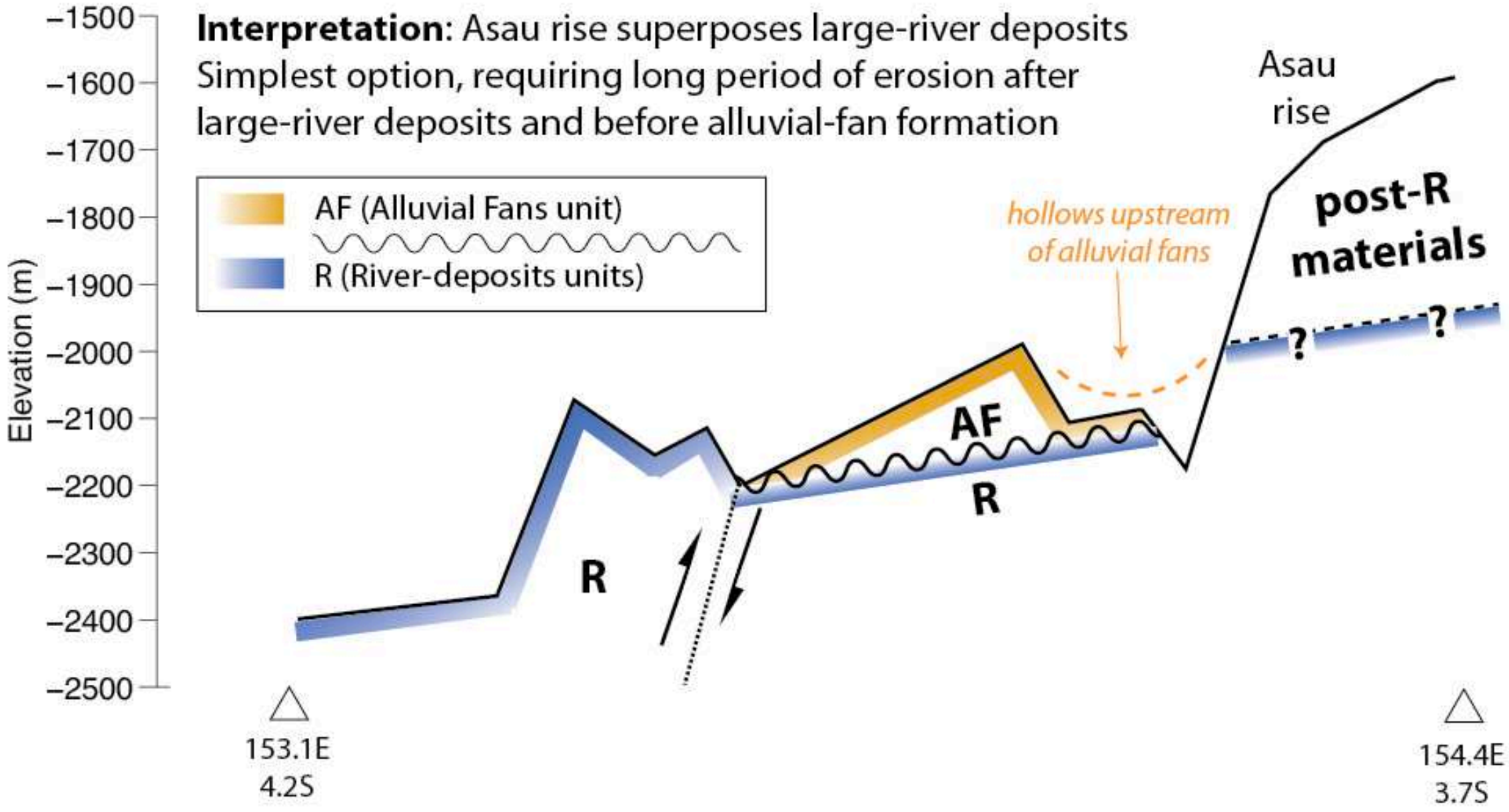

**Figure 18.** Interpreted stratigraphic relationship between Asau lobe and fluvial deposits. Fig. 4a shows line of section. Fig. S8 shows alternative interpretation.

---

[4] These assumptions are consistent with embedded-crater counts (Kite et al. 2013a), absence of major planetwide surface weathering at this time (Ehlmann et al. 2011), erosion modeling (Hoke et al. 2011), lithospheric stress-evolution modeling (Andrews-Hanna et al. 2008a), and a global geologic synthesis (Fassett & Head 2011).

[5] The probability distribution of the time gap between runoff events conditional on the change in faulting style occurring between the runoff episodes is a quadratic function of the time gap, such that the $2\sigma$ lower limit is ~13% of the duration of the Hesperian (nominally $(3.3\pm0.1) \times 10^8$ yr; Michael, 2013).

We do not know if this time gap is correlative underneath fan-shaped deposits across the basin. However, two arguments taken together suggest that alluvial fans around Aeolis Dorsa are broadly correlative with each other: (1) Aeolis Dorsa fan-shaped deposits are at similar elevations, always superpose river deposits, are never found beneath river deposits, and are always associated with upstream moats/hollows and alcoves cut back into high-standing deposits (e.g. Fig. 18, Fig. S7). This similarity of stratigraphic relationships suggests similar relative timing. (2) Although alluvial fans can result from intermittent localized triggers (Williams & Malin 2008, Goddard et al. 2014, Mangold et al. 2012, Kite et al. 2011), we did not find any evidence for localized water sources (such as volcanic fissures) in Aeolis Dorsa. We also did not find any evidence for localized impact-induced precipitation. This favors the alternative of alluvial-fan formation triggered by regional ("synoptic") climate change (Grant & Wilson 2012). Synoptic climate change would be correlative between fan-shaped deposits across Aeolis Dorsa.

## 3.4 The unconformity below the yardang-forming materials corresponds to a >1 × $10^8$ yr time gap.

On Earth, unconformities can represent >1 Gyr time gaps (Karlstrom & Timmins 2012, Peters & Gaines 2012). We care about the time gaps on unconformities both because of their chronological significance, and because erosion processes can be constrained by the time gaps and the relief on unconformity surfaces.

Many impact craters that formed in the river deposits (R) are now partly-covered by yardang-forming materials (Y) (Fig. 19a, Table S1). In order to allow time for these impact craters to form, there must have been a long time gap after the end of river-deposit deposition, and before the onset of deposition of yardang-forming materials. During this time gap, resurfacing was limited – if net resurfacing exceeded the depths of the now-partly-covered craters, then those craters would not have been preserved. The duration of the time gap can be quantified using the population of craters that are partly-covered by yardang-forming materials (i.e. that are embedded within the stratigraphy, at the unconformity). Embedded-crater constraints on the time gap would allow rate estimation of the paleo-erosion processes at the unconformity.

Because layers are commonly exposed around the edges of the outcrops of yardang-forming materials, Y was formerly more extensive. Therefore, we interpret the population of craters that are partly-covered by Y as being exhumed from beneath Y (Edgett 2005), and preserved because of a contrast in resistance to wind-induced saltation abrasion between R materials (more resistant) and Y materials (less resistant).

We want to convert a crater count $n(\phi)$ (where $n$ is the number of craters, and $\phi$ is minimum diameter) to a time gap (units yr). To do this we need estimates of crater flux ($n$ /km$^2$ /yr) and of count area (km$^2$). The details of our approach are given in the Supplementary Methods; the results are shown in Table 1.

We find that time gaps <$10^8$ yr are very unlikely. Such short time gaps can only be obtained if relatively small craters (1 km > $\phi$ > 0.5 km) are representative of the true time gap. It is very unlikely that the 1 km > $\phi$ > 0.5 km crater population is representative of the true time gap because of the high density of $\phi$ > 1km craters. The number of $\phi$ > 1km craters is 22, which is not small, and it is very improbable that this large number of $\phi$ > 1 km craters could have formed in the small time interval suggested by the 1 km > $\phi$ > 0.5 km population. On the other hand, there are many processes that can preferentially remove smaller craters from our counts. We conclude that the time gap on the unconformity is >$10^8$ yr. We emphasize that this is a lower limit. It assumes a high impact flux (corresponding to 3.7 Gya). If the unconformity is younger, then the impact flux was lower and so the time gap on the unconformity must have been longer than $10^8$ yr. Additionally, our estimation procedure linearizes the flux at the assumed start time of the unconformity. In reality the flux declines over time, so the $10^8$ yr time gap obtained using this linearization underestimates the time gap at the unconformity.

Relative crater densities at the sub-Y unconformity are consistent with the stratigraphy established by crosscutting relationships – in which a topographically high-standing early landscape (outliers of which form rock package I) predates river deposits (rock package II), which in turn predate alluvial fans (rock package III) (Figs. 20-21). Consistent with this history, the density of craters found embedded at the unconformity is highest where rock package IV drapes rock package I, intermediate where rock package IV drapes rock package II, and we found only 1 crater embedded at the unconformity where rock package IV drapes rock package III (Table S1).

Despite the small number of craters embedded at the unconformity where rock package IV drapes rock package III, there must have been a significant time gap between the end of deposition of fan-shaped deposits (rock package III) and the start of yardang-forming layered deposits (rock package IV) deposition. A significant time gap is required to explain the observation that outliers of rhythmite sit in hollows between inverted channels in the alluvial-fan deposits (Fig. 19b, 19c). Therefore, alluvial-fan flow ceased, and erosion of >20-30m took place to define the inverted channels and the elongated pits between the channels, before deposition of rhythmite in the inter-channel gaps. At modern Mars sedimentary rock wind-erosion rates of <1 μm/yr (Golombek et al. 2015), this would take >20 Myr.

There is also no doubt that rock package IV unconformably postdates R-2 (rather than just R-1). Channels in R-2 are oriented across lobes that are cut by ~200m troughs that contain outliers of Y (Figs. 4c, 4d). This requires ~200m of post-R-2 / pre-Y erosion, an order of magnitude greater than the minimum post-AF / pre-Y erosion and consistent with our inference of a long time gap between R-2 and Y (§3.3).

In summary, our results show nondeposition or erosion from pre-Y to Y all along the currently exposed pre-Y/Y contact. This suggests a period of regionwide nondeposition/erosion.

| $\phi$ | #craters > threshold $\phi$, all data (#craters > threshold $\phi$, lobes only) | ***N*** (per $10^6$ km$^2$)<br><br>*Details in Supplementary Methods* | | | **Time gap on sub-Y unconformity**<br><br>*Details in Supplementary Methods* | | |
|---|---|---|---|---|---|---|---|
| | | Buffered crater counting method | "Lobes only" method | **Inlier-outlier annulus normalization** | Buffered crater counting method | "Lobes only" method | **Inlier-outlier annulus normalization** |
| 0.5 km | 34 (19) | $1.5 \times 10^4$<br><br>(*Area*: $2.3 \times 10^3$ km$^2$) | $1.1 \times 10^4$<br><br>(*Area*: $1.7 \times 10^3$ km$^2$) | $2.7 \times 10^3$<br><br>(*Area*: $1.2 \times 10^4$ km$^2$) | $3 \times 10^9$ yr @3.0 Ga<br><br>$1 \times 10^8$ yr @3.7 Ga | $2 \times 10^9$ yr @3.0 Ga<br><br>$1 \times 10^8$ yr @3.7 Ga | $5 \times 10^8$ yr @3.0 Ga<br><br>$2 \times 10^7$ yr @3.7 Ga |
| **1 km** | 22 (10) | $4.9 \times 10^3$<br><br>(*Area*: $4.5 \times 10^3$ km$^2$) | $2.9 \times 10^3$<br><br>(*Area*: $3.4 \times 10^3$ km$^2$) | $1.8 \times 10^3$<br><br>(*Area*: $1.2 \times 10^4$ km$^2$) | $7 \times 10^9$ yr @3.0 Ga<br><br>$4 \times 10^8$ yr @3.7 Ga | $4 \times 10^9$ yr @3.0 Ga<br><br>$2 \times 10^8$ yr @3.7 Ga | **$3 \times 10^9$ yr @3.0 Ga**<br><br>**$1 \times 10^8$ yr @3.7 Ga** |
| 2 km | 11 (7) | $1.2 \times 10^3$<br><br>(*Area*: $8.7 \times 10^3$ km$^2$) | $1.0 \times 10^3$<br><br>*Area*: 6.6 $\times 10^3$ km$^2$) | $8.8 \times 10^2$<br><br>*Area*: 1.2 $\times 10^4$ km$^2$) | $9 \times 10^9$ yr @3.0 Ga<br><br>$5 \times 10^8$ yr @3.7 Ga | $8 \times 10^9$ yr @3.0 Ga<br><br>$4 \times 10^8$ yr @3.7 Ga | $6 \times 10^9$ yr @3.0 Ga<br><br>$3 \times 10^8$ yr @3.7 Ga |

**Table 1.** Constraints on the time gap at the sub-Y unconformity from counts of craters embedded at the unconformity. Only "definite" craters are considered. Bold border highlights preferred, conservative time-gap estimate (see text). Application of a recently-published recalibration of the lunar crater densities - radiometric age relationship (Robbins 2014) would alter these values, but would not affect our conclusions. See Table S1 for detailed crater data, and Supplementary Methods for details of methods.

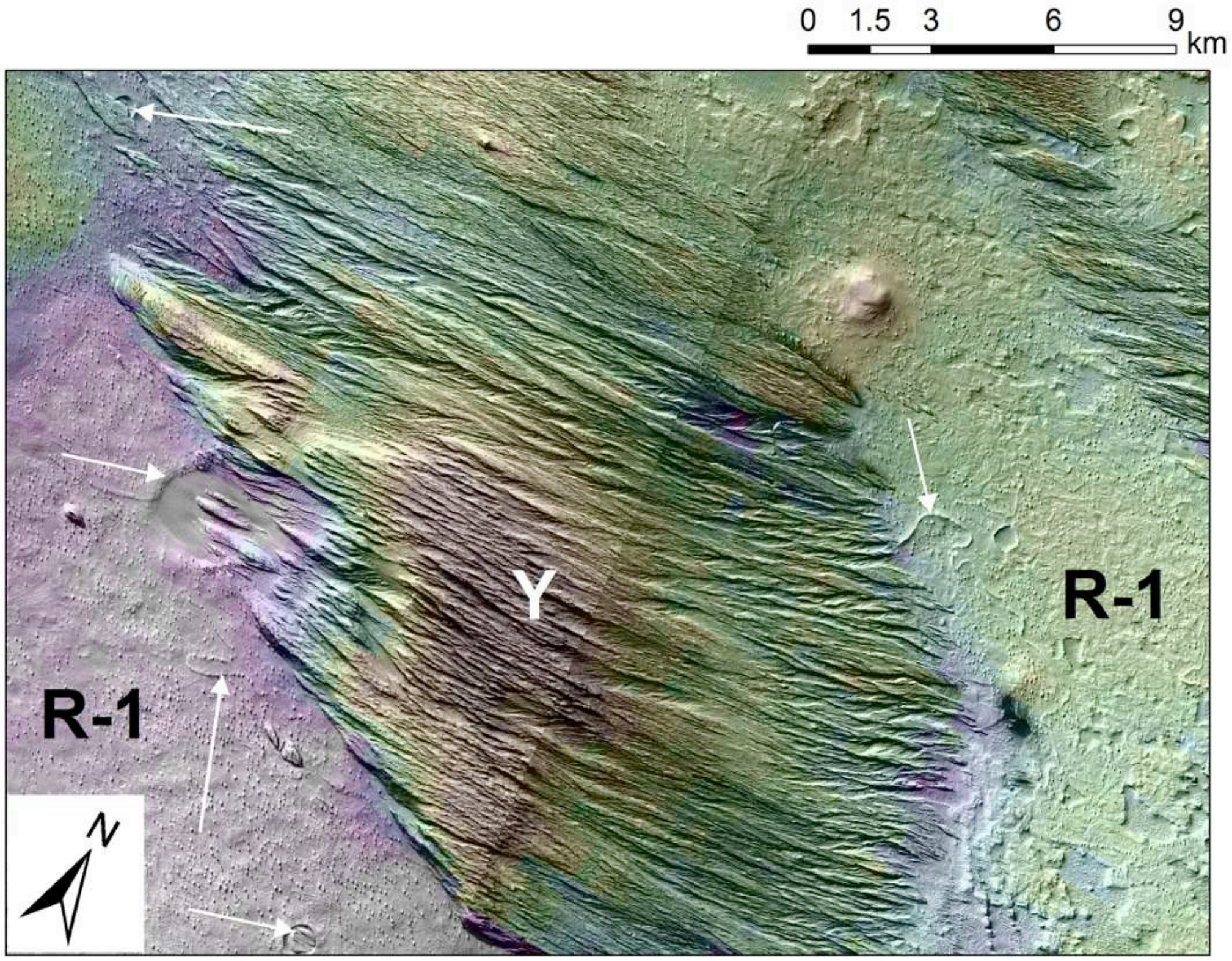

(a)

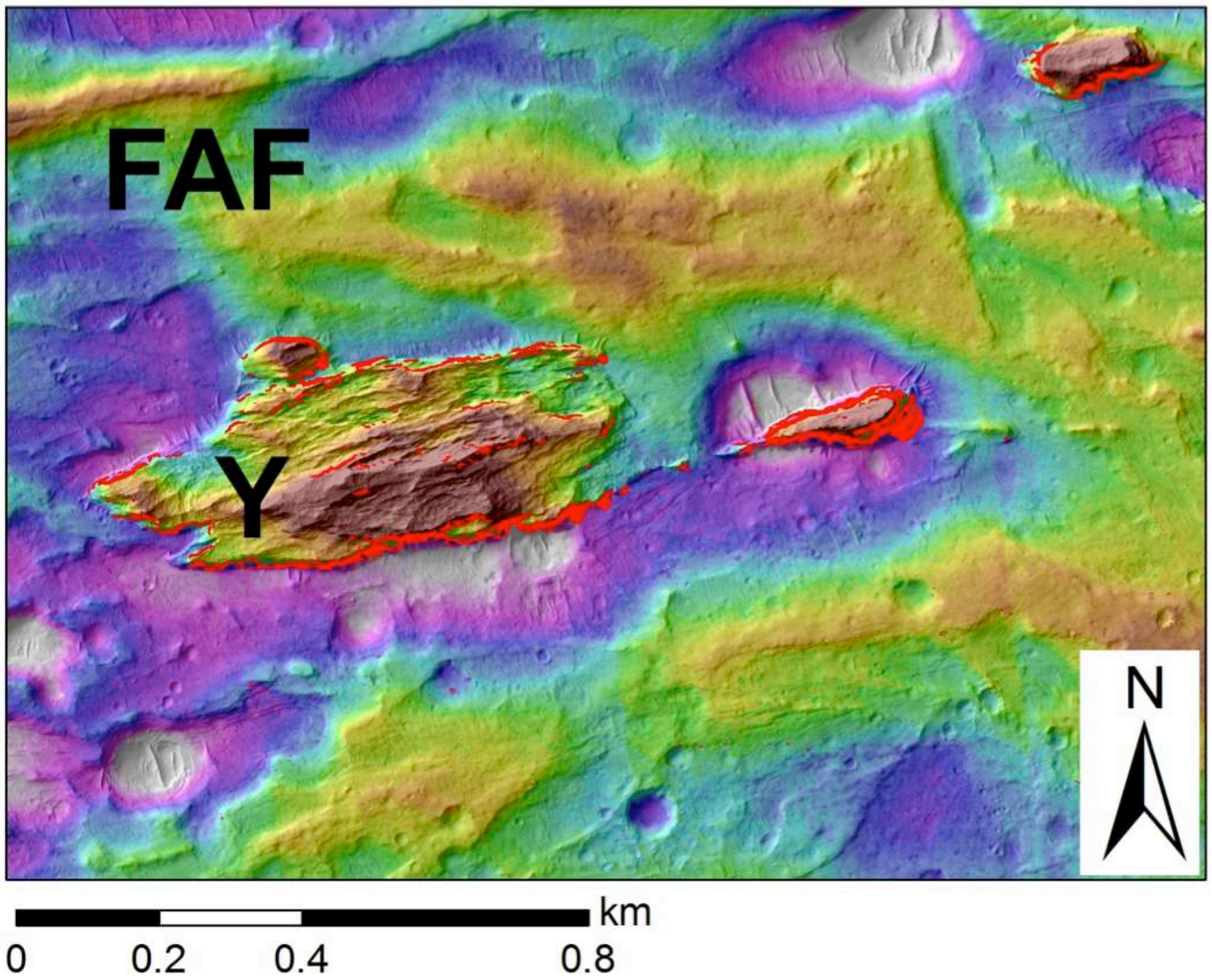

(b)

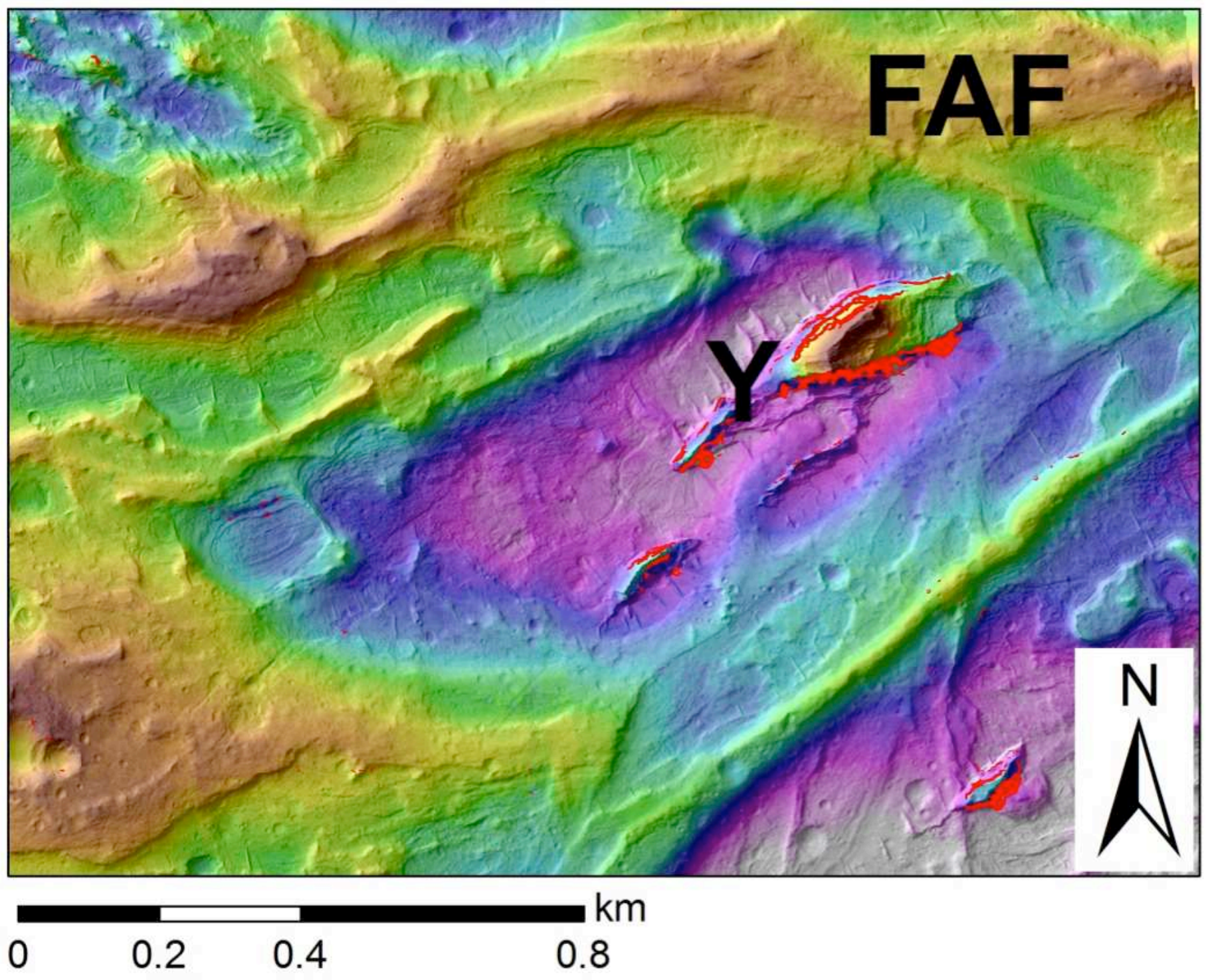

(c)

**Figure 19.** Yardang-forming layered deposits (Y) unconformably superpose both river deposits (R) and alluvial fans (AF). (a) Yardang-forming layered material (highstanding grooved terrain) superposes a surface showing both inverted channels and impact craters (white arrows highlight examples). Near 154.3°E, 5.1°S. (b) Steep-sided outliers of regularly-layered yardang-forming material in hollows on fan-shaped deposits. Opaque bright red tint corresponds to >40° slopes. The pit just S of the largest steep-sided outcrop is 20m below inverted-channel elevation. Near 154.85°E 4.72°S. (c) Steep-sided outcrops of regularly-layered yardang-forming material in hollows on fan-shaped deposits. Opaque bright red tint corresponds to >40° slopes. The drop from the north inverted channel to the pit is 30m; the drop from the south inverted channel to the pit is 20m. Near 154.90°E 4.72S. (b) and (c) are from HiRISE DTM PSP_009795_1755/PSP_009623_1755, available at `http://geosci.uchicago.edu/~kite/stereo`.

# 4. Summary of stratigraphic relationships.

Cross-cutting stratigraphic relationships permit relative time-ordering of stratal units in the Aeolis Dorsa region (Fig. 20).

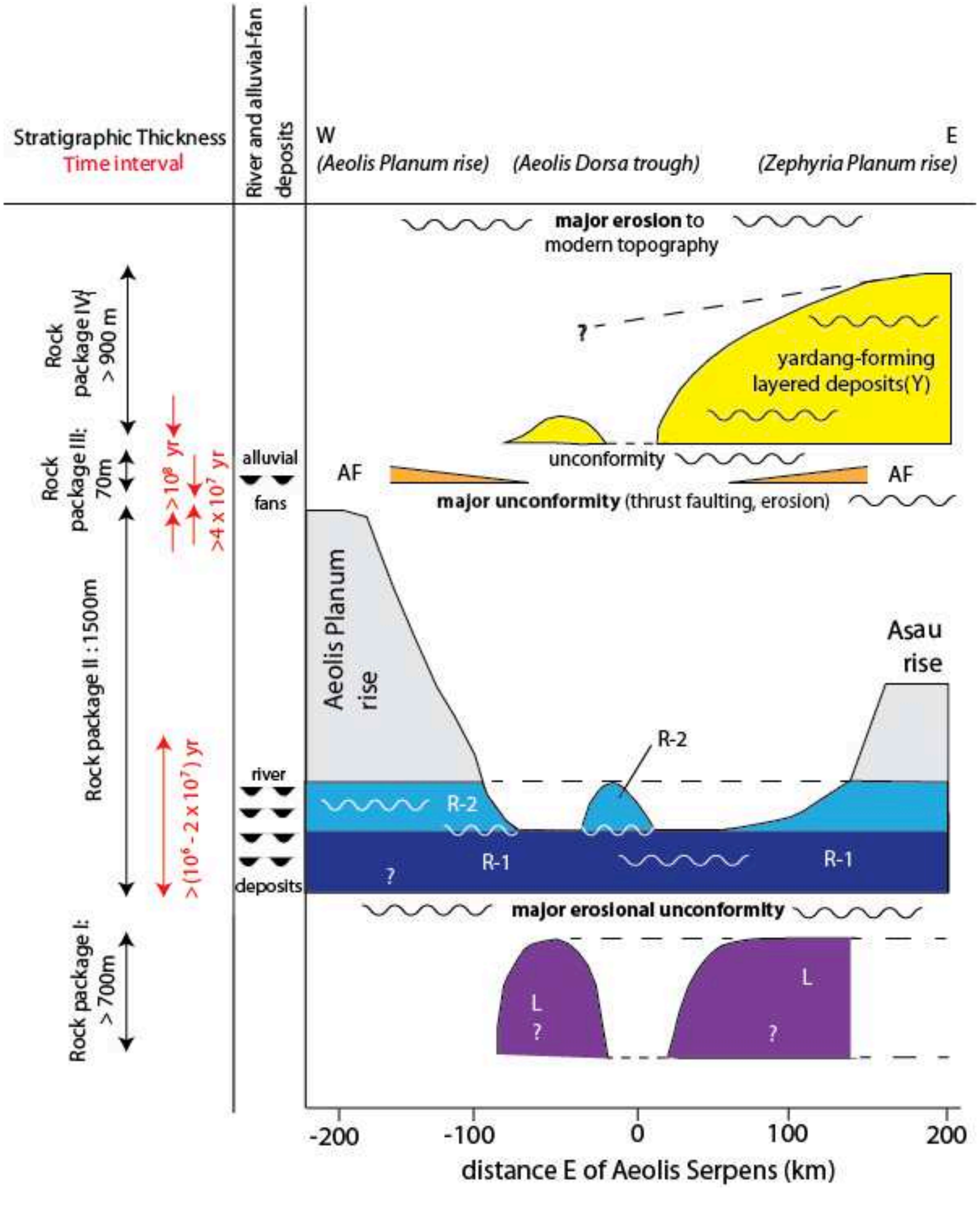

**Fig. 20.** (Relative time)-(stratigraphy) diagram showing stratigraphic context of river deposits and fan-shaped deposits in Aeolis Dorsa, and time constraints discussed in this paper. Wavy lines correspond to unconformities. Fan-shaped deposits are interpreted as alluvial fans. Filled "U" shapes in the column at left correspond to the stratigraphic levels at which river deposits and alluvial-fan deposits are found. R-SW (not shown in column) may correlate with R-1, or postdate R-2 (see discussion in §3.2).

The relationships described in §3.2 require that Aeolis Dorsa rock units stratigraphically encapsulate the river deposits (channel symbols shown in central column of Fig. 20).

It is possible that additional basinwide unconformities exist beyond those shown in Fig. 20 (Supplementary Discussion). Ruling out major unconformities in Mars sedimentary successions is difficult even with rover-scale data (Grotzinger et al. 2014).

Most unconformities have chronostratigraphic significance (Christie-Blick et al. 1988). However, because there is no "type section" where we can clearly see all the stratigraphic relationships vertically, it is possible that the different vertically-separated facies within rock package II are lateral equivalents of a prograding or retrograding system. Confusion between time-equivalence and lateral-equivalence is less likely for units from different rock packages, because of the evidence for significant time gaps and regional unconformities separating the rock packages (although time-transgressive unconformities are possible).

Topographic elevation correlates only weakly with geologic youth in Aeolis Dorsa. Erosion by wind, erosion by water, tectonism, differential compaction, and possibly ice removal and crustal flow have distorted the elevation-age relationship (Jacobsen & Burr 2012, Kite et al. 2012, Lefort et al. 2012, Lefort et al. 2015, Nimmo 2005).

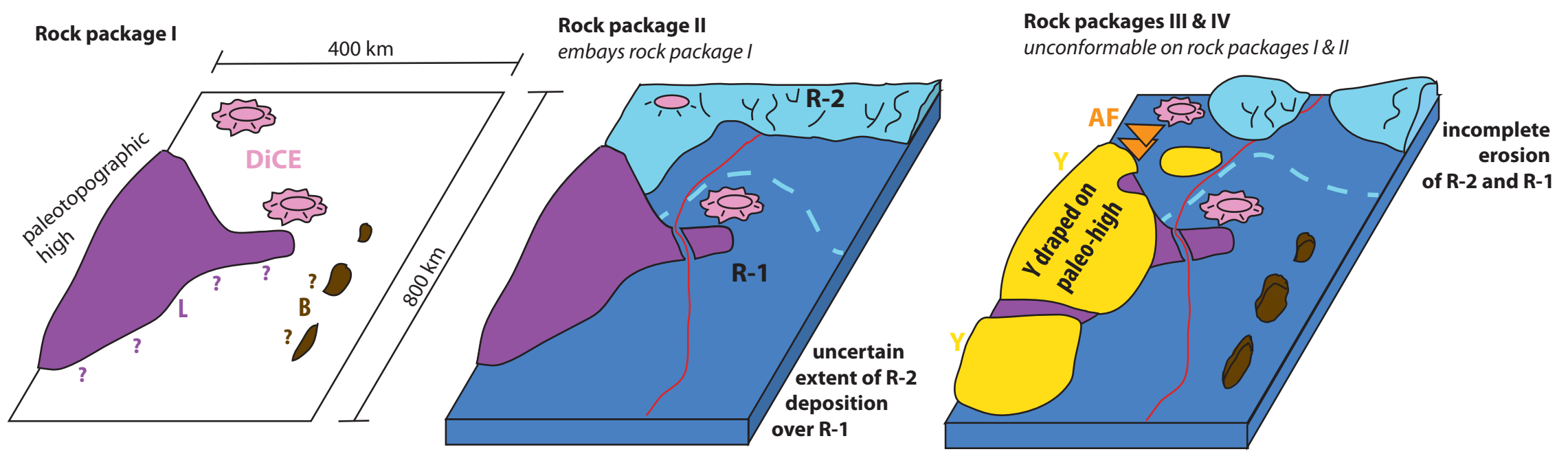

**Fig. 21.** Simplified paleogeographic evolution of Aeolis Dorsa and the area immediately to the E, showing pre-fluvial (left), fluvial (center), and post-fluvial (right) depositional episodes. Some unconformities omitted (see Fig. 20, and Supplementary Discussion).

# 5. Discussion.

## 5.1. Hypothesis: orbital forcing modulated erosion-deposition cycling in Aeolis.

Erosion and deposition alternated in the Aeolis Dorsa region (§2-§4). What caused these alternations? Our results suggest that the stratigraphy of Aeolis Dorsa is a record of liquid water availability (and aeolian sediment input), and because physical models indicate that liquid-water availability is strongly influenced by orbital forcing, we hypothesize that orbital forcing was a strong influence on the stratigraphy of Aeolis Dorsa. Alternatively, deposition may have been regulated by unsteady supply of easily-eroded sediment. Testing these hypotheses is possible through future work that correlates erosion-deposition cycles across the low-latitudes of Mars.

Deposition of river deposits, alluvial fans and rhythmite in Aeolis Dorsa indicates past periods of equatorial liquid water for runoff and sediment induration – requiring a regional climate different from today. On the other hand, the topography of the modern erosion surface, the sub-rhythmite paleosurface, and the sub-alluvial-fan paleosurface, all suggest periods of net wind erosion. These deposition-erosion cycles can be interpreted as controlled by the availability of liquid water. Widespread surface liquid water suppresses wind-induced aeolian abrasion because damp ground binds the tools (sand) that aeolian abrasion requires. A boost for the water-limited interpretation is that there is a straightforward mechanism allowing for wet-dry cycles in Aeolis: obliquity change. Liquid water at the equator requires a different climate from today – atmospheric pressure $\gtrsim 10^2$ mbar (to suppress evaporitic cooling) and obliquity $\gtrsim 40°$ (to drive ice and snow to the equator) (e.g. Jakosky & Carr 1985, Mischna et al. 2003, Forget et al. 2013, Kite et al. 2013b, Hecht 2002).[6] Under these conditions, melt can occur for infrequent, but expected orbital forcings (e.g., Jakosky & Carr 1985, Kite et al. 2013b). At low obliquities, similar to today's 25°, ice and snow is cold-trapped at mid-to-high latitudes and at depth, and low-elevation surfaces near the equator quickly dry out (Mellon & Jakosky 1995, Hudson & Aharonson 2008). At low obliquities, rivers and alluvial fans cannot form at low elevations near the equator even if temperatures are high enough for melting.

Our simulations of Mars long-term obliquity change (Figs. 22-23) (Armstrong et al. 2004) show chaotic, long-term alternations between obliquity >40° and obliquity <40°. Intervals with $>10^8$ yr continuously low mean obliquity are common. Fig. 22 suggests a hypothesis for the rock-package-scale stratigraphy of Aeolis Dorsa (Fig. 20): *at high obliquity, liquid water was intermittently available to move and indurate sediment, and at low obliquity, net wind erosion led to the development of unconformities*. In this hypothesis, other global factors (e.g. loss of the atmosphere via carbonate formation and via escape to space; Lammer et al. 2013) led to the drying-out recorded by successive intervals of deposition at high obliquity – from large rivers, to alluvial fans, to indurated rhythmite.

[6] Low-elevation equatorial liquid water at obliquity <40° is possible if the ancient climate was much warmer or wetter. Therefore, if the atmosphere was much warmer or wetter, then the importance of obliquity as a driver of erosion-deposition cycling would be reduced (Wordsworth 2014, Andrews-Hanna & Lewis 2011).

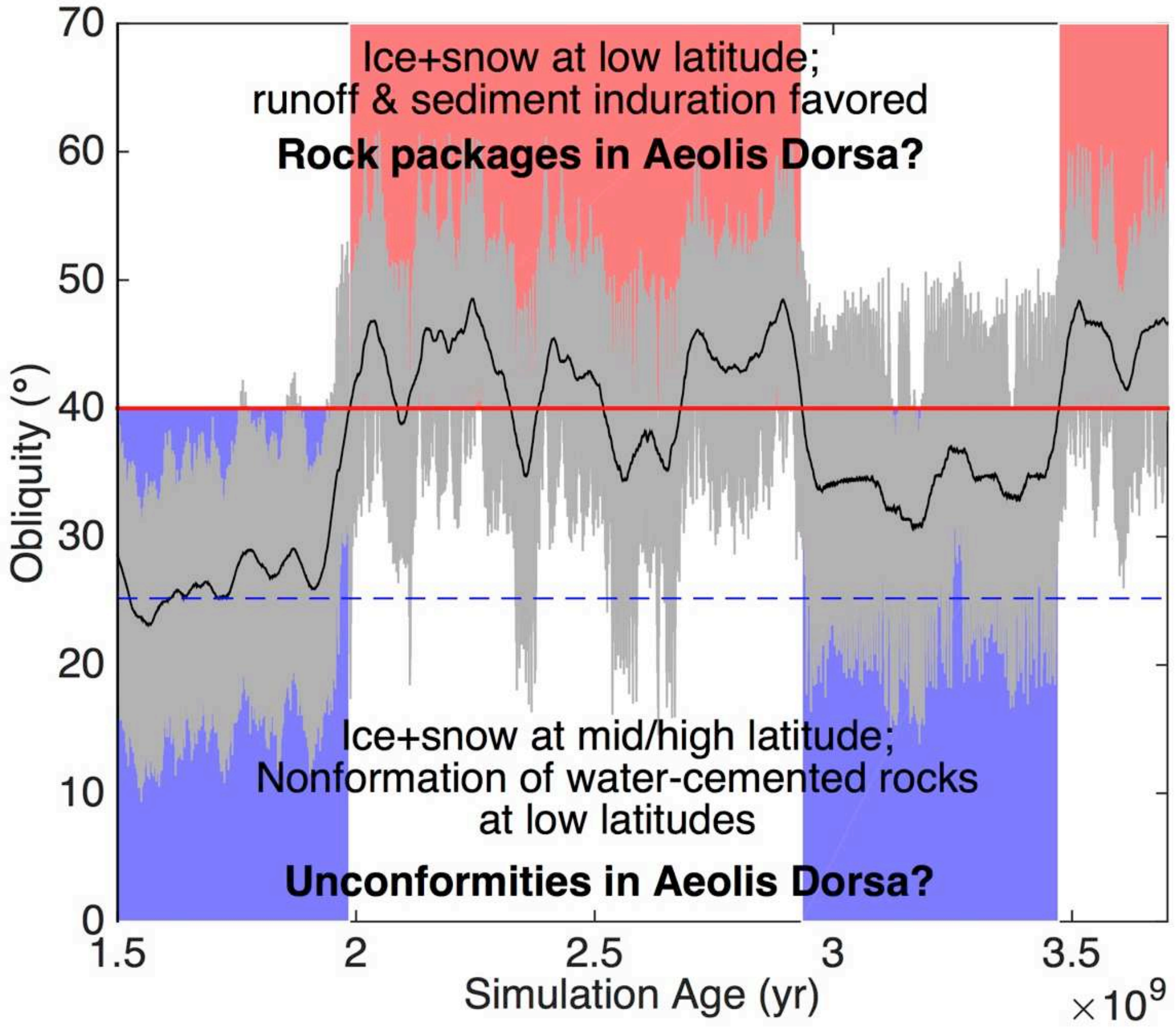

**Figure 22.** One possible Mars obliquity history, showing alternations between low and high mean obliquity. Gray swath shows obliquity output, and black line shows 50 Myr running-average of obliquity. Dashed blue line corresponds to present-day obliquity of Mars (25°). Red line corresponds to threshold for low-elevation equatorial ice and snow. This threshold can be slightly lower in some models (but no lower than 30°). Simulation by J.C. Armstrong, using the `mercury` integrator (Chambers 1999) and an obliquity model based on Armstrong et al. (2004). Because the Solar System is chaotic, it is impossible to recover the single true 4+ Gyr history of Mars' obliquity solely by reverse-integrating the solar system from inexact observations of its present state (Laskar et al. 2004).

High obliquity alone is insufficient for surface liquid water (Fig. 22). Mars is expected to spend >1 Gyr at >40° obliquity (Laskar et al. 2004), which at sediment accumulation rates of 10-100 μm/yr (Lewis & Aharonson 2014) would lead to 10-100 km stratigraphic thicknesses. This is greater than observed sedimentary-rock thicknesses (1-10 km). Therefore, some other factor is necessary for sedimentary rock formation and helps to limit the growth of sedimentary rock piles, e.g., water supply (e.g. Kite et al. 2013, Andrews-Hanna & Lewis 2011), supply of ions for cementation (Milliken et al. 2009), or sediment supply.

Alternative controls on stratigraphy are tectonics, eustasy, and unsteady sediment supply. Tectonics and eustasy are less plasusible than orbital forcing as pacemakers of Martian rock package stratigraphy. Tectonic forcing of alternations between erosion and deposition requires uplift or subsidence, but the basic tectonic fabric of Aeolis was set in place very early in Mars history (Watters et al. 2007, Phillips et al. 2001). In particular, the equatorial dichotomy boundary scarp was a 3km-high cliff well before Aeolis Dorsa formed (Andrews-Hanna et al. 2008b, Marinova et al. 2008, Nimmo et al. 2008, Irwin & Watters 2010, Andrews-Hanna 2012): it is still a 3km-high cliff today. Therefore, given sufficiently wet conditions, relief is unlikely to have limited aggradation. Sedimentation can be controlled by variation in true sea level (eustasy). Although other studies have inferred that the river units were deposited in conjunction with a large standing body of water thay may have fluctuated in level (DiBiase et al. 2013, Cardenas & Mohrig 2014), there is no positive evidence for a northern ocean on post-Noachian Mars (Malin & Edgett 1999, Ghatan & Zimbelman 2006).

Unsteady supply of sand, silt, and dust could have contributed to erosion-deposition cycling. Dust input may have been regulated by orbital parameters, coupling sediment supply to liquid water availability. Therefore, the outcome of sediment-supply control might be difficult to distinguish from the orbital-control hypothesis. Impact-ejecta input and ash input would also have been unsteady, and regionally variable (e.g. Mustard et al. 2009, Kerber et al. 2011, 2012, 2013). In particular, long periods of dormancy for individual Mars volcanoes are indicated by caldera-age dating (Robbins et al. 2011) and by magma-cooling timescales (Wilson 2001). Periods of volcanic repose would correspond to time gaps in the depositional record of explosive volcanism.

Future work on global correlation might allow these hypotheses to be tested. Because ash fall and impact ejecta are regionally variable, if ash input or ejecta input limited aggradation, then global correlation of the rock packages shown in Fig. 20 should not be possible. In contrast, orbital forcing is similar at similar latitudes (Schorghofer 2008). Therefore, the orbital-control hypothesis makes a testable prediction: that Aeolis Dorsa's sedimentary history (Fig. 20) should be echoed by the history of other Martian equatorial depocenters – Meridiani (Edgett 2005, Hynek & Phillips 2008, Wiseman et al. 2010, Zabrusky et al. 2012), Isidis (Jauman et al. 2010), Valles Marineris (Murchie et al. 2009, Weitz et al. 2010, Williams & Weitz 2014), and Gale. A more immediate test is described in the next section.

## 5.2. Test: Orbital simulations predict >$10^8$ yr intervals of low Mars obliquity, consistent with the orbital forcing hypothesis.

Combining the orbital forcing hypothesis (§5.1) with the >$10^8$ yr duration of the sub-Y unconformity in Aeolis Dorsa (§3.4) yields a clear prediction: the obliquity trajectory shown in Fig. 22, which has >$10^8$ yr intervals of continuously low obliquity, should be representative (or typical) of possible Mars histories. To test this hypothesis, we used the `mercury6` code (Chambers 1999) to generate dozens of different >3.5 Gyr integrations of the orbits of the eight planets, plus Pluto. Different initial conditions were generated by perturbing Mars' position by <100m relative to JPL ephemerides for 1/1/2000. Because the Solar System is chaotic with a Lyapunov timescale of << $10^8$ yr, these initially similar conditions diverged on a timescale much shorter than the length of the simulations. An mixed-variable symplectic integrator was used, with a timestep of 1.2 days. For successful integrations, we ran an obliquity code (Armstrong et al. 2004) to calculate Mars obliquity as a function of time for a spread of initial obliquities (24 obliquities per orbital integration). The initial obliquities were drawn randomly, with weights assigned using the analytical probability distribution function of Laskar et al. (2004) for 3.5 Gyr. Mars spin rate was assumed constant with time, and post-Newtonian corrections to the precession of Mercury's orbit were neglected. Output was sampled at 200 year intervals, which is sufficient to capture the relevant dynamics. The resulting ensemble contains ~4 trillion years of simulated Mars obliquity evolution.

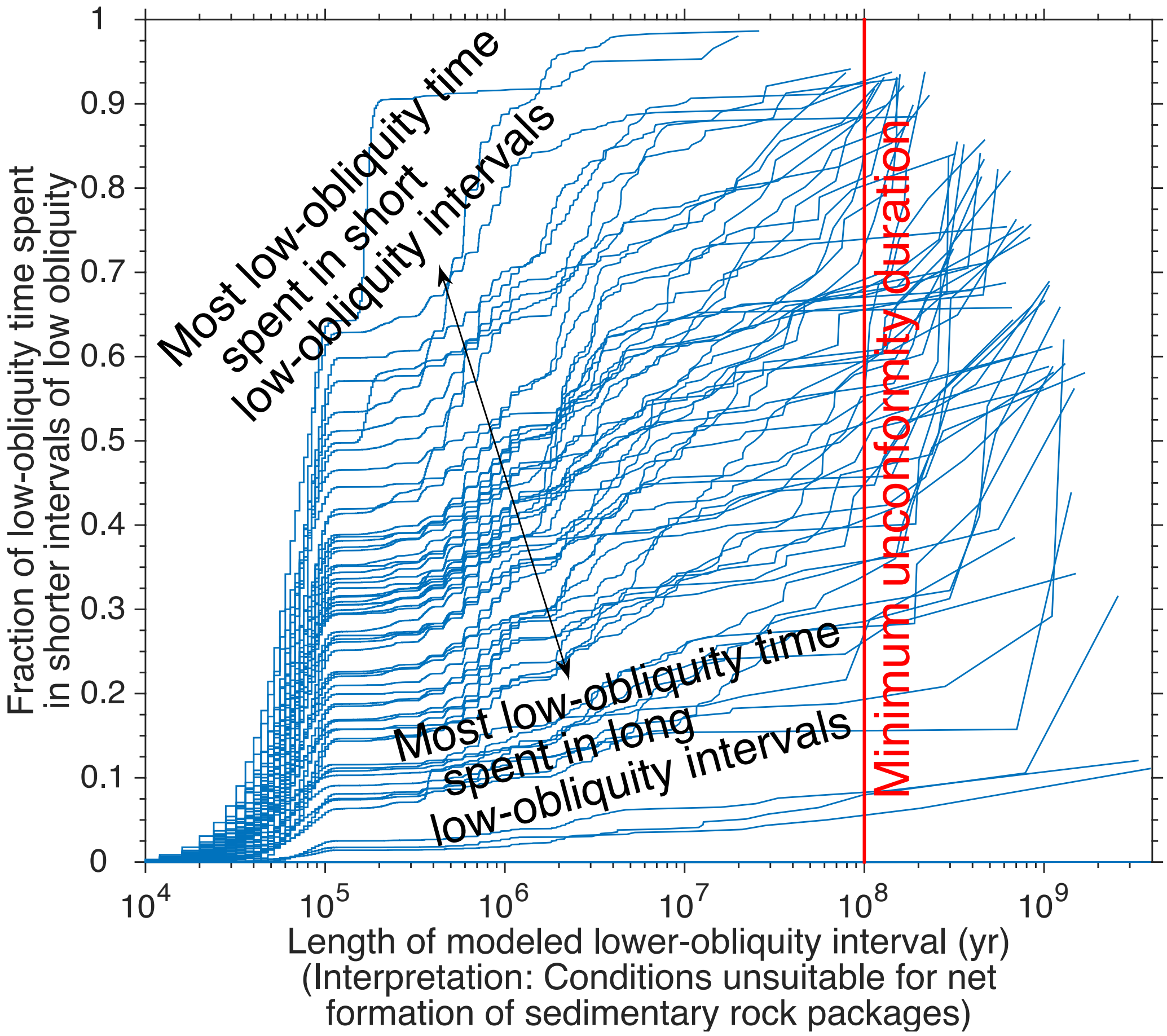

**Figure 23.** Statistics of low-obliquity (<40°) intervals calculated using the `mercury` integrator (Chambers 1999) and an obliquity model based on Armstrong et al. (2004). Each blue line shows the distribution of low-obliquity interval lengths for a different >3 Gyr long integration. Each integration represents a different, equally plausible a priori, Mars history similar to the track shown in Figure 22. The total duration of low obliquity varies between the integrations. The rightmost point on each line corresponds to the length of the longest continuous interval of low obliquity for that integration. The lower limit on the duration of the unconformity beneath the yardang-forming materials (§3.4) is shown by the red line. 57 out of 61 simulations show at least one continuous interval of low obliquity that is consistent with the duration of the unconformity.

Fig. 23 shows low obliquity interval statistics for 61 obliquity simulations drawn from the ensemble that each yielded low present-day obliquity, consistent with present-day Mars. 57 out of 61 simulations show at least one continuous interval of <40° obliquity that is >$10^8$ yr in duration. Therefore the hypothesis that the Aeolis Dorsa unconformity (>$10^8$ yr of net nondeposition) corresponds to an interval of continuously <40° obliquity is reasonable, and survives this test.

## 5.3. Correlation between Aeolis Dorsa and Aeolis Mons (Gale crater's mound, Mt. Sharp).

Correctly placing high-resolution stratigraphic sections from rover traverses in the global context of orbiter data increases the science value of both (Ehlmann et al. 2011, Grotzinger et al. 2012). This motivates correlating Gale crater (the *Curiosity* rover field site) to the wider Aeolis region. Our work suggests that yardang-forming layered material in Aeolis Dorsa correlates to the upper unit of Gale crater's mound (Aeolis Mons, Mt. Sharp), which is the destination of *Curiosity* (see also Zimbelman & Scheidt 2012). We also set out the temptations and dangers of further stratigraphic correlations between the rover field site and Aeolis Dorsa.

Aeolis Dorsa is 10°E of Gale, and like Gale it accumulated sedimentary rocks carrying evidence for overland flow and aqueous cementation (e.g. Malin & Edgett 2000). Models predict strong correlation between times when surface liquid water exists at Gale, and times when surface liquid water exists in the wider Aeolis region (Andrews-Hanna et al. 2012, Kite et al. 2013a). Therefore, we might expect the rock record of surface liquid water at Gale to correlate with the rock record of surface liquid water in Aeolis Dorsa (Kite et al. 2013a, Irwin et al. 2004, Zimbelman & Scheidt 2012).

*Yardang-forming layered materials*. Necessary conditions for lithostratigraphic correlation include multiple lines of geologic evidence that the same process formed the rocks being correlated, and a physical basis for simultaneous action of that process at the sites being correlated[7]. These conditions are satisfied for Aeolis Dorsa rhythmite and the upper Gale crater mound.

The upper unit of Mt. Sharp and the Aeolis Dorsa yardang-forming materials are both regularly bedded materials that unconformably drape underlying sediments (Malin & Edgett 2000, Anderson & Bell 2010, Thomson et al. 2011, Le Deit et al. 2013). Therefore, the sediment source was atmospherically-transported sediment – e.g., airfall. Geologically recent airfall deposits on Mars have >1000km extent (Bridges & Muhs 2012, Mangold et al. 2009). This broad extent suggests that ancient dust storms that deposited dust in Gale would also have deposited dust in

---

[7] There are obvious limits to lithostratigraphic correlations across gaps where no layers can be traced. For example, the Old Red Sandstone of Britain (Hutton 1788) formed ~150 Myr prior to the New Red Sandstone and at a paleolatitude differing by ~40° – but both are desert red-beds with river and dune deposits. Incorrect stratigraphic correlation has a directional bias: it is particularly easy to misread an unsteady or oscillatory time evolution as a monotonic change (for example, compare Bibring et al. 2006 with Wray et al. 2009, Ehlmann & Edwards 2014, Sun & Milliken 2014). This difficulty and this bias is one of the reasons why the superposition relationships that ordinate the Aeolis Dorsa river deposits are so valuable.

Aeolis Dorsa.

Fallout of ash is an alternative to fallout of dust. 3D ash-dispersal simulations suggest that distal-ash fallout in Gale crater would correlate with distal-ash fallout in Aeolis Dorsa (Kerber et al. 2011, 2012, 2013). Landscape-terrain feedbacks can explain preferential preservation of rhythmite and/or ash on preexisting local-to-regional topographic highs such as Mt. Sharp and Zephyria Planum (Brothers et al. 2013, Kite et al. 2013c).

Because of the strong evidence for correlation of the youngest deposits in Aeolis Dorsa to the youngest deposits at Aeolis Mons, we hypothesize that the >$10^8$ yr duration of the unconformity below the yardang-forming materials in Aeolis Dorsa (§3.4) is correlated to the major unconformity underlying the draping unit that forms the summit of Aeolis Mons (Malin & Edgett 2000, Anderson & Bell 2010, Thomson et al. 2011, Le Deit et al. 2013). This time-gap prediction may be tested by measuring the abundance of cosmogenic nuclides at the unconformity using the Sample Analysis at Mars (SAM) instrument on *Curiosity* (Farley et al. 2014).

*Rivers*. The large-river deposits in Aeolis Dorsa are not part of the same catchment as the river deltas in Gale. We do not know if the large-river deposits in Aeolis Dorsa correlate to river deltas in Gale. The nominal age of Gale ejecta is the same within statistical error (±0.1 Gyr) as the estimated age of the Aeolis Dorsa rivers (Le Deit et al. 2013, Kite et al. 2014), so correlation cannot be ruled out.

*Alluvial fans*. The water source for the alluvial fans is thought to be snow/ice melting (Palucis et al. 2014, Morgan et al. 2014, Grant & Wilson 2012). Peak melt rates depend on atmospheric pressure and orbital forcing, which set the surface energy balance (Clow 1987, Hecht 2002, Kite et al. 2013b). The surface energy balance of modestly-tilted water source regions is insensitive to longitude, but sensitive to time-varying orbital forcing, latitude, and elevation. The latitude and elevation of the Peace Vallis alluvial fan headwaters are identical to the latitude and elevation of the Aeolis Dorsa alluvial fan headwaters 10° to the E (Fig. 2e). This suggests that times of peak melt rates at Gale match up with times of peak melt rate at Aeolis Dorsa.

# 6. Conclusions.

We conclude that the sedimentary rocks of Aeolis Dorsa can be divided into four unconformity-bounded rock packages.

1. A >700m-thick sedimentary rock package lacking obvious river deposits was erosionally dissected prior to embayment by river deposits. The main evidence for this is the dissection of crater ejecta and the embayment of topographically high-standing early deposits by river deposits (§3.1).
2. River deposits can be subdivided into at least two units, which stratigraphically encapsulate a time series of river-deposit dimensions and channel-deposit proportions (§3.2). The total thickness of river deposits is >400m.
3. Fan-shaped deposits are deflected by thrust faults, and the thrust faults crosscut the river deposits. The fan-shaped deposits are interpreted as alluvial fans. The alluvial fan source regions cut back into high-standing deposits (§3.3).

4. All three earlier rock packages were unconformably draped by layered materials that contain rhythmite, and which have recently eroded to form yardangs (§3.4). Outliers and fingers of yardang-forming layered materials sit in hollows incised between inverted channels and inverted meander-belts. Based on the density of impact craters embedded at the unconformity, the time gap on this unconformity is >$10^8$ yr.
5. Following deposition of yardang-forming layered materials, erosion resumed and is currently active.

This stratigraphic framework clears the ground for future work using Aeolis Dorsa's river deposits to study competition between cratering and fluvial mass transport (Howard 2007), continental rock-package stratigraphy / macrostratigraphy (Sloss 1963, Shanley & McCabe 1994, Peters 2006, Hinnov 2013, Holt et al. 2010), river meandering in the absence of vegetation (Davies et al. 2011, Matsubara and Howard in press), putative deltas (e.g. DiBiase et al. 2013), alluvial architecture and stacking patterns (Mohrig et al. 2000, Straub et al. 2009, Hajek et al. 2010, Reijeinstein et al. 2011), and paleohydrology and climate versus time (Hajek & Wolinsky 2012, Amundson et al. 2012, Foreman et al. 2012, Foreman 2014, Zachos et al. 2001).

We hypothesize that orbital forcing modulated the erosion-deposition cycles that we have documented here (§5.1). We showed that orbital simulations predict >$10^8$ yr intervals of continuously-low obliquity, consistent with the orbital forcing hypothesis (§5.2).

## Acknowledgements.

We thank Devon Burr, Lynn Carter, Laura Kerber, Robert Jacobsen, Misha Kreslavsky, Caleb Fassett, David Mohrig, Ben Cardenas, Stephen Scheidt, and Bethany Ehlmann, for sharing their ideas about Aeolis Dorsa in many discussions that have made this a stronger paper. We thank Noah Finnegan, Jonathan Stock and Ross Irwin for discussions on paleohydrology, and Gary Kocurek, Dick Heermance and Paul Olsen for discussions on aeolian-fluvial interactions. We thank all the participants in the Caltech Mars Geomorphology Reading Group, especially Katie Stack, Kirsten Siebach, Roman DiBiase, Ajay Limaye, Joel Scheingross, Woody Fischer, Lauren Edgar, and Jeff Prancevic. Kevin Lewis, Frederik Simons, and Or Bialik provided useful feedback on an early draft. Brian Hynek shared an enlightening pre-publication map of Meridiani. Jenny Blue coordinated the naming of Obock, Neves, Kalba, Asau and Gunjur craters. We thank Ross Beyer, Sarah Mattson, Audrie Fennema, Annie Howington-Kraus, and especially Cynthia Phillips for help with DTM generation. We thank Sanjeev Gupta, and another reviewer who chose to remain anonymous, for suggestions that improved the paper. We are particularly grateful to the HiRISE team for maintaining the HiWish program, which led to the acquisition of multiple images that were essential to this study. This work was completed in part with resources provided by the University of Chicago Research Computing Center (MIDWAY cluster). This work was financially supported by the US taxpayer through NASA grant NNX11AF51G (to O.A.), by Caltech through the award of an O.K. Earl Fellowship (to E.S.K.), and by Princeton University through the award of a Harry Hess Fellowship (to E.S.K.).

## References.

Amundson, R., et al. 2012, Geomorphologic evidence for the late Pliocene onset of hyperaridity in the Atacama desert, Geol. Soc. Am. Bulletin 124, 1048-1070.

Anderson, R. B., Bell, J. F., III 2010. Geologic mapping and characterization of Gale Crater and implications for its potential as a Mars Science Laboratory landing site. International Journal of Mars Science and Exploration (Mars Journal) 5, 76-128.

Andrews-Hanna, J. C. 2012. The formation of Valles Marineris: 2. Stress focusing along the buried dichotomy boundary. Journal of Geophysical Research (Planets) 117, 4009.

Andrews-Hanna, J. C., Lewis, K. W. 2011. Early Mars hydrology: 2. Hydrological evolution in the Noachian and Hesperian epochs. Journal of Geophysical Research (Planets), 116, 2007.

Andrews-Hanna, J. C., Zuber, M. T., Hauck, S. A. 2008a. Strike-slip faults on Mars: Observations and implications for global tectonics and geodynamics. Journal of Geophysical Research (Planets) 113, 8002.

Andrews-Hanna, J. C., Zuber, M. T., Banerdt, W. B. 2008b. The Borealis basin and the origin of the martian crustal dichotomy. Nature 453, 1212-1215.

Armitage, J. J., Duller, R. A., Whittaker, A. C., Allen, P. A. 2011. Transformation of tectonic and climatic signals from source to sedimentary archive. Nature Geoscience 4, 231-235.

Armstrong, J.C., Leovy, C.B., and Quinn, T., 2004, A 1 Gyr climate model for Mars: new orbital statistics and the importance of seasonally resolved polar processes, Icarus 171, 255-271.

Bibring, J.-P., et al. 2006. Global Mineralogical and Aqueous Mars History Derived from OMEGA/Mars Express Data. Science 312, 400-404.

Blakey, R. C., 1994, Paleogeographic and tectonic controls on some Lower and Middle Jurassic erg deposits, Colorado Plateau and vicinity, in Mesozoic systems of the Rocky Mountain region, USA, Caputo, M. V., Peterson, J. A., and Franczyk, editors, Denver, Colorado: Rocky Mountain Section of the Society for Sedimentary Geology, p. 273-298.

Blakey, R.C., & W. Ranney, 2008, Ancient landscapes of the Colorado Plateau, Grand Canyon Association.

Boyce, J. M., Wilson, L., Mouginis-Mark, P. J., Hamilton, C. W., Tornabene, L. L. 2012. Origin of small pits in martian impact craters. Icarus 221, 262-275.

Bridge, J.S., 2003, Rivers and floodplains: Forms, processes, and sedimentary record, Wiley-Blackwell.

Bridges, N.T. & D.R. Muhs, 2012, Duststones on Mars: source, transport, deposition, and erosion, pp. 169-182 in Grotzinger, J.P., & R.E. Milliken (eds.) Sedimentary Geology of Mars, SEPM Special Publication No. 102.

Brothers, T. C., & Holt, J. W, 2013, Korolev Crater, Mars: Growth of a 2-km Thick Ice-Rich Dome Independent of, but Possibly Linked to, the North Polar Layered Deposits, 44th Lunar and Planetary Science Conference, held March 18-22, 2013 in The Woodlands, Texas. LPI Contribution No. 1719, p.3022.

Brothers, T. C., Holt, J. W., Spiga, A. 2013. Orbital radar, imagery, and atmospheric modeling reveal an aeolian origin for Abalos Mensa, Mars. Geophysical Research Letters 40, 1334-1339.

Buczkowski, D.L., et al. 2012, Giant polygons and circular graben in western Utopia basin, Mars: Exploring possible formation mechanisms, J. Geophys. Res. 117, E08010, doi:10.1029/2011JE003934.

Burbank, D.W. & R.S. Anderson, 2011, Tectonic geomorphology 2$^{nd}$ edition. John Wiley & Sons.

Burr, D. M., Enga, M.-T., Williams, R. M. E., Zimbelman, J. R., Howard, A. D., Brennand, T. A. 2009. Pervasive aqueous paleoflow features in the Aeolis/Zephyria Plana region, Mars. Icarus 200, 52-76.

Burr, D. M., Williams, R. M. E., Wendell, K. D., Chojnacki, M., Emery, J. P. 2010. Inverted fluvial features in the Aeolis/Zephyria Plana region, Mars: Formation mechanism and initial paleodischarge estimates. Journal of Geophysical Research (Planets) 115, 7011.

Byrne S, Murray BC. 2002. North polar stratigraphy and the paleo-erg of Mars. J. Geophys. Res. 107:5044.

Cardenas, B. T.; Mohrig, D., 2014, Evidence for Shoreline-Controlled Changes in Baselevel from Fluvial Deposits at Aeolis Dorsa, Mars, 45th Lunar and Planetary Science Conference, held 17-21 March, 2014 at The Woodlands, Texas. LPI Contribution No. 1777, p.1632

Carter, L., et al., 2009, Shallow radar (SHARAD) sounding observations of the Medusae Fossae Formation, Mars, Icarus 199, 295-302.

Chambers, J.E., 1999, A hybrid symplectic integrator that permit close encounters between massive bodies, Monthly Notices of the Royal Astronomical Society 304, 793-799.

Christie-Blick, N., Grotzinger, J.P., and von der Borch, C.C., 1988, Sequence stratigraphy in Proterozoic successions, Geology, 16, 100-.

Christie-Blick, N., von der Borch, C.C., DiBona, P.A., 1990, Working hypotheses for the origin of the Wonoka canyons (Neoproterozoic), South Australia: American Journal of Science, 290-A (Cloud volume), 295-332.

Clow, G. D. 1987. Generation of liquid water on Mars through the melting of a dusty snowpack. Icarus 72, 95-127.

Davies, N.S., Gibling, M.R., Rygel, M.C., 2011, Alluvial facies during the Palaeozoic greening of the land: case studies, conceptual models and modern analogues. Sedimentology (Special Decadal Issue), 58, 220-258.

Dibiase, R.A.; Limaye, A.B.; Scheingross, J.S.; Fischer, W.W.; Lamb, M.P. (2013), Deltaic deposits at Aeolis Dorsa: Sedimentary evidence for a standing body of water on the northern plains of Mars, Journal of Geophysical Research: Planets, Volume 118, Issue 6, pp. 1285-1302.

Edgett, K. S., Malin, M. C. 2001. Rock Stratigraphy in Gale Crater, Mars. Lunar and Planetary Science Conference 32, 1005.

Edgett, K. S. 2005. The sedimentary rocks of Sinus Meridiani: Five key observations from data acquired by the Mars Global Surveyor and Mars Odyssey orbiters. International Journal of Mars Science and Exploration 1, 5-58.

Ehlmann, B. L., Mustard, J. F., Murchie, S. L., Bibring, J.-P., Meunier, A., Fraeman, A. A., Langevin, Y. 2011. Subsurface water and clay mineral formation during the early history of Mars. Nature 479, 53-60.

Ehlmann, B.L. & C.E. Edwards, 2014, Mineralogy of the Martian surface, Annual Reviews of Earth and Planetary Science, volume 42, Review in Advance.

Ewing, S. A., Sutter, B., Amundson, R., Owen, J., Nishiizumi, K., Sharp, W., Cliff, S. S., Perry, K., Dietrich, W. E. and McKay, C. P., 2006, A threshold in soil formation at Earth's arid-hyperarid transition. Geochimica et Cosmochimica Acta 70(21), 5293-5322, doi: 10.1016/j.gca.2006.08.020.

Farley, K.A., et al., 2014, In Situ Radiometric and Exposure Age Dating of the Martian Surface, Science 343 (6169): DOI: 10.1126/science.1247166

Fassett, C. I., Head, J. W. 2008. The timing of martian valley network activity: Constraints from buffered crater counting. Icarus 195, 61-89.

Fassett, C.I., Head, J.W. 2011, Sequence and timing of conditions on early Mars, Icarus 211, 1204-1214.

Foreman, B. Z., Heller, P. L., Clementz, M. T. 2012. Fluvial response to abrupt global warming at the Palaeocene/Eocene boundary. Nature 491, 92-95.

Foreman, B.Z., 2014. Climate-driven generation of a fluvial sheet sand body at the Paleocene–Eocene boundary in north-west Wyoming (USA), Basin Research 26, 225-241.

Forget, F., Wordsworth, R., Millour, E., Madeleine, J.-B., Kerber, L., Leconte, J., Marcq, E., Haberle, R. M. 2013. 3D modelling of the early martian climate under a denser $CO_2$ atmosphere: Temperatures and $CO_2$ ice clouds. Icarus 222, 81-99.

Ghatan, G.J., & Zimbelman J.R., 2006, Paucity of candidate coastal constructional landforms along proposed shorelines on Mars: Implications for a northern lowlands-filling ocean, Icarus 185, 171-196.

Goddard, K., Warner, N.H., Gupta, S., Kim, J.-R., 2014, Mechanisms and timescales of fluvial activity at Mojave and other young Martian craters, J. Geophys. Res. 119, 604—634.

Golombek, M.P., 2001, Martian wrinkle ridge topography: Evidence for subsurface faults from MOLA, J. Geophys. Res. – Planets, 106(E10), 23811-23821.

Golombek, M. P., et al., 2015, Small crater modification on Meridiani Planum and implications for erosion rates and climate change on Mars, J. Geophys. Res. – Planets, DOI: 10.1002/2014JE004658.

Grant, J.A., & Wilson, S.A., 2012, A possible synoptic source of water for alluvial fan formation in southern Margaritifer Terra, Mars, 72, 44-52.

Grotzinger, J., et al., 2006, Sedimentary textures formed by aqueous processes, Erebus crater, Meridiani Planum, Mars, Geology 34, 1085-1088.

Grotzinger, J., et al., 2011, Mars Sedimentary Geology: Key Concepts and Outstanding Questions, Astrobiology, 11, 77-87.

Grotzinger, J.P. & R.E Milliken 2012a, The Sedimentary Rock Record of Mars: Distribution, origins, and global stratigraphy, pp. 1-48 in Grotzinger, J.P., & R.E. Milliken (eds.) Sedimentary Geology of Mars, SEPM Special Publication No. 102.

Grotzinger, J.P., & R.E. Milliken (eds.) 2012b, Sedimentary Geology of Mars, SEPM Special Publication No. 102.

Grotzinger, J.P. et al., 2014, A habitable fluvio-lacustrine environment at Yellowknife Bay, Gale crater, Mars, Science 343(6169), doi: 10.1126/science.1242777

Haberlah, D., Williams, M. A. J., Halverson, G., McTainsh, G. H., Hill, S. M., Hrstka, T., Jaime, P., Butcher, A. R., Glasby, P. 2010. Loess and floods: High-resolution multi- proxy data of Last Glacial Maximum (LGM) slackwater deposition in the Flinders Ranges, semi-arid South Australia. Quaternary Science Reviews 29, 2673-2693.

Hajek, E. A., Heller, P. L., Sheets, B. A. 2010. Significance of channel-belt clustering in alluvial basins. Geology 38, 535-538.

Hajek, E. A., Wolinsky, M. A. 2012. Simplified process modeling of river avulsion and alluvial architecture: Connecting models and field data. Sedimentary Geology 257, 1-30.

Harrison, S.K., et al., 2010. Mapping Medusae Fossae Formation materials in the southern highlands of Mars, Icarus 209, 405-415.

Harrison, S. K., Balme, M. R., Hagermann, A., Murray, J. B., Muller, J.-P., Wilson, A. 2013. A branching, positive relief network in the middle member of the Medusae Fossae Formation, equatorial Mars - Evidence for sapping? Planet. & Space Sci. 85, 142-163.

Hartmann, W.K., 2005, Martian cratering 8: Isochron refinement and the chronology of Mars, Icarus, 174, 294–320.

Head, J. W., Kreslavsky, M. A. 2001. Medusae Fossae Formation as volatile-rich sediments deposited during high obliquity: an hypothesis and tests. Conference on the Geophysical Detection of Subsurface Water on Mars 7053.

Hecht, M. H. 2002, Metastability of Liquid Water on Mars. Icarus 156, 373-386.

Hinnov, L.A., 2013, Cyclostratigraphy and its revolutionizing applications in the earth and planetary sciences, Geological Society of America Bulletin 125, 1703-1734.

Holt, J.W., et al., 2010, The construction of Chasma Boreale on Mars, Nature 465, 446-449.

Hoke, M.R.T., Hynek, B.M., and Tucker, G.E., 2011, Formation timescales of large Martian valley networks, Earth & Planetary Science Letters 312, 1-12.

Howard, A. D. 2007. Simulating the development of Martian highland landscapes through the interaction of impact cratering, fluvial erosion, and variable hydrologic forcing. Geomorphology 91, 332-363.

Howard, A.D., 2009, How to make a meandering river. Proceedings of the National Academy of Sciences 106, 41, 17245-17246.

Hudson, T.L., & Aharonson, O. 2008, Diffusion barriers at Mars surface conditions: Salt crusts, particle size mixtures, and dust, J. Geophys. Res. – Planets, 113(E9), E09008.

Hutton, J., 1788. p. 458 in Theory of the Earth.

Hynek, B.M., & Phillips, R.J., 2008, The stratigraphy of Meridiani Planum, Mars, and implications for the layered deposits' origin, Earth and Planetary Science Letters 274, 214-220

Irwin, R. P., Watters, T. R. 2010. Geology of the Martian crustal dichotomy boundary: Age, modifications, and implications for modeling efforts. Journal of Geophysical Research (Planets) 115, 11006.

Irwin, R. P., Watters, T. R., Howard, A. D., Zimbelman, J. R. 2004. Sedimentary resurfacing and fretted terrain development along the crustal dichotomy boundary, Aeolis Mensae, Mars. Journal of Geophysical Research (Planets) 109, 9011.

Irwin, Rossman P.; Tanaka, Kenneth L.; Robbins, Stuart J. 2013, Distribution of Early, Middle, and Late Noachian cratered surfaces in the Martian highlands: Implications for resurfacing events and processes, Journal of Geophysical Research: Planets 118, 278-291.

Jacobsen, R.P., and Burr, D., 2012, Paleo-Fluvial Features in the Western Medusae Fossae Formation, Aeolis and Zephyria Plana, Mars: Elevations and Implications, 43rd Lunar and Planetary Science Conference, held March 19-23, 2012 at The Woodlands, Texas. LPI Contribution No. 1659, id.2398.

Jakosky, B. M., Carr, M. H. 1985. Possible precipitation of ice at low latitudes of Mars during periods of high obliquity. Nature 315, 559-561.

Jaumann, R.; Nass, A.; Tirsch, D.; Reiss, D.; Neukum, G., 2010, The Western Libya Montes Valley System on Mars: Evidence for episodic and multi-genetic erosion events during the Martian history, Earth and Planetary Science Letters 294, 272-290.

Jerolmack, D. J., Paola, C. 2007. Complexity in a cellular model of river avulsion. Geomorphology 91, 259-270.

Jerolmack, D.J., Sadler, P., 2007, Transience and persistence in the depositional record of continental margins, J. Geophys. Res. – Earth Surface 112(R-SW), F03S13.

Johnson, D., 2009, Birth of modern Australia, p. 145-166 in Johnson, D., The geology of Australia, Cambridge University Press.

Karlstrom. KE, and JM Timmons, 2012, Many unconformities make one Great Unconformity. in JM Timmons and KE Karlstrom, eds., pp. 73-79, Grand Canyon Geology: Two Billion Years of Earth's History. Special Paper no. 489. Geological Society of America, Boulder, Coloprado.

Kerber, L., Head, J. W. 2010. The age of the Medusae Fossae Formation: Evidence of Hesperian emplacement from crater morphology, stratigraphy, and ancient lava contacts. Icarus 206, 669-684.

Kerber, L., et al. 2011, The dispersal of pyroclasts from Apollinaris Patera, Mars: Implications for the origin of the Medusae Fossae Formation, Icarus, 216, 212-220.

Kerber, L., et al., 2012, The dispersal of pyroclasts from ancient explosive volcanoes on Mars: Implications for the friable layered deposits, Icarus, 219, 358-381.

Kerber, L., Forget, F., Madeleine, J.-B., Wordsworth, R., Head, J. W., Wilson, L. 2013. The effect of atmospheric pressure on the dispersal of pyroclasts from martian volcanoes. Icarus 223, 149-156.

Kite, E. S., Michaels, T. I., Rafkin, S., Manga, M., Dietrich, W. E. 2011. Localized precipitation and runoff on Mars. Journal of Geophysical Research (Planets) 116, 7002.

Kite, E.S., 2012, Evidence for Melt-Fed Meandering Rivers in the Gale-Aeolis-Zephyria Region, Mars, 43rd Lunar and Planetary Science Conference, held March 19-23, 2012 at The Woodlands, Texas. LPI Contribution No. 1659, id.2778.

Kite, E. S., Lucas, A., Fassett, C. I. 2013a. Pacing early Mars river activity: embedded craters in the Aeolis Dorsa region imply river activity spanned ≳(1–20) Myr, Icarus 225, 850-855.

Kite, E. S., Halevy, I., Kahre, M. A., Wolff, M. J., Manga, M. 2013b. Seasonal melting and the formation of sedimentary rocks on Mars, with predictions for the Gale Crater mound. Icarus 223, 181-210.

Kite, E. S., Lewis, K. W., Lamb, M. P., Newman, C. E., Richardson, M. I. 2013c. Growth and form of the mound in Gale Crater, Mars: Slope wind enhanced erosion and transport. Geology 41, 543-546.

Kite, E. S., Williams, J.-P., Lucas, A., Aharonson, O. 2014. Low paleopressure of the martian atmosphere estimated from the size distribution of ancient craters, Nature Geoscience 7, 335-339. doi:10.1038/ngeo2137.

Kite, E.S., Howard, A., Lucas, A.S., & Lewis, K.W., Resolving the era of river-forming climates on Mars using stratigraphic logs of river-deposit dimensions, submitted to Earth & Planetary Science Letters.

Lammer, H., et al., 2013, Outgassing history and escape of the Martian atmosphere and water inventory, Space Sci. Rev. 174, 113-154.

Laskar, J., Correia, A. C. M., Gastineau, M., Joutel, F., Levrard, B., Robutel, P., 2004, Long term evolution and chaotic diffusion of the insolation quantities of Mars, Icarus 170, 343-364.

Le Deit, L., et al. 2013, Sequence of infilling events in Gale Crater, Mars: Results from morphology, stratigraphy, and mineralogy, J. Geophys. Res: Planets 118, 12, 2439-2473.

Lefort, A., Burr, D. M., Beyer, R. A., Howard, A. D. 2012. Inverted fluvial features in the Aeolis-Zephyria Plana, western Medusae Fossae Formation, Mars: Evidence for post-formation modification. Journal of Geophysical Research (Planets) 117, 3007.

Lefort, A., Burr, D.M., Nimmo, F., and Jacobsen, R.E., 2015, Channel slope reversal near the Martian dichotomy boundary: Testing tectonic hypotheses, Geomorphology, doi:10.1016/j.geomorph.2014.09.028.

Lewis, K.W., 2009, The rock record of Mars: structure, sedimentology and stratigraphy. Dissertation (Ph.D.), California Institute of Technology. http://resolver.caltech.edu/CaltechETD:etd-06062009-150120

Lewis, K. W., and O. Aharonson, 2014, Occurrence and Origin of Rhythmic Sedimentary Rocks on Mars. J. Geophys Res., J. Geophys. Res., doi: 10.1002/2013JE004404.

Lewis, K. W., Aharonson, O., Grotzinger, J. P., Kirk, R. L., McEwen, A. S., Suer, T.-A. 2008. Quasi-Periodic Bedding in the Sedimentary Rock Record of Mars. Science 322, 1532.

Malin, M.C., Edgett K.S. 1999, Oceans or seas in the Martian northern lowlands: High resolution imaging tests of proposed coastlines, Geophys. Res. Lett. 26, 3049-3052.

Malin, M. C., Edgett, K. S. 2000. Sedimentary Rocks of Early Mars. Science 290, 1927-1937.

Mangold, N., et al., 2009, Estimate of aeolian dust thickness in Arabia Terra, Mars, Geomorphologie: relief, processus, environnement, url:http://geomorphologie.revues.org/7472 ; doi:10.4000/geomorphologie.7472.

Mangold, N., et al., 2012a, The origin and timing of fluvial activity at Eberswalde crater, Mars, Icarus 220, 530-551.

Marinova, M. M., Aharonson, O., Asphaug, E. 2008. Mega-impact formation of the Mars hemispheric dichotomy. Nature 453, 1216-1219.

Matsubara, Y., & A.D. Howard, in press, Modeling planform evolution of a mud-dominated meandering river: Quinn River, Nevada, USA, Earth Surface Processes and Landforms, doi:10.1002/esp.3588.

Mellon, M.T., and Jakosky, B.M. 1995, The distribution and behavior of Martian ground ice during past and present epochs, J. Geophys. Res. 100(E6), 11781-11799.

Milkovich, S.M. & Plaut, J.J., Martian South Polar Layered Deposit stratigraphy and implications for accumulation history, J. Geophys. Res 113, CiteID E06007.

Milliken, R. E., Fischer, W. W., Hurowitz, J. A. 2009. Missing salts on early Mars. Geophysical Research Letters 36, 11202.

Michael, G.G. 2013, Planetary surface dating from crater size-frequency distribution measurements: Multiple resurfacing episodes and differential isochron fitting, Icarus 226(1), 885-890.

Milliken, R. E.; Fischer, W. W.; Hurowitz, J. A., 2009, Missing salts on early Mars, Geophysical Research Letters 36, 11, CiteID L11202.

Milliken, R. E.; Grotzinger, J. P.; Thomson, B. J., 2010, Paleoclimate of Mars as captured by the stratigraphic record in Gale Crater, Geophysical Research Letters, Volume 37, Issue 4, CiteID L04201.

Mischna, M. A., Richardson, M. I., Wilson, R. J., McCleese, D. J. 2003. On the orbital forcing of Martian water and $CO_2$ cycles: A general circulation model study with simplified volatile schemes. Journal of Geophysical Research (Planets) 108, 5062.

Mohrig, D., Heller, P.L., Paola, C., Lyons, W.J., 2000, Interpreting avulsion process from ancient alluvial sequences: Guadalope-Matarranya system, northern Spain and Wasatch Formation, western Colorado: Geological Society of America Bulletin 112, 1787-1803.

Moore, J. M. 1990. Nature of the mantling deposit in the heavily cratered terrain of northeastern Arabia, Mars. Journal of Geophysical Research 95, 14279-14289.

Moore, J. M., Howard, A. D. 2005. Large alluvial fans on Mars. Journal of Geophysical Research (Planets) 110, 4005.

Morgan, G. A., Head, J. W., Forget, F., Madeleine, J.-B., Spiga, A. 2010. Gully formation on Mars: Two recent phases of formation suggested by links between morphology, slope orientation and insolation history. Icarus 208, 658-666.

Morgan, A.M., et al., 2014. Sedimentology and climatic environment of alluvial fans in the martian Saheki crater and a comparison with terrestrial fans in the Atacama Desert, Icarus 229, 131-156.

Mouginot, J.; Pommerol, A.; Kofman, W.; Beck, P.; Schmitt, B.; Herique, A.; Grima, C.; Safaeinili, A.; Plaut, J. J., 2010, The 3-5 MHz global reflectivity map of Mars by MARSIS/Mars Express: Implications for the current inventory of subsurface $H_2O$, Icarus, Volume 210, Issue 2, p. 612-625.

Murchie, S., et al., Evidence for the origin of layered deposits in Candor Chasma, Mars, from mineral composition and hydrologic modeling, Journal of Geophysical Research 114, CiteID E00D05.

Mustard, J.F., et al. 2009, Composition, Morphology, and Stratigraphy of Noachian Crust around the Isidis basin, J. Geophys. Res. 114, CiteID E00D12.

Nahm, A. L., Schultz, R. A. 2011. Magnitude of global contraction on Mars from analysis of surface faults: Implications for martian thermal history. Icarus 211, 389-400.

Nickling, W. G. 1984. The stabilizing role of bonding agents on the entrainment of sediment by wind. Sedimentology 31, 111-117.

Nimmo, F. 2005. Tectonic consequences of Martian dichotomy modification by lower- crustal flow and erosion. Geology 33, 533.

Nimmo, F., Hart, S. D., Korycansky, D. G., Agnor, C. B. 2008. Implications of an impact origin for the martian hemispheric dichotomy. Nature 453, 1220-1223.

Palucis, M.C. et al., 2014, The origin and evolution of the Peace Vallis fan system that drains to the Curiosity landing area, Gale Crater, Mars, J. Geophys. Res. doi:10.1002/2013JE004583.

Pederson, G.B.M., & Head, J.W. 2011, Chaos formation by sublimation of volatile-rich substrate: Evidence from Galaxias Chaos, Mars, Icarus 211, 316-329.

Peters, S.E., 2006, Macrostratigraphy of North America, Journal of Geology 114, 391-412.

Peters, S.E., and R.R. Gaines, 2012, Formation of the 'Great Unconformity' as a trigger fro the Cambrian Explosion, Nature 484, 363-366.

Phillips, R.J., et al. 2001, Ancient geodynamics and global-scale hydrology on Mars, Science 291, 2587-2591.

Reijeinstein, H.M., Posamentier, H.W., and Bhattacharaya, J.P., 2011, Seismic geomorphology and high-resolution seismic stratigraphy of inner-shelf fluvial, estuarine, deltaic and marine sequences, Gulf of Thailand, American Association of Petroleum Geologists Bulletin 95(11), doi:10.1306/03151110134.

Robbins, S.J., Achille, G.D., and Hynek, B.M., 2011, The volcanic history of Mars: High-resolution crater-based studies of the calderas of 20 volcanoes, Icarus, 211, 1179-1203.

Robbins, S.J., 2014, New crater calibrations for the lunar crater-age chronology, Earth and Planet. Sci. Lett. 403, 188-198.

Rose, K.C., F. Ferraccioli, S.S.R. Jamieson, R.E. Bell, H. Corr, T.T. Creyts, D. Braaten, T.A. Jordan, P. Fretwell, D. Damaske, 2013, Early East Antarctic Ice Sheet growth recorded in the landscape of the Gamburtsev subglacial mountains. Earth and Planetary Science Letters 375, 1-12.

Schiller, M. et al. 2014, Rapid soil accumulation in a frozen landscape, Geology, doi: 10.1130/G35450.1

Schorghofer, N. 2008, Temperature response of Mars to Milankovitch cycles, Geophys. Res. Lett. 35, CiteID L18201.

Segura, T.L., Zahnle, K., Toon, O.B., & McKay, C.P. 2013, The effects of impacts on the climates of terrestrial planets, p. 417-438 in Mackwell, S.J., et al., eds., Comparative Climatology of Terrestrial Planets, U. Arizona Press.

Shanley, K.W., & McCabe, P.J. 1994, Perspectives on the sequence stratigraphy of continental strata, American Association of Petroleum Geologists Bulletin 78(4), 544-568.

Slingerland, R., Smith, N. D. 1998. Necessary conditions for a meandering-river avulsion. Geology 26, 435-438.

Sloss, L.L., 1963, Sequences in the cratonic interior of North America, Geol. Soc. Am. Bull. 74, 93-114.

Smith, M. R., Gillespie, A. R., Montgomery, D. R., Batbaatar, J. 2009. Crater-fault interactions: A metric for dating fault zones on planetary surfaces. Earth and Planetary Science Letters 284, 151-156.

Solomon, S. C., et al., 2005. New perspectives on ancient Mars. Science 307, 1214-1220.

Straub, K.M., et al., 2009, Compensational stacking of channelized sedimentary deposits, J. Sedimentary Research 79, 673-688.

Sun, V.Z., & R.E. Milliken, 2014, The geology and mineralogy of Ritchey crater, Mars: evidence for post-Noachian clay formation, J. Geophys. Res. – Planets, doi: 10.1002/2013JE004602.

Thomson, B. J., Bridges, N. T., Milliken, R., Baldridge, A., Hook, S. J., Crowley, J. K., Marion, G. M., de Souza Filho, C. R., Brown, A. J., Weitz, C. M. 2011. Constraints on the origin and evolution of the layered mound in Gale Crater, Mars using Mars Reconnaissance Orbiter data. Icarus 214, 413-432.

Tornabene, L., et al., 2013, A Revised Global Depth-Diameter Scaling Relationship for Mars Based on Pitted Impact Melt-Bearing Craters, 44th Lunar and Planetary Science Conference, held March 18-22, 2013 in The Woodlands, Texas. LPI Contribution No. 1719, abstract #2592.

Ward, A.W., 1979, Yardangs on Mars: Evidence of recent wind erosion, J. Geophys. Res. 84(B14), 8147-8166.

Watters, T.R., McGovern, P.J., and Irwin, R.P., Hemispheres apart: the crustal dichotomy on Mars, Annual Reviews of Earth and Planetary Sciences 35, 621-652.

Weitz, C. M.; Milliken, R. E.; Grant, J. A.; McEwen, A. S.; Williams, R. M. E.; Bishop, J. L.; Thomson, B. J., 2010, Mars Reconnaissance Orbiter observations of light-toned layered deposits and associated fluvial landforms on the plateaus adjacent to Valles Marineris, Icarus 205, 73-102.

Wesnousky, S.G., 2008, Displacement and geometrical characteristics of earthquake surface ruptures, Bulletin of the Seismological Society of America, 98, 1609–1632, doi: 10.1785/012007011156.

Willett, S.D., et al., 2014, Dynamic Reorganization of River Basins, Science 343, doi: 10.1126/science.1248765.

Williams, R.M.E.; Malin, Michael C., 2008, Sub-kilometer fans in Mojave Crater, Mars. Icarus 198, 365-383.

Williams, R.M.E., et al., 2011, Evidence for episodic alluvial fan formation in far western Terra Tyrrhena, Mars, Icarus 211, 222-237.

Williams, R.M.E., Irwin, R. P., Burr, D. M., Harrison, T., McClelland, P. 2013a. Variability in martian sinuous ridge form: Case study of Aeolis Serpens in the Aeolis Dorsa, Mars, and insight from the Mirackina paleoriver, South Australia. Icarus 225, 308-324.

Williams, R.M.E., et al., 2013b, Martian Fluvial Conglomerates at Gale Crater, Science, Volume 340, Issue 6136, pp. 1068-1072 (2013).

Williams, R. M. E. & Weitz, C. M., 2014, Reconstructing the aqueous history within the Southwestern Melas Basin, Mars: Clues from Stratigraphic and Morphometric Analyses of Fans, Icarus, in press, July 2014.

Wilson, L., 2001, Evidence for episodicity in the magma supply to the large Tharsis volcanoes, J. Geophys. Res., 106, 1423-1434.

Wiseman, S.M., et al., 2010, Spectral and stratigraphic mapping of hydrated sulfate and phyllosilicate-bearing deposits in northern Sinus Meridiani, Mars, Journal of Geophysical Research 115, CiteID E00D18.

Wordsworth, R., et al., 2014, Surface ice migration and transient melting events on early Mars, Fifth international workshop on the Mars atmosphere: Modelling and observations, Oxford, UK.

Wray, James J.; Murchie, Scott L.; Squyres, Steven W.; Seelos, Frank P.; Tornabene, Livio L., 2009, Diverse aqueous environments on ancient Mars revealed in the southern highlands, Geology 37, 1043-1046.

Wray, J. J. 2013. Gale crater: the Mars Science Laboratory/Curiosity Rover Landing Site. International Journal of Astrobiology 12, 25-38.

Zabrusky, K., et al., 2012, Reconstructing the distribution and depositional history of the sedimentary deposits of Arabia Terra, Mars, Icarus 220, 311-330.

Zachos, J., Pagani, M., Sloan, L., Thomas, E., Billups, K. 2001. Trends, Rhythms, and Aberrations in Global Climate 65 Ma to Present. Science 292, 686-693.

Zimbelman, J. R., Scheidt, S. P. 2012. Hesperian Age for Western Medusae Fossae Formation, Mars. Science 336, 1683.

# Supplementary Methods:

*Definition of lithofacies*. We created a base-map by mosaicking publicly available CTX images. We mapped the boundaries of lithofacies (that were defined using HiRISE and CTX images) onto the CTX base-map. Where multiple CTX images of the same terrain were available, we chose the CTX image with the smallest emission angle. Terrain data were obtained from MOLA MEGDR, MOLA PEDR, an inverse-distance weighted raster interpolating between MOLA PEDR elevation data, and our own CTX stereo DTMs and HiRISE stereo DTMs. These additional stereo DTMs were constructed in SOCET SET (BAE Systems). We did not map mobile cover (aeolian bedforms).

*Crater counting at unconformity*. Several crater-flux chronology functions have been proposed for Mars (Michael 2013). All chronology functions feature impact fluxes that are high early in Mars history, decline more than tenfold between 4.0 Gya ago and 3.0 Gyr ago, and remain relatively constant after 3.0 Gya. This decline in the impact flux introduces an uncertainty into the calculation of the time gap on the unconformity, because we do not know the age of the unconformity. An older age for the unconformity leads to a larger crater flux, and thus a shorter time gap on the unconformity for a given $n(\phi)$. This age-induced uncertainty is larger than the uncertainty introduced by choosing a chronology function and production function (we use the chronology function and production function given by Michael 2013 and based on Hartmann 2005). This age-induced uncertainty is also larger than the uncertainty due to shot noise (statistical error). Therefore, we calculate the time gap for a range of assumed unconformity ages (guided by the chronology in Kite et al. 2014), but we do not track statistical uncertainty nor uncertainty introduced by choice of chronology functions or production functions.

We also need to define a count area ($n$ /km$^2$) – the area where craters embedded at the unconformity would have been detected, if they existed. We obtained this count area using 3 different methods (Fig. S9):

- **BCC – Buffered crater counting** (Fassett & Head 2008, Smith et al. 2009). In our implementation of this approach, the embedded-crater counts are divided by the area of a strip bracketing the contact and twice as wide as the crater diameter of interest (Fig. S9). This assumes no correlation between the position of the contact and the presence/absence of an embedded crater. It also assumes that the overlying unit is very thick, so that the rims of craters formed in the river deposits never poke up through the yardang-forming layered deposits.
- **LO – Lobes-only**. This approach is similar to buffered crater counting, but considers only the craters that intersect the contact bounding those 3 main lobes of Y. The lobes-only method ignores craters containing outliers of Y that are not connected to the main lobes of Y and the "buffer" areas for those outliers (Fig. S9, Fig. 2e). The count area is defined as a buffer region around the main lobes of Y. Buffer regions around small outliers of Y are not considered when defining the count area. Craters partly-covered by small outliers of Y are not included in the count of embedded craters.

- **ION – Inlier-Outlier annulus Normalization**. In this approach, we consider each of the 3 main lobes of Y separately (Fig S9). For each lobe, we draw a polygon that is tightly wrapped around neighboring outliers of Y – connecting their outermost apices with straight lines. This polygon defines the outer boundary of the annulus of inliers and outliers for that lobe. Next, we draw a polygon that connects the innermost apex of all inliers of sub-Y materials with straight lines. This polygon defines the inner boundary of the annulus of inliers and outliers for that lobe. We define the annulus between the polygons as the inlier-outlier annulus. We use the combined area of the 3 inlier-outlier annuli (one annulus for each lobe) to calculate the crater densities.

Count areas obtained using methods **LO** and **ION** are more realistic than using count method **BCC**. We can see this by first noting that outliers of Y are preferentially preserved in embedded-crater interiors (e.g. Fig. 19a), where they are protected against the wind[8]. Imagine (as a thought experiment) that the large lobes of Y were entirely removed, so that the only Y remaining was the material preferentially preserved in embedded-crater interiors. Then, because every part of the perimeter of Y would partly cover a crater, buffered-crater counting would return a fractional crater cover close to 100% (crater saturation). The resulting (over-)estimate for the time gap on the unconformity is the oldest time gap possible. The long-gap bias will be less severe for a real-world count than the absurdly high age resulting from this thought experiment, but the bias will still be in the direction of overestimating the time gap at the unconformity. Methods **ION** and **LO** attack this time-gap-overestimation by taking account of correlations between the position of the Y/sub-Y contact and the location of embedded craters (Fig. S9). In method **ION**, the current area spanned by the farthest-flung outliers of Y is taken to correspond to the area within which some Y would be preserved, if there were embedded craters present to protect them from obliteration by the wind. In method **LO**, we count only craters that intersect the margins of one of the three main lobes of Y. These margins are in erosional retreat, but the current position of the margins is not obviously correlated with the locations of embedded craters. Therefore such craters should not suffer from the age-overestimation bias adhering to the **BCC** method at this location.

The results (Table 1) are consistent with our expectations. The **BCC** method produces an impossibly large time gap (larger than the age of Mars). The **LO** and **ION** methods produce more realistic (lower) time gaps. There are fewer small craters embedded at the unconformity than expected for a pristine population of craters. This is consistent with preferential undercounting of smaller craters due to geometric effects (Lewis & Aharonson, 2014), survey incompleteness, obliteration by erosional processes (Jerolmack & Sadler 2007), or some combination.

---

[8] Preferential *formation* of layered deposits within craters has also been suggested (Brothers & Holt 2013; Figure DR3 in Kite et al. 2013c). The craters considered in those studies are generally larger than the craters considered in this study, and we assume for the purposes of this study that $\phi \lesssim 10$ km craters in Aeolis Dorsa aided preservation (not formation) of Y.

# Supplementary Discussion:

Crater-river interactions indicate one or more additional unconformity involving the river deposits (shown by wavy lines *within* R-1 and R-2 in Fig. 20), although we are not sure if these unconformities are basinwide. The evidence for this additional unconformity is as follows. Intra-crater sinuous ridges exist within Kalba, Neves, Obock, and the unnamed crater at 153.5°E 3.8°S. These ridges are best seen within Kalba crater (ESP_034545_1740). Because the ridges are sinuous and parallel to the paleo-drainage direction defined by sinuous ridges and meander-belts outside the crater, we interpret them as inverted channels (Burr et al. 2009). The inverted channels are 400m lower than the rim of Kalba and only 3 km from the rim. Remarkably, the inverted channels do not correspond to breaches in Kalba's rim. Some knicks are found in the N and the S rim of Kalba (parallel to paleoflow and suggesting fluvial incision), but they do not extend as deep as the inverted channels within Kalba, nor are they aligned with the interior channels. This lack of deep rim breaching presents a paradox: how could water have flowed over the rim of Kalba (as required to form the inverted channels in Kalba's interior) without deeply incising the rim? The resolution of the paradox is that the fluvial deposits were laid down at a time when the landscape had aggraded above the level of the crater rim. Therefore, the river deposits formed at a paleo-elevation above the crater rim. (This also explains why the sinuous ridges are aligned with the regional drainage direction and do not respond to the topography of the rim). Still further aggradation led to compaction (Lefort et al. 2012). The thickness of compactible material within Kalba was greater than the thickness of compactible material outside Kalba, because a $\phi$ = 14 km crater (Kalba's diameter) forms a ~1.66 km deep hole (Tornabene et al. 2013). This contrast in thickness of compactible material drove differential compaction (Lefort et al. 2012, Buczkowski et al. 2012), such that the river deposits which formed topographically above the rim are now below the river deposits on the surrounding plains. Other locations showing knicked rims preferentially in regional upstream-downstream directions and/or intra-crater river deposits are Obock, Neves (ESP_035600_1765), and the unnamed craters at 153.1°E 5.8°S, 153.5°E 3.8°S (ESP_027807_1765/ ESP_035112_1765 stereopair), and 152.6°E 6.3°S. The existence of knicks in the crater rims and of intra-crater river deposits points to an interval of aggradation to above the level of the crater rims, and an interval of incision to cut the crater rims. The relative timing of the knicks is unknown. It is also unknown whether the incision was restricted to the vicinity of preexisting crater rims, or represents an additional basinwide unconformity. However, a thick stack of now-vanished material above the current level of the crater rims is required to explain the current elevations of the inter-crater sinuous ridges.

# Supplementary Figures:

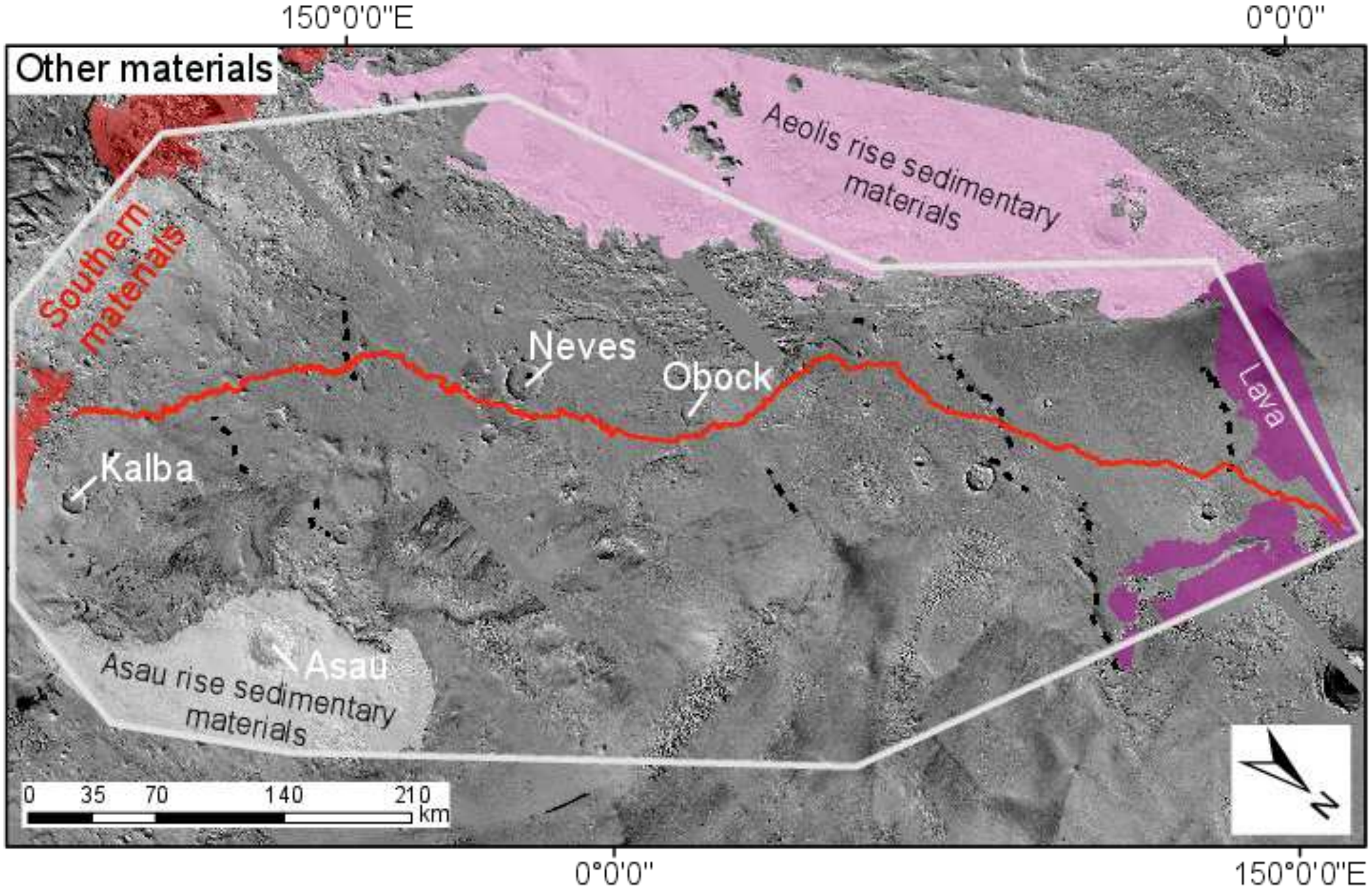

**Figure S1.** Materials outcropping in our study area, but not assigned to one of the sedimentary rock packages in Figure 2. Aeolis rise sedimentary materials and Asau rise sedimentary materials probably postdate river deposits (Zimbelman & Scheidt 2012), but may unconformably predate them. Figs. 17 and S8 show these alternatives. Lava in the N of the study region embays Aeolis Serpens river deposits (Kerber & Head, 2010).

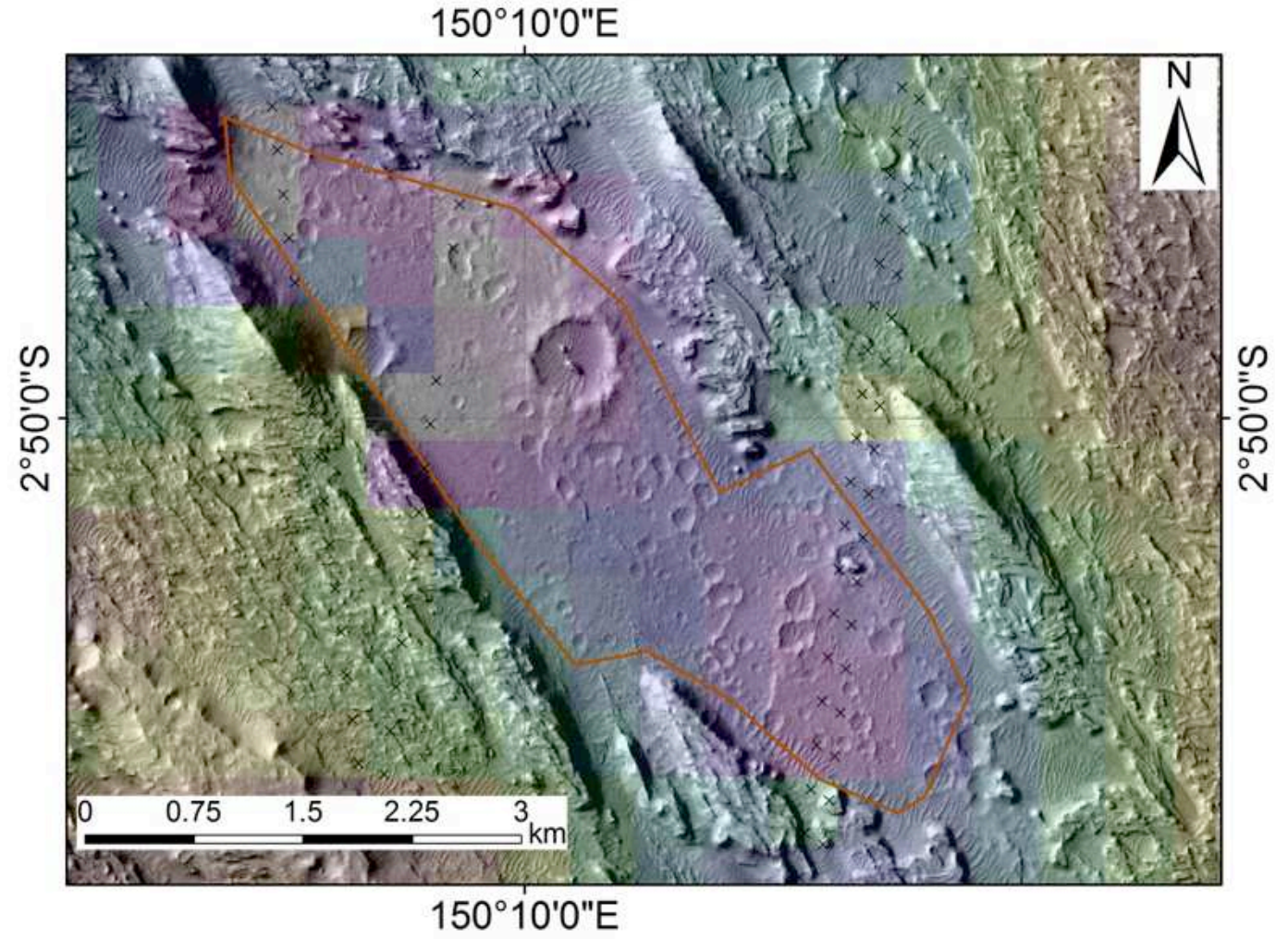

a)

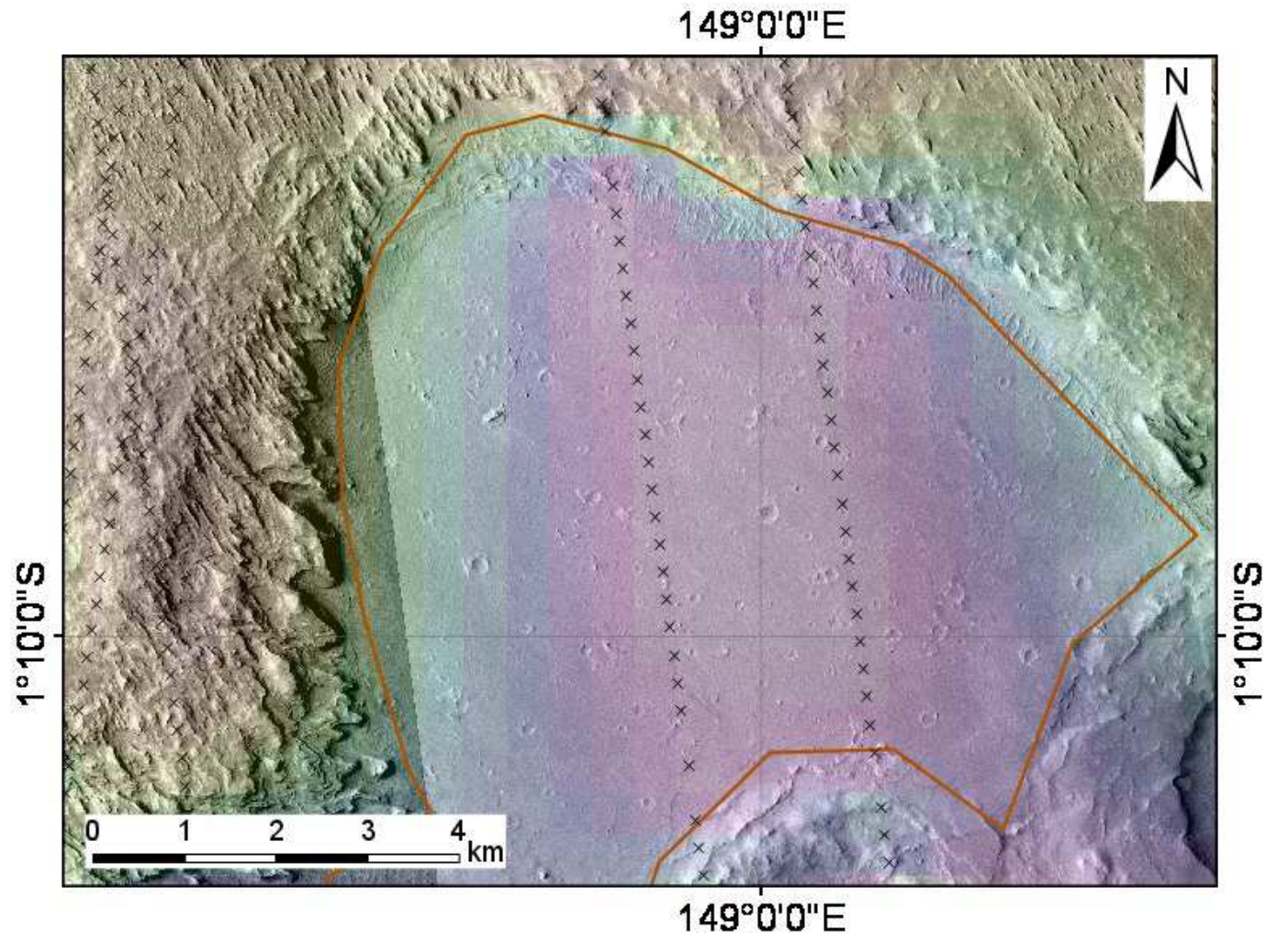

b)

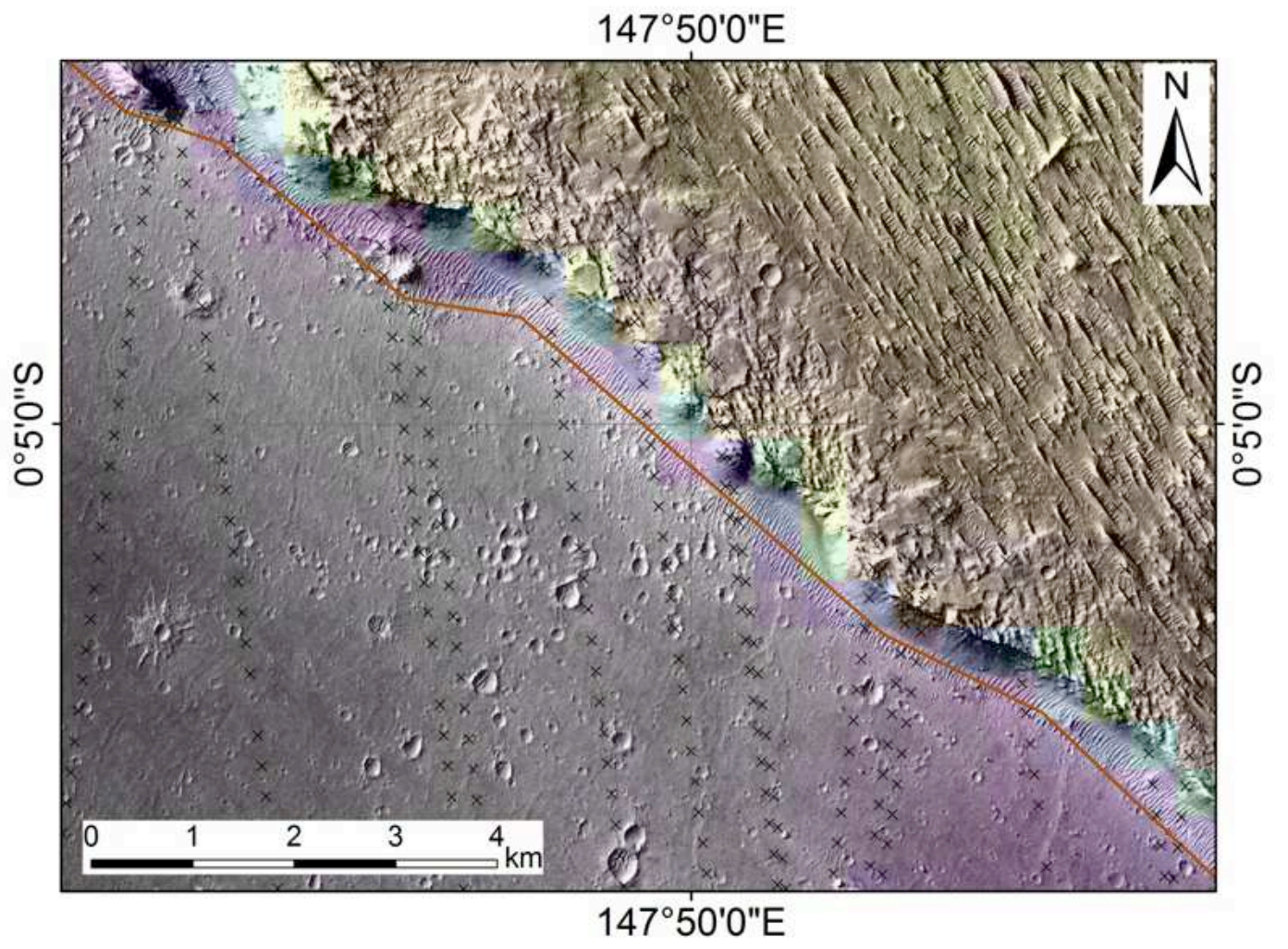

c)

**Figure S2.** (a) Southernmost inlier of B (low-lying, heavily cratered material). Boundary scarp is ~100m high. Lowermost floor elevation is ~-2760m (from PEDR). Note wide spacing of PEDR tracks (columns of black crosses). (b) Part of central large inlier of B. Boundary scarp is ~150 m high. Lowermost floor elevation is ~-2700m. (c) Part of northernmost inlier of B. Floor elevation is -2840m. Boundary scarp is ~200m high.

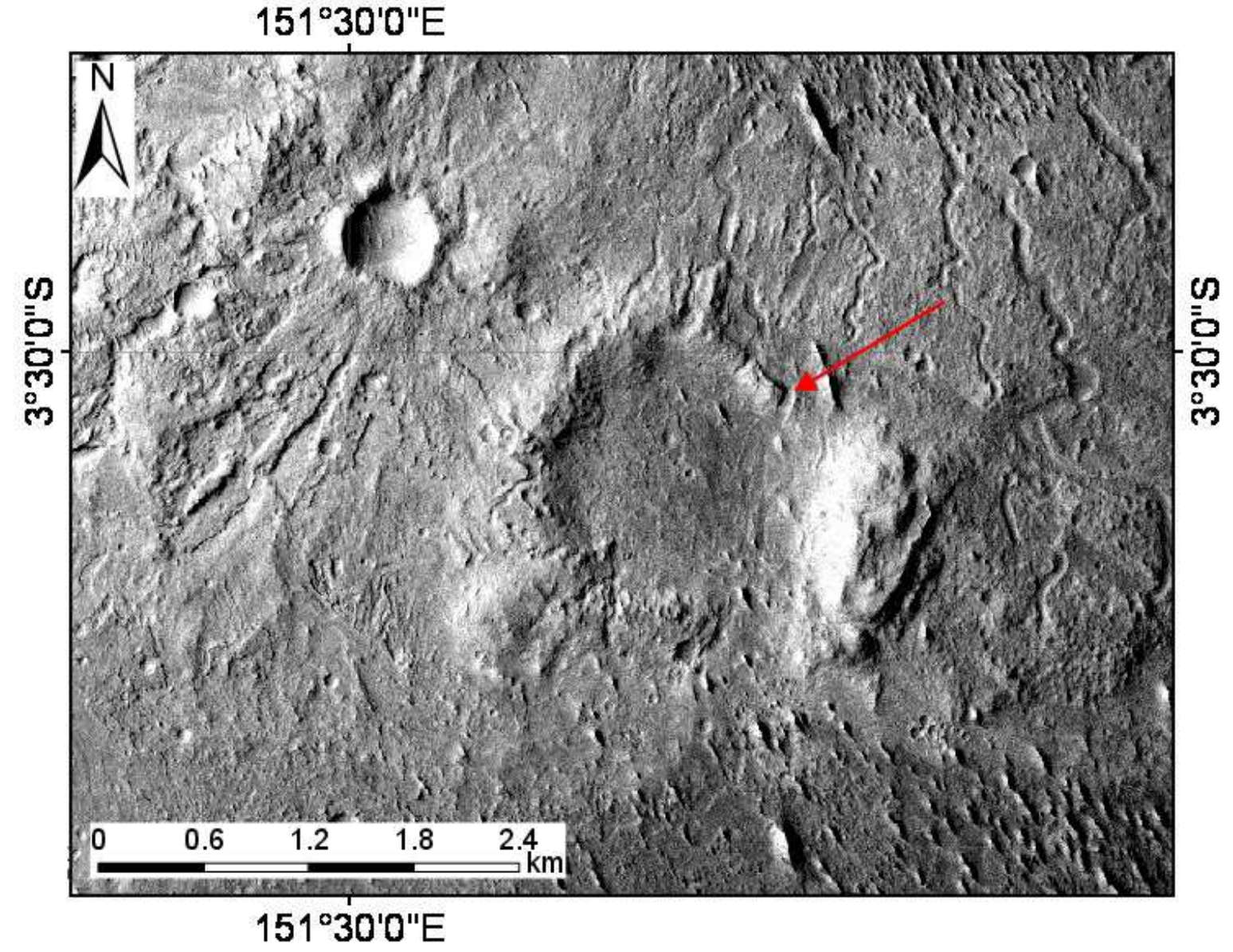

a)

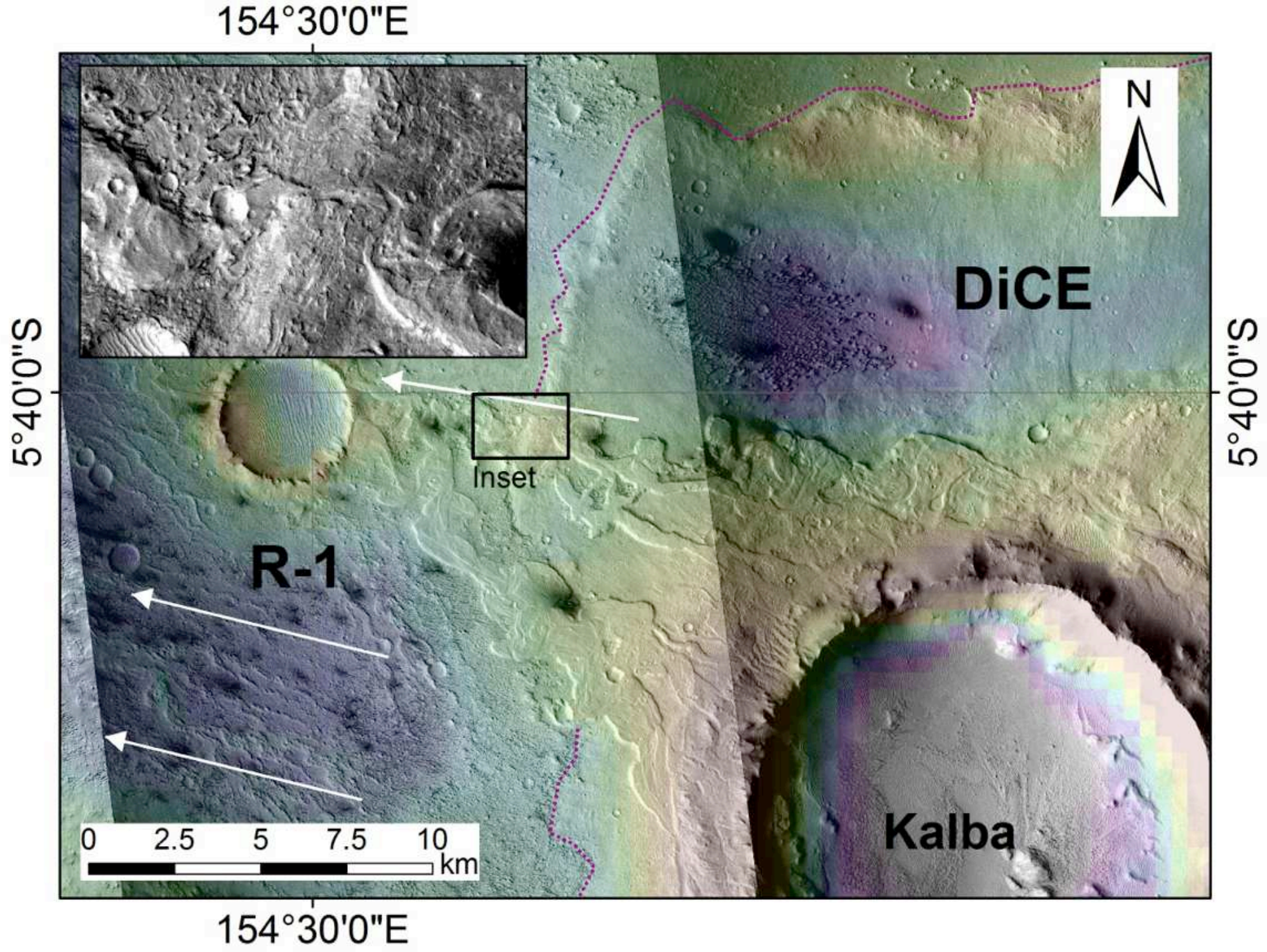

b)

**Figure S3.** (**a**) Channels crosscut a 1.5km-diameter crater formed in Neves ejecta, suggesting $\gtrsim 3 \times 10^8$ yr elapsed between the Neves impact and channel incision. Example shown by red arrow. These channels cannot result from localized runoff triggered by the Neves impact, because localized impact-triggered runoff shuts down much faster ($10^{-5}$-$10^3$ yr; Kite et al. 2011, Mangold et al. 2012) than the expected wait time for $\phi \geq 1$ km diameter impacts on the 270 $km^2$ area of remnant Neves ejecta (>$3 \times 10^8$ yr using a typical Hesperian impact flux; Michael 2013). There is also no evidence for point sources of surface water near Neves, such as volcanic vents or chaos terrain. Because a local trigger is unlikely, we infer that these channels record regional or global climate processes.Image is CTX P15_006894_1744_XI_05S208W. (**b**) Kalba's ejecta is onlapped to the W by fluvial deposits (Burr et al. 2009). White arrows show general trend of the rivers. Meander-belt deposits encircle the Kalba rim, less than 1km from the rim. One meandering river cut a valley in the distal rampart of Kalba ejecta (inset). (P06_003215_1752_XI_04S205W, B11_014014_1746_XI_05S205W).

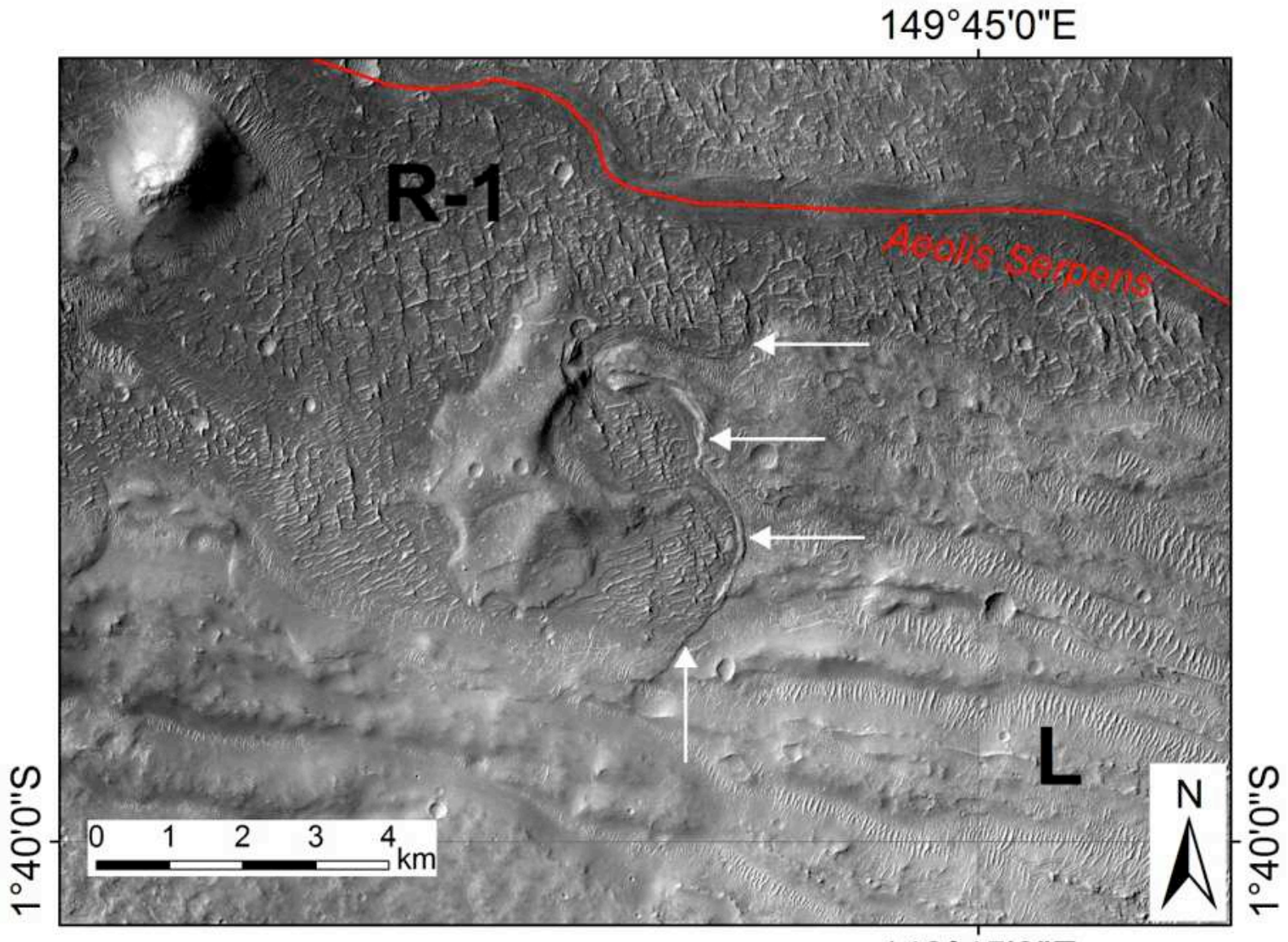

(a)

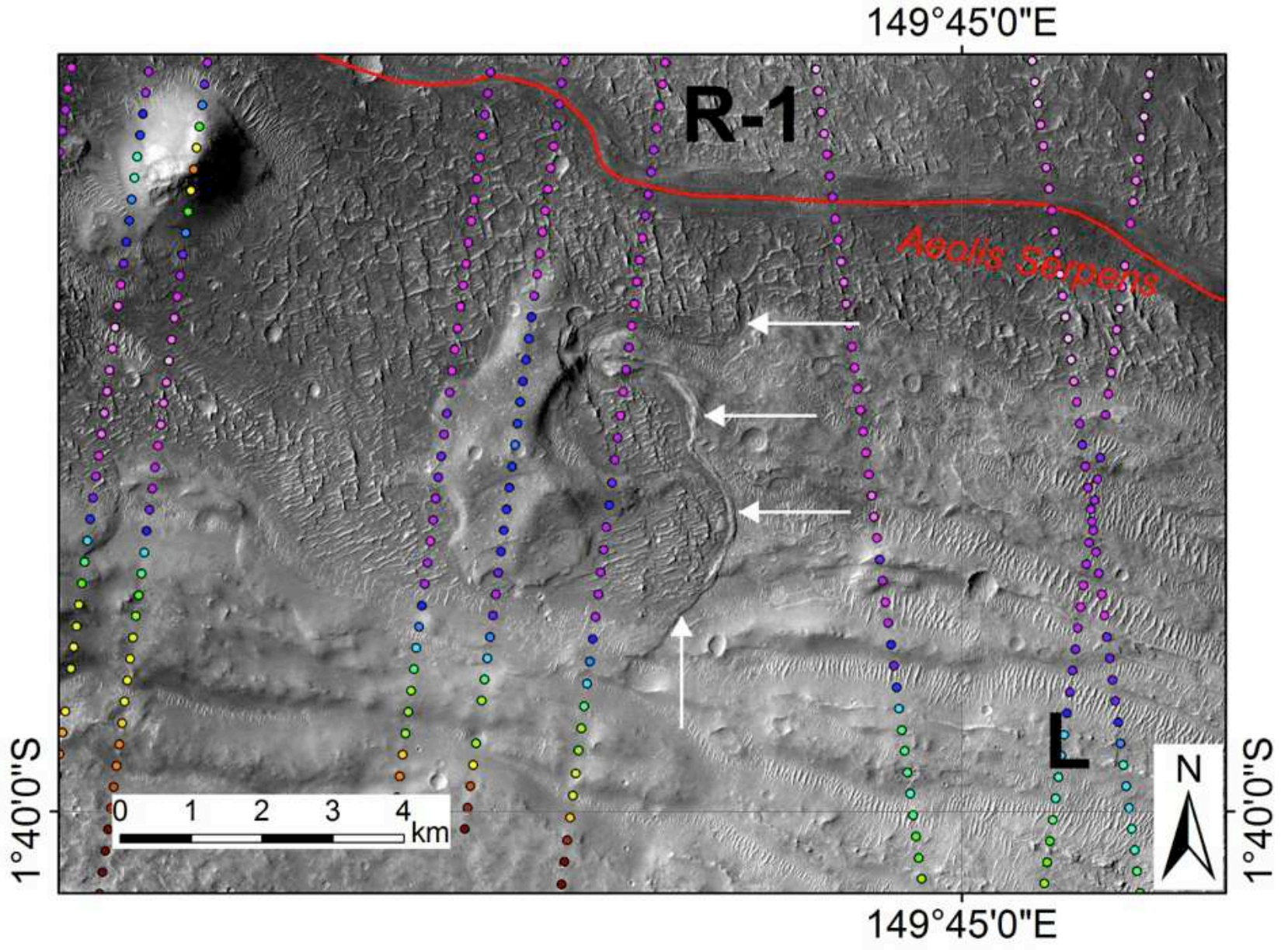

(b)

**Figure S4.** (a) shows contact between L and R-1, with R-1 embaying L, 70km to the SW of the contact shown in Fig. 11. P16_007461_1784_XI_01S210W; Fig. S6 shows context. Illumination from the W highlights locations where R-1 overrides adjacent L (white arrows). Red trace corresponds to Aeolis Serpens. (b) With MOLA PEDR altimetry data to highlight topographically highstanding outcrop of L embayed by R-1. Color range for MOLA PEDR spots is -2040m (red) to -2500m (light pink).

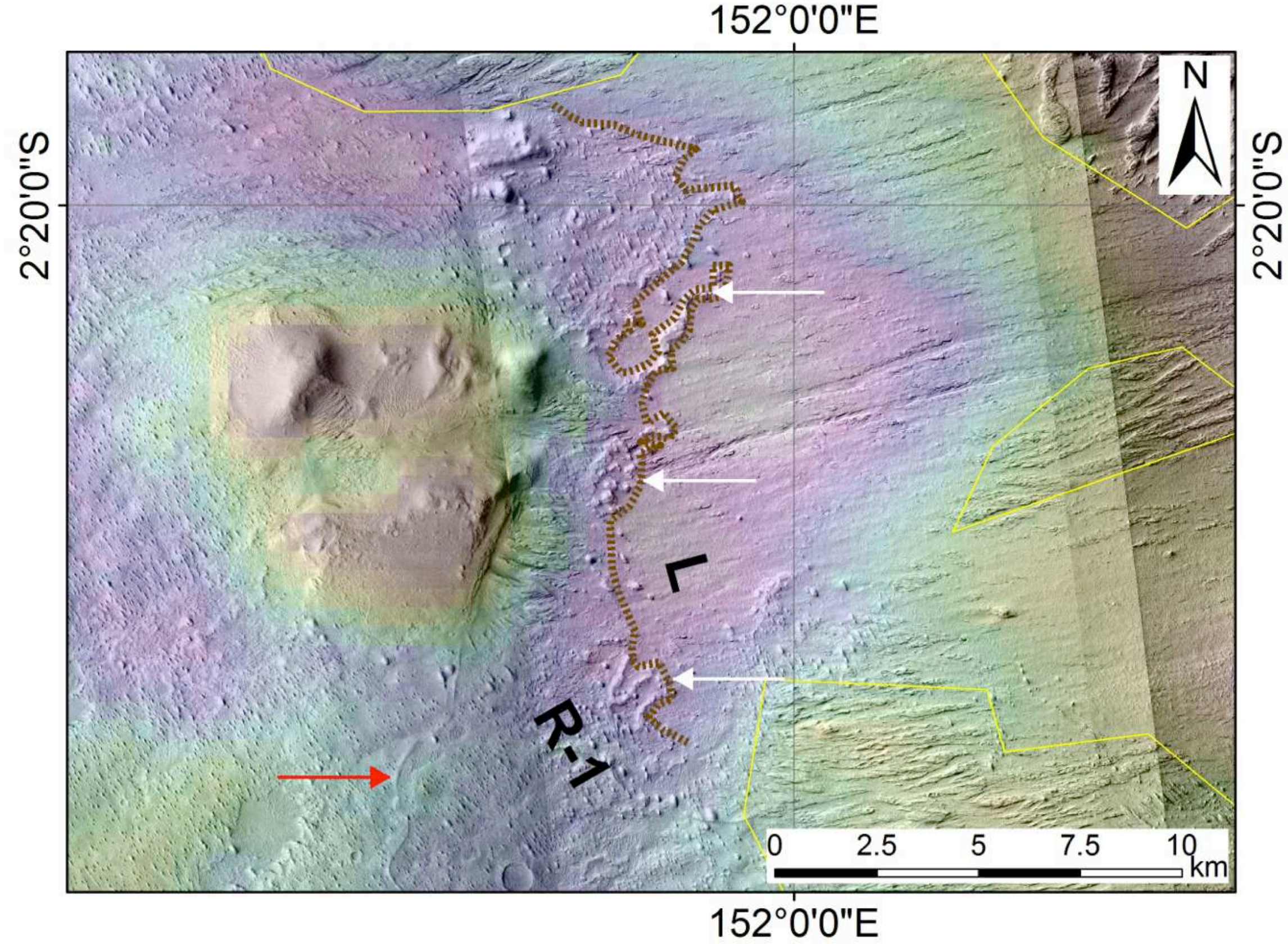

**Figure S5.** Showing a location where R-1 onlaps L (Fig. S6 shows context), 130 km to the SSE of the contact shown in Fig. 11. Brown dotted line shows contact between river-deposit-containing materials (R-1) to W and deeply grooved materials (L) to E. One river deposit is highlighted by red arrow. River-deposit-containing materials overlie (white arrows) the deeply grooved materials. The deeply grooved early-stage materials rise steadily to the E (~1° slope). River-channel-deposits within R-1 are erosionally resistant and are locally highstanding at the contact between R-1 and L. Yardang-forming materials (topography above solid yellow lines) drape both R-1 and L. Range of topography is from -2100m (red) to -2600m (white). Inspection of elevation data along MOLA tracks (not shown) confirms the onlap relationship described here.

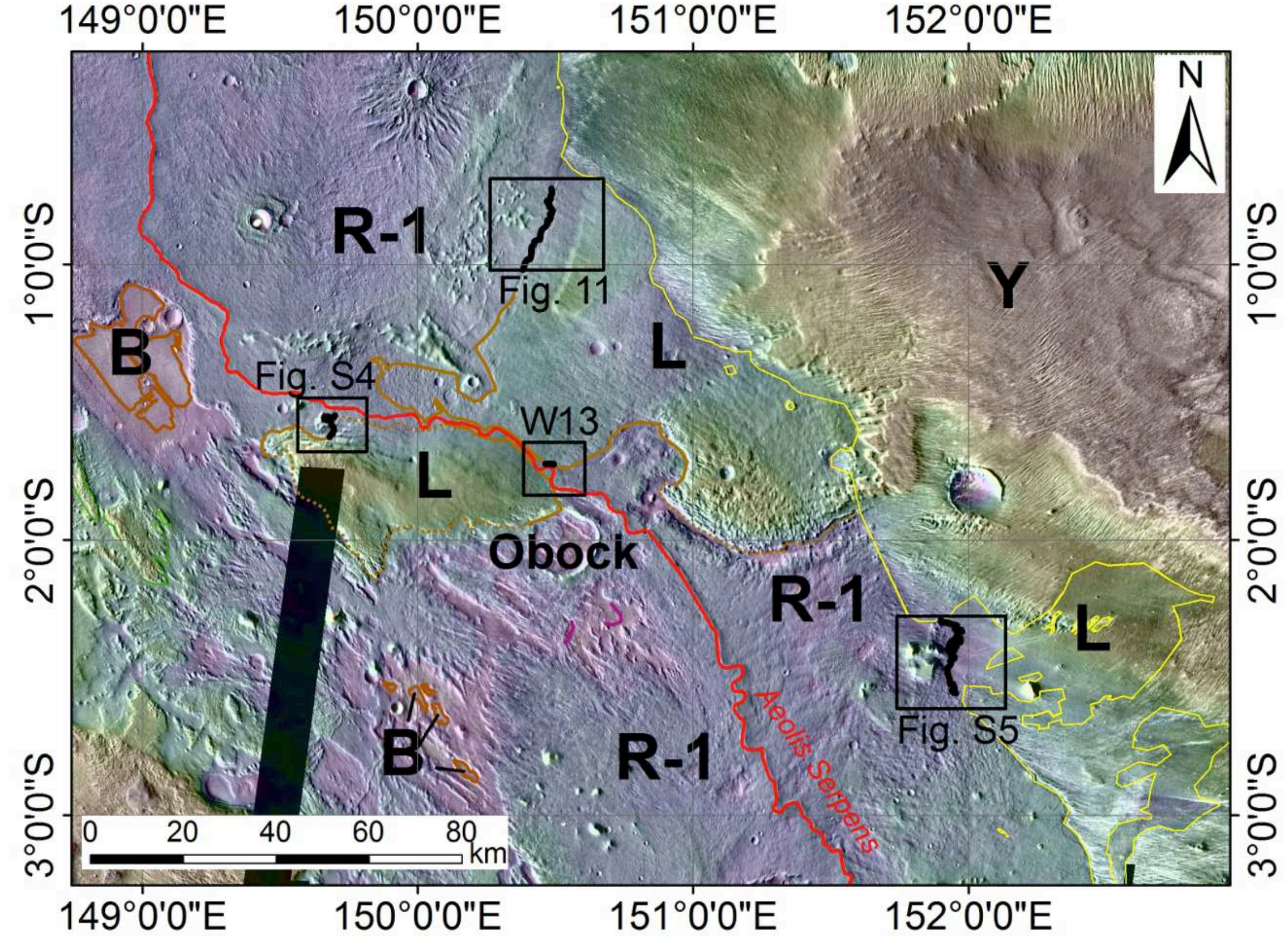

**Figure S6**. Context for locations where R-1 (river deposits) onlaps L (locally high-standing early sedimentary rocks). "W13" refers to area shown in Williams et al. (2013a), their Figure 15. L is generally higher-standing than R-1. However, at the highlighted locations, L outcrop is less resistant to late-stage erosion than the river-deposit-containing material, and it is topographically lower than the river-channel-containing material at the contact. We interpret this as an onlap relationship (Fig. 11). Yardang-forming materials (Y) unconformably drape both L and R-1.

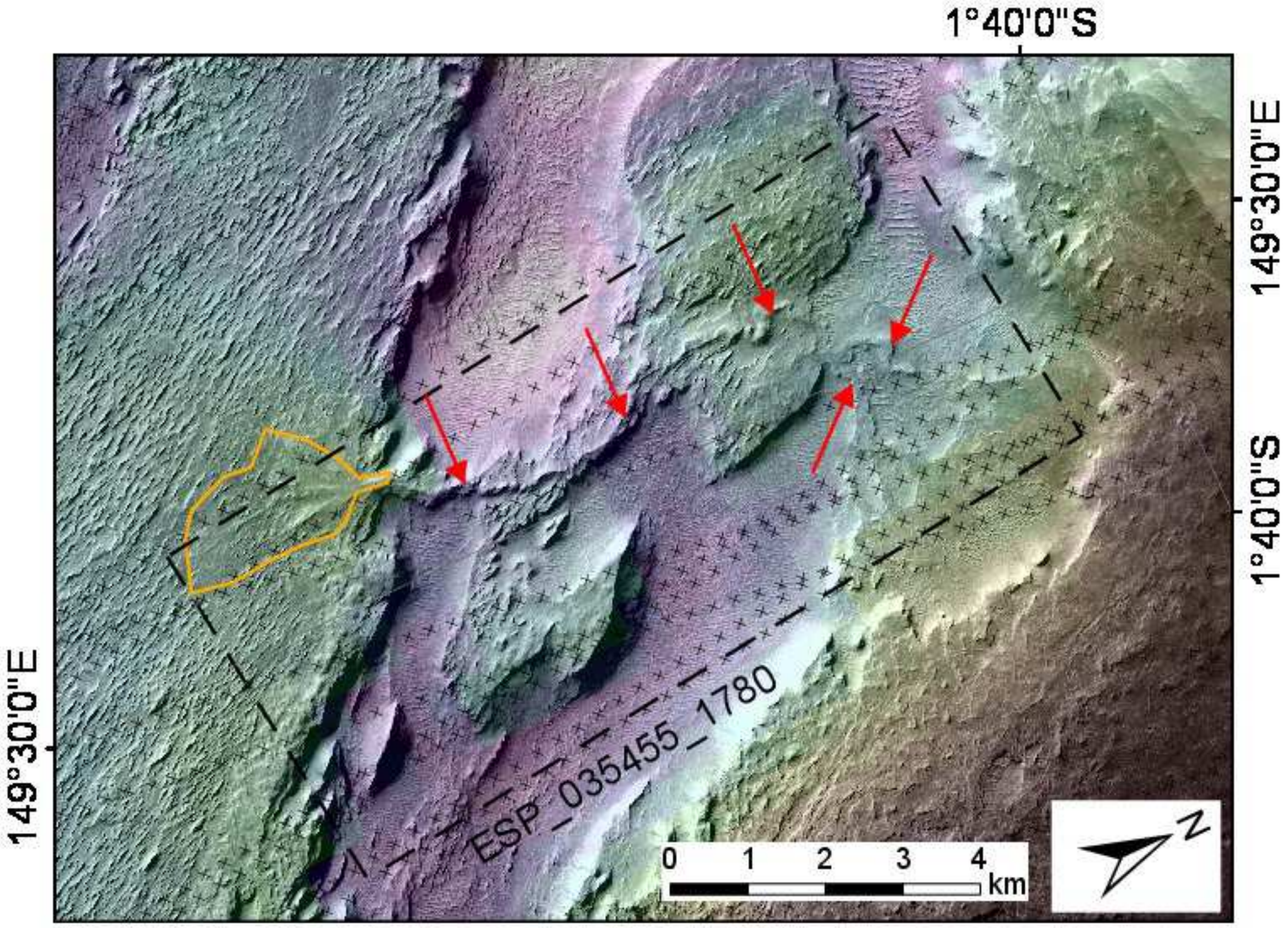

**Figure S7.** Source-to-sink sediment routing system preserved in inverted relief: erosion of L (high ground to E) to form an alluvial fan. Alcove formed by erosion of L contributes sediment to inverted channel (highlighted by red arrows) which terminates in alluvial fan (orange outline). Some elevations within alcove are now lower than the fan apex, which may correspond to undermining or differential compaction after the network was active. Black crosses show the location of MOLA PEDR elevation data, which form the basis for the MOLA gridded terrain (color ramp). P18_008107_1800_XI_00N210W; outline of HiRISE ESP_035455_1780 shown by dashed black line.

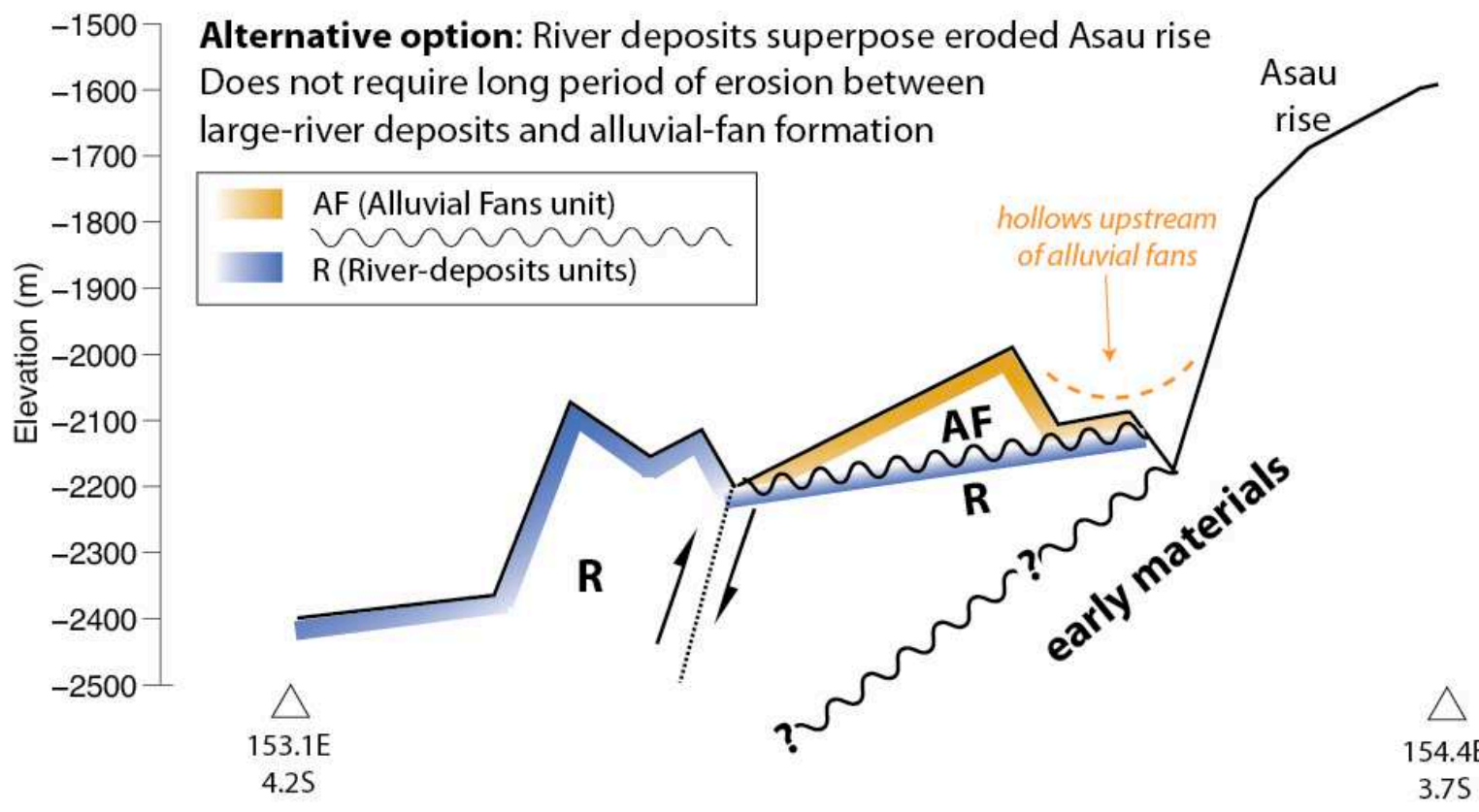

**Figure S8**. "Alternative option" for the sub-alluvial-fan stratigraphy corresponding to the cross-section on Fig. 4c and Fig. 17. "Preferred option" is shown in Fig. 18.

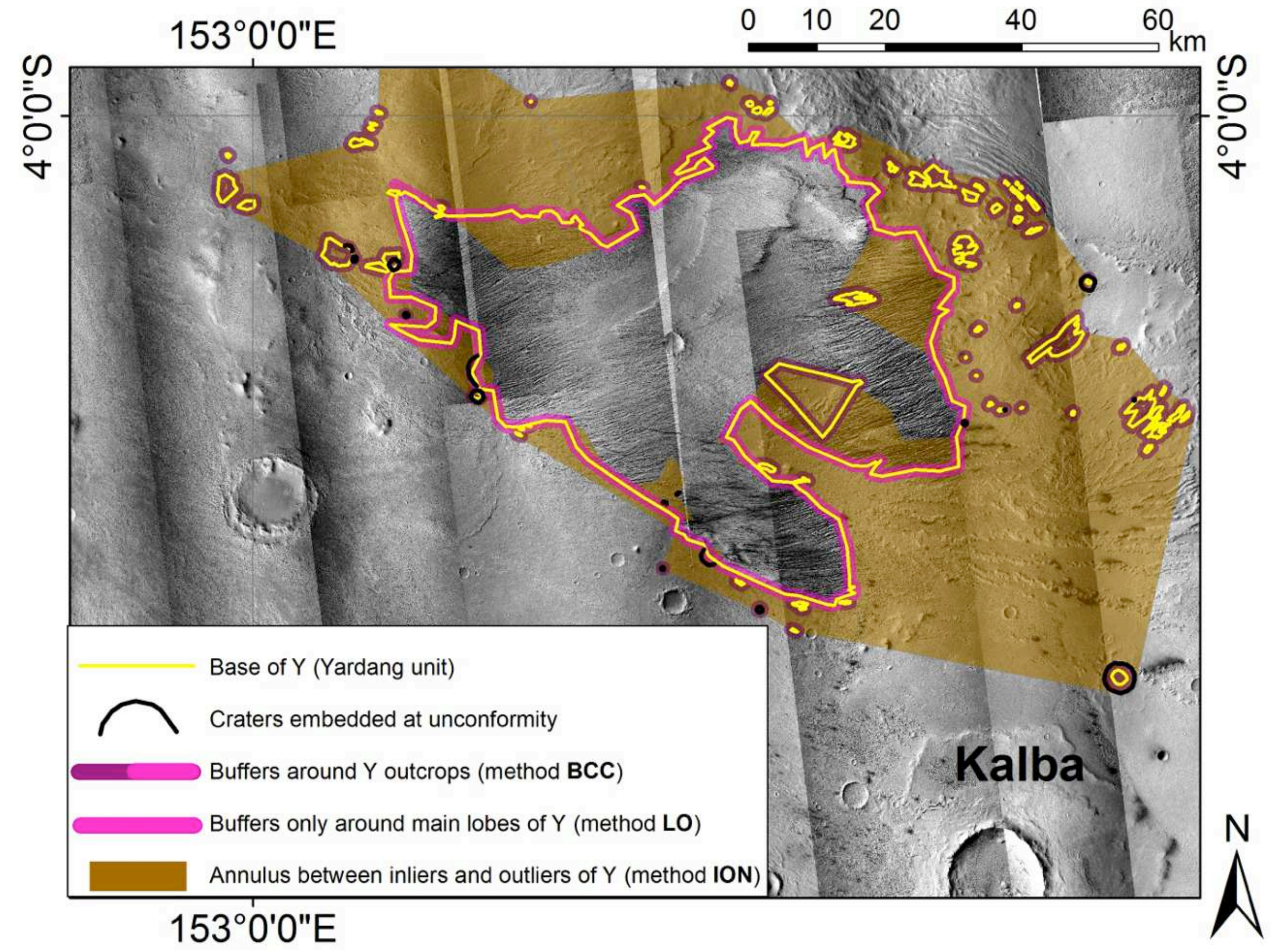

**Figure S9**. To show approaches to area normalization for measuring the time gap at the unconformity. A 1-km buffer is shown, although for each crater diameter we define a new buffer area (Fassett & Head 2008).

# Supplementary Tables

| Lat (°N) | Lon (°E) | Best-fit diam. (km) | Rock package underlying Y/Rock package IV at this location |
|---|---|---|---|
| -5.099 | 154.110 | 2.62 | II |
| -5.226 | 154.250 | 0.83 | II |
| -4.761 | 154.760 | 0.66 | II |
| -4.953 | 154.018 | 0.61 | II |
| -4.931 | 154.047 | 0.58 | II |
| -4.628 | 153.532 | 5.03 | II |
| -4.706 | 153.560 | 1.76 | II |
| -4.495 | 153.383 | 0.65 | II |
| -4.355 | 153.256 | 0.77 | II |
| -4.369 | 153.361 | 1.54 | II |
| -3.377 | 152.408 | 1.42 | II |
| -3.680 | 152.287 | 0.53 | II |
| -3.613 | 152.973 | 0.36 | II |
| -3.685 | 153.238 | 0.33 | II |
| -0.363 | 150.512 | 0.79 | II |
| 0.974 | 149.901 | 1.91 | I |
| 1.093 | 149.734 | 1.40 | I |
| 1.189 | 149.544 | 1.26 | I |
| 0.975 | 150.985 | 0.72 | I |
| 1.345 | 150.711 | 7.85 | I |
| 1.004 | 150.555 | 3.10 | I |
| -1.401 | 151.142 | 2.19 | I |
| -1.742 | 151.985 | 11.01 | I |
| -1.522 | 151.364 | 1.98 | I |
| -1.443 | 151.377 | 0.37 | I |
| -1.759 | 151.525 | 7.36 | I |
| -1.819 | 151.437 | 1.11 | I |
| -4.429 | 155.064 | 2.12 | II |
| -4.702 | 155.175 | 0.21 | II |
| -2.846 | 153.915 | 1.00 | II |
| -3.564 | 153.443 | 0.36 | II |
| -0.133 | 150.515 | 0.18 | II |
| -0.090 | 150.485 | 0.27 | II |
| 0.994 | 150.933 | 0.25 | I |
| 1.281 | 150.588 | 0.74 | I |
| -4.321 | 153.240 | 1.39 | II |
| -4.725 | 154.854 | 0.20 | III |
| -5.120 | 154.011 | 0.37 | II |
| -3.427 | 151.908 | 10.05 | II |
| 1.358 | 151.452 | 2.66 | I |
| 1.231 | 151.289 | 1.89 | I |
| -0.093 | 150.159 | 0.75 | II |
| -5.420 | 155.154 | 4.44 | II |
| 1.334 | 151.860 | 0.98 | I |

**Table S1.** List of impact craters that must have formed after Rock Package I, but before Rock Package IV, based on superposition relationships (i.e., craters that are currently embedded at the sub-Y unconformity.)